\documentclass[twocolumns,traditabstract]{aa}
\usepackage{graphicx}
\usepackage{txfonts}
\usepackage{natbib}
\usepackage{enumerate}

\def\gsimeq  
{\hbox{\raise0.5ex\hbox{$>\lower1.06ex\hbox{$\kern-1.07em{\sim}$}$}}}
\def\lsimeq
{\hbox{\raise0.5ex\hbox{$<\lower1.06ex\hbox{$\kern-1.07em{\sim}$}$}}}

\begin{document}

   \title{Anatomy of the AGN in NGC~5548}
   \subtitle{VIII. \emph{XMM-Newton}'s EPIC detailed view of an unexpected variable multilayer absorber}
   
   \author{ 
   M. Cappi \inst{1}
\and   
   B. De Marco\inst{2}
\and      
   G. Ponti\inst{2,1}
\and      
   F. Ursini \inst{3,4}
\and   
   P.-O. Petrucci \inst{4,5}
\and   
   S. Bianchi \inst{3}
\and   
   J.S. Kaastra \inst{6,7,8}
\and   
   G.A. Kriss \inst{9,10}
\and   
   M. Mehdipour \inst{6,11}
\and   
   M. Whewell \inst{11}
\and   
   N. Arav \inst{12}
\and   
   E. Behar \inst{13}
\and   
   R. Boissay \inst{14}
\and   
   G. Branduardi-Raymont \inst{11}
\and   
   E. Costantini \inst{8}
\and   
   J. Ebrero \inst{15}
\and   
   L. Di Gesu \inst{8}
\and   
   F.A. Harrison \inst{16}
\and   
   S. Kaspi \inst{13}
\and   
   G. Matt \inst{3}
\and   
   S. Paltani \inst{14}
\and   
   B.M. Peterson \inst{17,18}
%   \and   
%   F. Pozo Nu\~{n}ez \inst{21},
%   \and   
%   A. De Rosa \inst{22}
\and   
   K.C. Steenbrugge \inst{19}
%    \and   
%   C.P. de Vries \inst{8}
\and   
   D.J. Walton \inst{16,20}
  }

   \offprints{M. Cappi \\ \email{massimo.cappi@inaf.it}}
   
   \institute{
INAF-IASF Bologna, Via Gobetti 101, I-40129 Bologna, Italy 
\and
Max-Planck-Institut f\"{u}r extraterrestrische Physik, Giessenbachstrasse, D-85748 Garching, Germany
\and
Dipartimento di Matematica e Fisica, Universit\`{a} degli Studi Roma Tre, via della Vasca Navale 84, 00146 Roma, Italy
\and
Univ. Grenoble Alpes, IPAG, F-38000 Grenoble, France
\and
CNRS, IPAG, F-38000 Grenoble, France
\and
SRON Netherlands Institute for Space Research, Sorbonnelaan 2, 3584 CA Utrecht, the Netherlands
\and
Department of Physics and Astronomy, Universiteit Utrecht, P.O. Box 80000, 3508 TA Utrecht, the Netherlands
\and
Leiden Observatory, Leiden University, PO Box 9513, 2300 RA Leiden, the Netherlands
\and
Space Telescope Science Institute, 3700 San Martin Drive, Baltimore, MD 21218, USA
\and
Department of Physics and Astronomy, The Johns Hopkins University, Baltimore, MD 21218, USA
\and
Mullard Space Science Laboratory, University College London, Holmbury St. Mary, Dorking, Surrey, RH5 6NT, UK
\and
Department of Physics, Virginia Tech, Blacksburg, VA 24061, USA
\and
Department of Physics, Technion-Israel Institute of Technology, 32000 Haifa, Israel
\and
Department of Astronomy, University of Geneva, 16 Ch. d'Ecogia, 1290 Versoix, Switzerland
\and
European Space Astronomy Centre, P.O. Box 78, E-28691 Villanueva de la Ca\~{n}ada, Madrid, Spain
\and
Cahill Center for Astronomy and Astrophysics, California Institute of Technology, Pasadena, CA 91125, USA
\and
Department of Astronomy, The Ohio State University, 140 W 18th Avenue, Columbus, OH 43210, USA
\and
Center for Cosmology \& AstroParticle Physics, The Ohio State University, 191 West Woodruff Ave., Columbus, OH 43210, USA
%\and
%INAF/IAPS - Via Fosso del Cavaliere 100, I-00133 Roma, Italy
\and
Instituto de Astronom\'{i}a, Universidad Cat\'{o}lica del Norte, Avenida Angamos 0610, Casilla 1280, Antofagasta, Chile
\and
%%Department of Physics, University of Oxford, Keble Road, Oxford, OX1 3RH, UK
%%\and
%%\and
%%Astronomisches Institut, Ruhr-Universit\"{a}t Bochum, Universit\"{a}tsstra\ss e 150, 44801, Bochum, Germany
%\and
Jet Propulsion Laboratory, California Institute of Technology, 4800 Oak Grove Drive, Pasadena, CA 91109, USA
}

\abstract{
%{\it Context.}
In 2013 we conducted a large multi-wavelength campaign on the archetypical Seyfert 1 galaxy NGC~5548. Unexpectedly, this usually unobscured source appeared strongly absorbed 
in the soft X-rays during the entire campaign, and signatures of new and strong outflows were present in the almost simultaneous UV HST/COS data. 
%\pn
%{\it Aims.}
%We aim here at modeling the new obscurer seen along the line of sight (LOS) of this otherwise archetypical type-1 (and usually unobscured) source.
%\pn
%{\it Methods.} 
Here we carry out a comprehensive spectral analysis of all available \emph{XMM-Newton} observations of NGC~5548 (precisely 14 observations from our campaign plus 3 from the archive, for a total of $\sim$763 ks) 
in combination with three simultaneous \emph{NuSTAR} observations.
%, which are instrumental to best understand and constrain the underlying intrinsic 
%continuum and the reflection component seen in the broad-band data. 
%We aim at modeling and characterizing the short (mostly $\sim$ks-to-days) timescale spectral variability properties of the new obscurer seen along the line of sight (LOS) 
%of this usually unobscured source.
We obtain a best-fit underlying continuum model composed by i) a weakly varying flat ($\Gamma$$\sim$1.5-1.7) power-law component; ii) a constant, cold reflection (FeK + continuum) component; 
iii) a soft excess, possibly due to thermal Comptonization; and iv) a constant, ionized scattered emission-line dominated component.
%; v) a de-ionized multi-temperature warm absorber; and vi) a multilayer obscurer at mild-to-neutral ionization state. 
Our main findings are that, during the 2013 campaign, the first three of these components appear to be partially covered by a heavy and variable obscurer located along the line of sight (LOS)
that is consistent with a multilayer of cold and mildly ionized gas.
%, and could also be responsible for the broad absorption line outflowing components found in the UV. 
We characterize in detail the short timescale (mostly $\sim$ks-to-days) spectral variability of this new obscurer, and find it is mostly due to a combination of column density and covering factor variations, on top of 
intrinsic power-law (flux and slope) variations. 
In addition, our best-fit spectrum is left with several (but marginal) absorption features at rest-frame energies $\sim$6.7-6.9~keV and $\sim$8~keV, as well as a weak broad emission line feature redwards of the 6.4 keV emission line. 
These could indicate a more complex underlying model, e.g. a P-Cygni-type emission profile if allowing for a large velocity and wide-angle outflow.
%which are consistent with being produced by H-like iron K$\alpha$ and K$\beta$ shell absorptions associated with an outflow with either a small, or a mildly relativistic velocity of $\sim$0.13~c. 
%Finally we also find marginal evidence for a weak emission line broad feature red ward of the 6.4 keV emission line, possibly indicating an additional reflection component, but relativistic, 
%or P-Cygni type of emission from the outflow itself.
%{\it Conclusions.}
These findings are consistent with a picture where the obscurer represents the manifestation along the LOS of a multilayer of gas, also in multiphase, that is likely outflowing at high speed, and simultaneously producing 
heavy obscuration and scattering in the X-rays, and broad absorption features in the UV. 
}
%Unfortunately, we find no obvious clue 
%in the source properties that could help understand the physical mechanism which activated such an outflowing wind.
%Unfortunately, the pn data quality does not allow us to 
%directly measure the velocity of this outflowing component(s), nor is it detected in the RGS instrument, thus we are unable to verify its (plausible) association with the 
%large velocity (1000-5000 km/s) found for the new UV BAL.}
%
%\item A best-fit is found, first based on the 3 \emph{XMM-Newton}  + \emph{NuSTAR} observations, and then applied to all 17 \emph{XMM-Newton}  observations which includes: 
%i) a power law continuum with $\Gamma$ $\sim$ 1.5-1.7; ii) a cold and constant reflection component that produces a narrow Fe K emission line; iii)  a soft-excess modeled by thermal Comptonization and contributing only below $\sim$ 1 keV; iv) a scattered emission line component dominating the RGS energy band; v) a constant, multi temperature warm absorber, consistent with its historical values, after accounting for de-ionization due to the obscurer itself; and vi) a multilayer obscurer at mild-to-neutral ionization state, and with at least two components which partially cover the source.


\authorrunning{M. Cappi et al.}

\titlerunning{X. An unexpected, heavy, multilayer and variable absorber in NGC~5548}

\maketitle

%\date{Received / Accepted}

\section{Introduction}

%Understanding the properties of the accretion and ejection flows in AGNs remains one of the greatest challenges of modern astronomy. 
%Key outstanding questions include: What are the properties and geometry of the accretion disk, the corona, the disk jet, the disk winds, the broad and narrow line regions, the molecular torus? 
%Are inflows related to outflows, and how? Do AGN outflows have a relevant impact on how galaxies form stars and assemble?

Unified models of active galactic nuclei (AGN) have ``historically" been proposed as a static unifying view based on a putative dusty molecular torus where type 1 AGN are seen face-on, thereby typically unabsorbed and bright, while type 2 sources are seen edge-on, thus typically obscured and faint (Antonucci \& Miller 1985).
However in recent years, an ever increasing number of type 1 Seyfert galaxies/AGN (up to 20-30\% of limited samples) have shown a significant amount of absorption clearly at odds with the source's optical classifications (Bassani et al. 1999; Cappi et al. 2006; 
Panessa et al. 2009; Merloni et al. 2014). Such complex, often variable, absorbing structures seem to call for a revision of the static version of Unified models into a more complex, maybe dynamical, structure 
(e.g. Murray et al. 1995; Elvis 2000; Proga 2007).

On the one hand, ionized absorbers (so-called warm absorbers, WA) are nowadays routinely observed as blue-shifted narrow and broadened absorption lines in the UV and X-ray spectra of a substantial (certainly greater than 30\%) fraction of AGN and quasars (e.g. Blustin et al. 2005; Piconcelli et al. 2005; McKernan et al. 2007; Ganguly \& Brotherton 2008). These absorption systems span a wide range of velocities and physical conditions (distance, density, ionization state), and may have their origin in an AGN-driven wind, sweeping up the interstellar medium, or thermally driven from the molecular torus (Blustin et al 2005) and outflowing at hundreds to few thousands km/s (see Crenshaw, Kraemer \& George 2003 and Costantini 2010 for reviews on the subject). Even more powerful outflows (the so-called ultra fast outflows, UFOs), so highly ionized that their only bound transitions left are for Hydrogen- and Helium-like iron detectable only at X-ray energies, seem to be present in 30-40\% of local radio-quiet and radio-loud AGN, with outflow speeds of up to $\sim$0.3c (Pounds et al. 2003a; Tombesi et al. 2010, 2013, 2014; Gofford et al. 2013). 
Both WAs and UFOs phenomena have been seen to vary on both short (hours-days) and long (months-years) time-scales (Cappi 2006) for various plausible reasons such as variations in either the photoionization balance, or the absorption column density and/or covering fraction (Risaliti et al. 2005; Reeves et al. 2014).

On the other hand, there is also mounting evidence for the presence of large columns of {\it additional neutral or mildly ionized gas} along the LOS to not only type 2 Seyfert galaxies (e.g. Turner et al. 1997, 1998) but also type 1 and intermediate Seyferts 
(e.g., Malizia et al. 1997, Pounds et al. 2004; Miller et al. 2007; Bianchi et al. 2009; Turner et al. 2009; Risaliti et al. 2010; Lobban et al. 2011; Marchese et al. 2012; Longinotti et al. 2009, 2013; Reeves et al., 2013; Walton et al. 2014; Miniutti et al. 2014; Rivers et al. 2015).
In some cases, the columns are so large that astronomers have called these sources ``changing look" (i.e. sources changing from being absorbed by a Compton thin to a Compton thick column density).
From the pioneering works of Risaliti et al. (Risaliti, Elvis \& Nicastro, 2002; Risaliti et al. 2005) on a few interesting sources such as the Seyfert 1 NGC 1365, to detailed studies on an increasing number of sources and on systematic studies of larger samples (Markowitz, Krumpe \& Nikutta 2014; Torricelli-Ciamponi et al. 2014; Tatum et al. 2013), evidence has accumulated for the importance of large, complex, and variable {\it neutral} absorption also in type 1 Seyfert galaxies. Current interpretations for this cold circumnuclear gas are clumpy molecular tori (Krolik \& Begelman 1988, Markowitz et al. 2014; H{\"o}nig 2013),  
absorption from inner and/or outer BLR clouds (Risaliti et al. 2009), or accretion disc outflows (Elitzur \& Shlosman '06; Proga 2007; Sim et al. 2008). 

NGC 5548 (z=0.017) is one of the X-ray brightest (2--10~keV flux of $\sim$2--5$\times10^{-11}$~erg~cm$^{-2}$~s$^{-1}$), and
most luminous (L$_{2-10{\rm keV}}$ $\sim 1-3 \times 10^{44}$ erg/s) Seyfert 1 galaxies known. As such, it is one of the best-studied example of a Seyfert 1 galaxy and has been observed by all major satellites since it entered the Ariel V high galactic latitude source catalog(Cooke et al. 1978). NGC 5548 exhibits all the typical components seen in type 1 Seyfert galaxies, that is: 
a steep ($\Gamma \sim$1.8-1.9) power-law spectrum plus a reflection component with associated FeK line, a soft-excess emerging below a few keV, 
evidence for warm ionized gas along the LOS, and typical soft X-ray variability as shown in the CAIXAvar sample by Ponti et al. (2012a). Recently, detailed studies of either the time-average or variability properties of the WA, 
the reflection component, and the soft-excess component have been made using either low- (CCD-type) or higher-(grating-type) energy resolution instrumentation available from \emph{Chandra}, \emph{XMM-Newton}
and \emph{Suzaku} (Pounds et al. 2003; Steenbrugge et al., 2003, 2005; Crenshaw et al., 2003; Andrade-Velazquez et al. 2010; Krongold et al. 2010; Liu et al. 2010; Brenneman et al. 2012; McHardy et al. 2014).

In 2013, our team conducted a long multi-satellite observing campaign on NGC 5548. The campaign is introduced in detail in Mehdipour et al. (2015) (hereafter Paper I). The main results from the campaign, 
the simultaneous appearance of an exceptional obscuring event in the X-rays and a broad absorption line (BAL) structure in the UV, are presented in Kaastra et al. (2014) (hereafter Paper 0). 
The shadowing effects of this UV BAL and X-ray obscurer on the larger-scale ``historical"  UV narrow absorption lines and WA are presented in Arav et al. 2014 (hereafter Paper II). 
The high energy properties of the source during, and before, the campaign are shown in Ursini et al. (2015) (hereafter Paper III), while the study of a short-timescale flaring event that happened
 in September 2013 and which triggered a \emph{Chandra} LETG observation is addressed in Di Gesu et al. (2015) (hereafter Paper IV), together with the analysis of the WA short-term variability during this event.
 The time-averaged soft X-ray line-dominated RGS spectrum is presented and modelled in detail in Whewell et al. (2015) (hereafter Paper V).  
 For the longer-term timescales, the historical source behavior is presented in Ebrero et al. (2016) (hereafter Paper VI), while the full {\it Swift} multi-year monitoring is presented in 
 Mehdipour et al (2016) (hereafter Paper VII).
% and compared to the other XMM-Newton observations.

Here, we focus on the detailed model characterization of the obscurer, and of its spectral variability, by making use of all available EPIC (pn+MOS) spectra (during and before the campaign) 
and simultaneous \emph{NuSTAR} observations (Sect. 2). We also incorporate in our modeling the results from the papers mentioned above obtained during the campaign and which were based 
on other instruments and satellites, such as RGS spectra, HST/COS spectra, and {\it Swift} long-term light-curves. This allows us to draw a comprehensive and self-consistent understanding of 
the obscurer properties (Sect. 4). We then attempt to place constraints on the location and physical origin of the various components required by our model (Sect. 5).
Values of H$_0$=70 km s$^{-1}$ Mpc$^{-1}$, and $\Omega$$_{\Lambda}$=0.73 are assumed throughout, and errors are quoted at the 90\% confidence ($\Delta \chi^2$ = 2.71) 
for 1 parameter of interest, unless otherwise stated.

\section{Observations and Data Reduction}

NGC~5548 was observed by the \emph{XMM-Newton} EPIC instruments on 17 separate 
occasions, for a total cleaned exposure time of $\sim$763~ks (see Table 1). 
The first three (archival) observations were performed once in December 2000 and twice in July 2001.
The remaining 14 observations were all part of the 2013 multi wavelength campaign (Paper 0 and Paper I).
12 observations were performed during the summer 2013, one five months later, in December 2013, and the last one in February 2014. 
The spacing was intended so in order to sample different timescales in a rough logarithmic spacing, i.e. multiple on short 
(days) timescales, and more sparse on longer (months) timescales. A full description of the numerous observations 
performed during the campaign, from space and ground observatories, is given in Paper I.

%\begin{onecolumn}
\begin{table*}[!]
\caption{Summary of NGC~5548 \emph{XMM-Newton} (pn) and \emph{NuSTAR} observations}
\label{tab:1}     
\begin{center}
%\centering          
\begin{tabular}{l l r c c c c}
\hline\hline             
 Obs. & Obs. ID & Start--End Date$^a$ & Exposure$^b$ & F$_{(0.5-2)}^c$ &  F$_{(2-10)}^d$ & F$_{(10-30)}^e$ \\
\hline                  
\multicolumn{7}{l}{\it Archival Data (2000-2001)}\\
 A1 & 0109960101 & 2000 Dec 24--25 & 16.1 & 17.2 & 32.8 & - \\

 A2 & 0089960301 & 2001 Jul 09--10 & 56.7 &  20.9 & 40.3 & -\\

 A3 & 0089960401 & 2001 Jul 12--12 &  18.9 & 28.7 & 50.8 & -\\
\hline
\multicolumn{7}{l}{\it Multiwavelength Campaign (2013)}\\

 M1& 0720110301 & 2013 Jun 22--22 & 35.4 & 1.14 & 15.3& -\\
 
 M2 & 0720110401 & 2013 Jun 30--30& 38.1 & 3.59 & 33.0& -\\

 M3 & 0720110501 & 2013 Jul 07--08 & 38.9 & 2.16 & 23.9& -\\
 
 M4N  & 0720110601 & 2013 Jul 11--12 & 37.4 & 3.66 & 35.8& -\\
 \dotfill    & 60002044002$/$3$^*$ & 2013 Jul 11--12 & 51.4 & 3.66 & 35.8 & 51.0\\
 
M5 & 0720110701 & 2013 Jul 15--16 & 37.5 & 2.62 & 29.9& -\\
   
 M6 & 0720110801 & 2013 Jul 19--20 & 37.1 & 2.38 & 30.5& -\\
    
M7 & 0720110901 & 2013 Jul 21--22 & 38.9 & 2.28 & 26.3& -\\
     
M8N  & 0720111001 & 2013 Jul 23--24 & 37.5 & 2.21 & 27.9& -\\
 \dotfill & 60002044005$^*$ & 2013 Jul 23--24 & 49.5 & 2.21 & 27.9 & 45.0\\
      
M9 & 0720111101 & 2013 Jul 25--26 & 32.1 & 3.14 & 33.2 & -\\

M10 & 0720111201 & 2013 Jul 27--28 & 38.9 & 3.01 & 32.4& -\\

M11 & 0720111301 & 2013 Jul 29--30 & 35.3 & 2.76& 29.6& -\\

M12 & 0720111401 & 2013 Jul 31--Aug 01 & 36.1 & 2.28 & 26.2& 42.8\\

M13N & 0720111501 & 2013 Dec 20--21 & 38.2 & 2.08 & 24.9& -\\
 \dotfill & 60002044008$^*$ & 2013 Dec 20--21 & 50.1 & 2.08 & 24.9 & -\\

M14 & 0720111601 & 2014 Feb 04--05 & 38.8 & 3.96 & 27.2 & -\\
\hline     

 \multicolumn{7}{l}{$^a$ Observation Start--End Dates}\\
 \multicolumn{7}{l}{$^b$ Net exposure time, after corrections for screening and deadtime, in ks.}\\
 \multicolumn{7}{l}{$^c$ Observed flux in the 0.5-2 keV band, in units of 10$^{-12}$ erg cm$^{-2}$ s$^{-1}$}\\
 \multicolumn{7}{l}{$^d$ Observed flux in the 2-10 keV band, in units of 10$^{-12}$ erg cm$^{-2}$ s$^{-1}$}\\
 \multicolumn{7}{l}{$^e$ Observed flux in the 10-30 keV band, in units of 10$^{-12}$ erg cm$^{-2}$ s$^{-1}$}\\                 
  \multicolumn{7}{l}{$^*$ \emph{NuSTAR} observations simultaneous to the \emph{XMM-Newton} observations}\\
  \end{tabular}
\end{center}
\small
\end{table*}
%\end{onecolumn}

   \begin{figure*}[htb]
   \centering
    \includegraphics[width=19cm,height=12cm,angle=0]{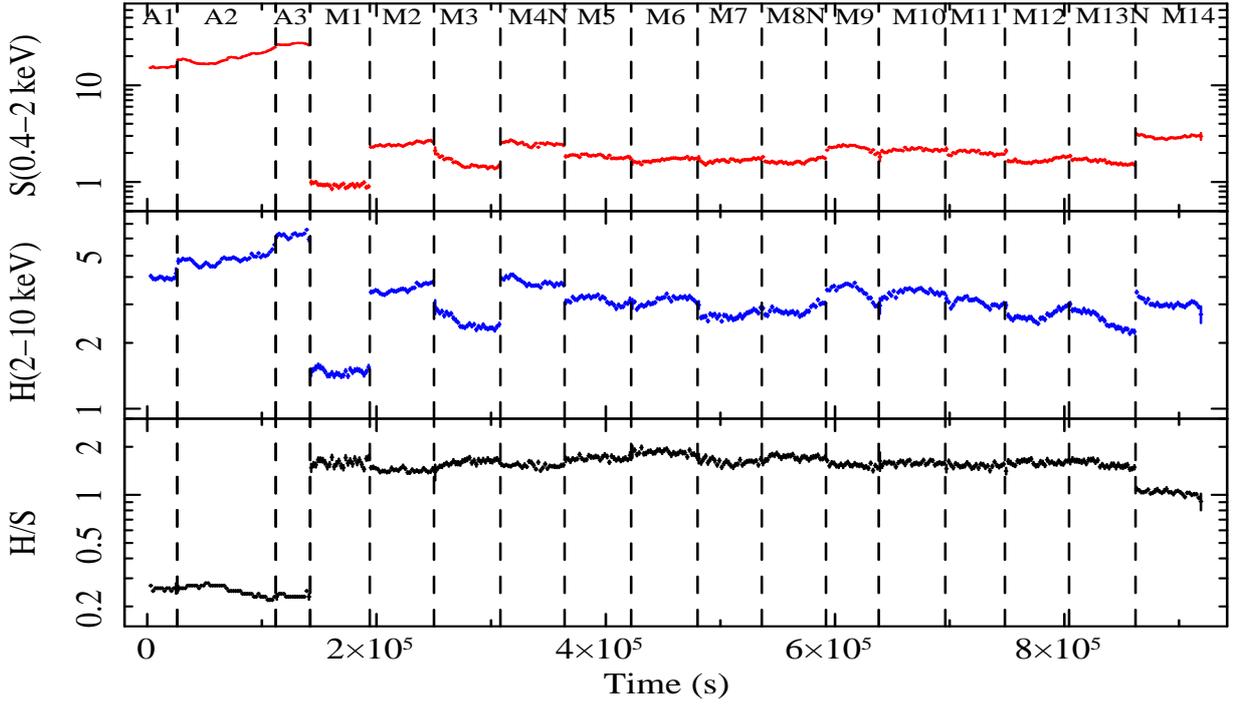}
   \caption{Background subtracted \emph{XMM-Newton} pn light curves of NGC 5548 calculated for the soft (S) 0.4-2 keV energy band (top), the hard (H) 2-10 keV energy band (middle) and their hardness ratios H/S (bottom). 
   Vertical dashed lines indicate when observations were interrupted along the years.}
             \label{f1}
   \end{figure*}

The EPIC data were reduced using the standard software SAS v.~14 (de la Calle, 2014)
and the analysis was carried out using the HEASoft v.~6.14 package\footnote{http://heasarc.gsfc.nasa.gov/docs/software/lheasoft/}.

After filtering for times of high background rate, and correction for the live time fraction (up to 0.7 for the pn in small window mode), 
the useful exposure times were typically between 30 and 50 ks per observation (see Table \ref{tab:1}). 
The EPIC pn and MOS cameras were operated in the ``small window'' mode with the thin filter applied for all observations, except for the first three archival 
observations (A1-3) for which the MOS 1 camera was operated in timing mode, thus the MOS1 was not used for this analysis.

Using the {\tt epatplot} command of the SAS, we checked the pattern distributions of the collected events and found that: 
i) pile-up was not significant (less than few \%) in the pn data at this level of source flux (and notably during the campaign when the source was heavily 
absorbed), thus we considered both single and double events as per standard procedure and calibrations;
ii) the pattern distribution in the pn deviates significantly below 0.4 keV w.r.t. model predictions, thus we did not consider the data below 0.4 keV in the present analysis. We note that 
this is slightly different from our previous analysis reported in Papers 0, III, and IV, in which we always used data down to 0.3 keV. 
We estimate that the effect of this slightly different choice of low-energy cut-off should be nevertheless largely within statistical errors of the present and previous analysis.
Given the significantly lower effective area of the MOS detectors, and for clarity and simplicity, we only report here the results obtained from the pn
data and used the MOS data for consistency checks only.

Source counts were extracted from a circular region of 40 arcsec radius, while the background counts were extracted from a nearby 
source-free and gap-free area of the detector of the same size. The size of the small window being less than 4.5$^\prime$$\times$4.5$^\prime$, both source and background were similarly 
affected by the instrumental Ni and Cu K$\alpha$ lines at 7.3-7.6 keV and 7.8-8.2 keV, respectively, thereby their effects are cancelled. 

Finally, we anticipate that all the Fe~K line measurements (see Sect. 4.1.2 below) obtained with the pn spectra during the campaign, but not in the earlier archival data, presented a systematic (and more or less constant) blueshift and broadening of the energy scale response by $\Delta E$$\sim$+30-40 eV and $\sigma(E)\sim$40-50 eV, respectively, when compared to the MOS and/or \emph{NuSTAR} independent results.
These shifts, albeit being modest, were nevertheless statistically significant and present with either single and double or single-only events, and despite our use of the latest CTI correction files (CCFv45, released in November 2014) and analysis procedure as described in Smith et al. (2014). To investigate this further, we performed a detailed analysis of the spectral energy scale by comparing our results for different instruments (MOS1, MOS2, pn, and \emph{NuSTAR}), using different pattern selections (from single-only to single+double events), and applying different CTI response correction files (from CCF v27 to v45) which were released in 2013 and 2014. We also compared our results with the energy and width obtained for the AlK line (at 1486 eV) and the MnK doublet (at 5895 and 6489 eV) of the calibration source during a $\sim$70 ks CALCLOSED observation performed in 2013, August 31st, i.e. shortly after the campaign and thus representative of the absolute energy scale of the instrument at that time. From this analysis, we attributed 
both blueshift and broadening to a remaining (admittedly small) uncertainty in the long-term degradation of the EPIC pn CTI as mentioned in the calibration technical notes of Guainazzi et al. (2014) and Smith et al. (2014).
In the following of the analysis, we thus took into account these systematic shifts and broadenings by using the {\tt zashift} and {\tt gsmooth} command in {\tt XSPEC} in order to correct for this remaining uncertainty (see also Sect. 4.1.2).

We also used data from three (out of 4) {\it NuSTAR} observations, those which were simultaneous to the {\it XMM-Newton} observations, namely the 4$^{th}$, 8$^{th}$ and 13$^{th}$ observations of the \emph{XMM-Newton} campaign (see Table 1).
A comprehensive analysis of the high-energy continuum, combining all available {\it NuSTAR}, {\it INTEGRAL} and previous {\it BeppoSAX} and {\it Suzaku} observations is presented in Paper III. We followed their same procedure for the {\it NuSTAR} data reduction that is we used the standard pipeline ({\tt nupipeline}) of the {\it NuSTAR} Data Analysis Software ({\tt nustardas}, v1.3.1; part of the {\tt heasoft} distribution as of version 6.14) to reduce the data, and used the calibration files from the {\it NuSTAR} {\tt caldb} v20130710. The spectra were then extracted from the cleaned event files using the standard tool {\tt nuproducts} for each of the two hard X-ray telescopes aboard {\it NuSTAR}, which have corresponding focal plane modules A and B (FPMA and FPMB). The spectra from FPMA and FPMB were analyzed jointly, but were not combined. Finally, the spectra were grouped such that each spectral bin contains at least 50 counts. 

\section{Timing Analysis}

Soft (0.4-2 keV) and hard (2-10 keV) band light curves of NGC 5548 are shown in Fig. \ref{f1} (top and middle panels, respectively) with a time bin of 2000 s. Hardness ratios (bottom panel) are also reported to highlight any spectral variations of the source. The light curves include the 3 archival (year 2000-2001) observations, and the 14 observations of the multifrequency campaign (year 2013-2014), with each observation separated by vertical dotted lines.
Within the single observations (including time scales of $<$ few tens of ks) we observe only low/moderate flux variations (less than 20-30 percent of the average flux). However, much stronger flux variations ($>$50 percent) occur between the different observations, on the time scales sampled by the campaign and longer (i.e. up to years, if we consider also the three archival \emph{XMM-Newton} observations).

As already presented in Papers 0 and I, the hardness ratios show a significant spectral hardening which characterizes the 2013-2014 campaign, when compared to the 2000-2001 archival data. The source does not show any significant spectral variation within each ($\sim$ half a day long) observation, besides a hint of spectral softening during the last observation. The lack of significant short-term spectral variability justifies our spectral analysis approach below (Sect. 4) of averaging the spectra observation by observation.

\begin{figure}[!]
   \centering
    \includegraphics[width=10cm,height=9cm,angle=0]{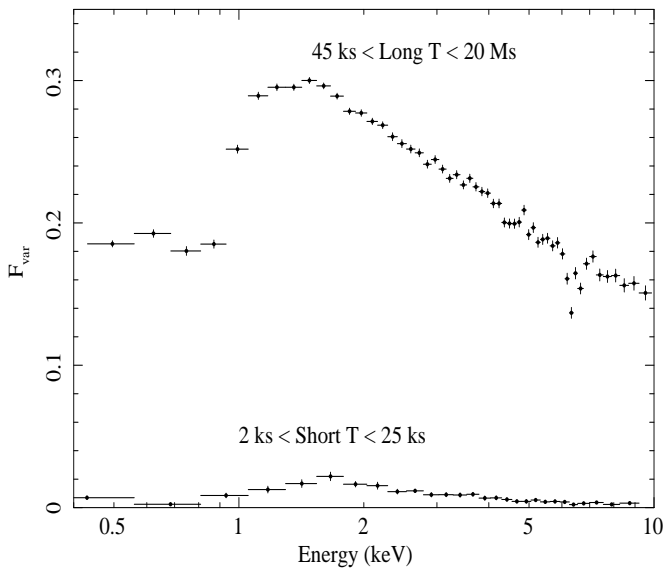}
   \caption{Fractional variability (rms) spectra of NGC 5548 calculated over long (45-2000 ks, top curve) timescales, i.e. longer than observations durations, but shorter than overall campaign, and over short (2-25 ks, bottom curve) timescales, i.e. longer than minimum time bin but shorter than shortest observation duration.}
                \label{f2}
   \end{figure}

To better characterize the distribution of variability power as a function of energy band we computed the fractional root mean squared (rms) variability amplitude (F$_{\rm var}$; Nandra et al. 1997, Vaughan et al. 2003, Ponti et al. 2004, Ponti 2007) on both short (hours) and long (days-to-months) timescales. The former is obtained by extracting light curves of the single pn observations, with a time bin of 2 ks, and averaging the F$_{\rm var}$ computed from each of them. The latter is obtained by taking the average count rates of each observation and computing the F$_{\rm var}$ over the entire campaign, thus sampling timescales going from $\sim$ 45 ks up to 20 Ms.
The light curves are extracted in different energy channels, the width of the channels ensuring the number of counts $\gg$20 per time and energy bin.
Results are shown in Fig. \ref{f2}. The F$_{\rm var}$ shows, in a clear and model-independent way, that the peak of variability power ($\sim$30 percent) lies in the energy band between 1 and 5 keV, and most of the variability occurs on the long-timescales. 
This shape is very similar to what is typically observed also in other Seyfert galaxies (Ponti 2007). Remarkably, a narrow feature is very clearly observed here at the energy of the 
Fe K$\alpha$ line, strongly suggesting the presence of a constant (on long-timescales) reflection component.

It is worth noting that, despite being very low (few percents), the F$_{\rm var}$ of the short-timescales is significantly different from zero. We verified that this residual variability is most likely to be ascribed to red-noise leakage effects (e.g. Uttley et al. 2002), rather than to intrinsic, short-term variability of the source. In other words, variability over timescales slightly longer than the maximum sampled timescale (limited by the duration of the observation) has introduced slow rising and falling trends across the light curve of the single observations, which have contributed to its short-timescale variance. As a consequence, variability power from long-timescales has been transferred to the short-timescales.
%This is probably a consequence of the obscuration of the inner regions, which precludes us from further exploring the short-term variability properties of the source using more advanced timing techniques. 
Variability studies on even longer timescales than those presented here (using \emph{Swift} light curves) are presented in McHardy et al (2014) and Paper VII.

  \begin{figure*}[!]
   \centering
   \vspace{-2.5cm}
    \includegraphics[width=19cm,height=12cm,angle=0]{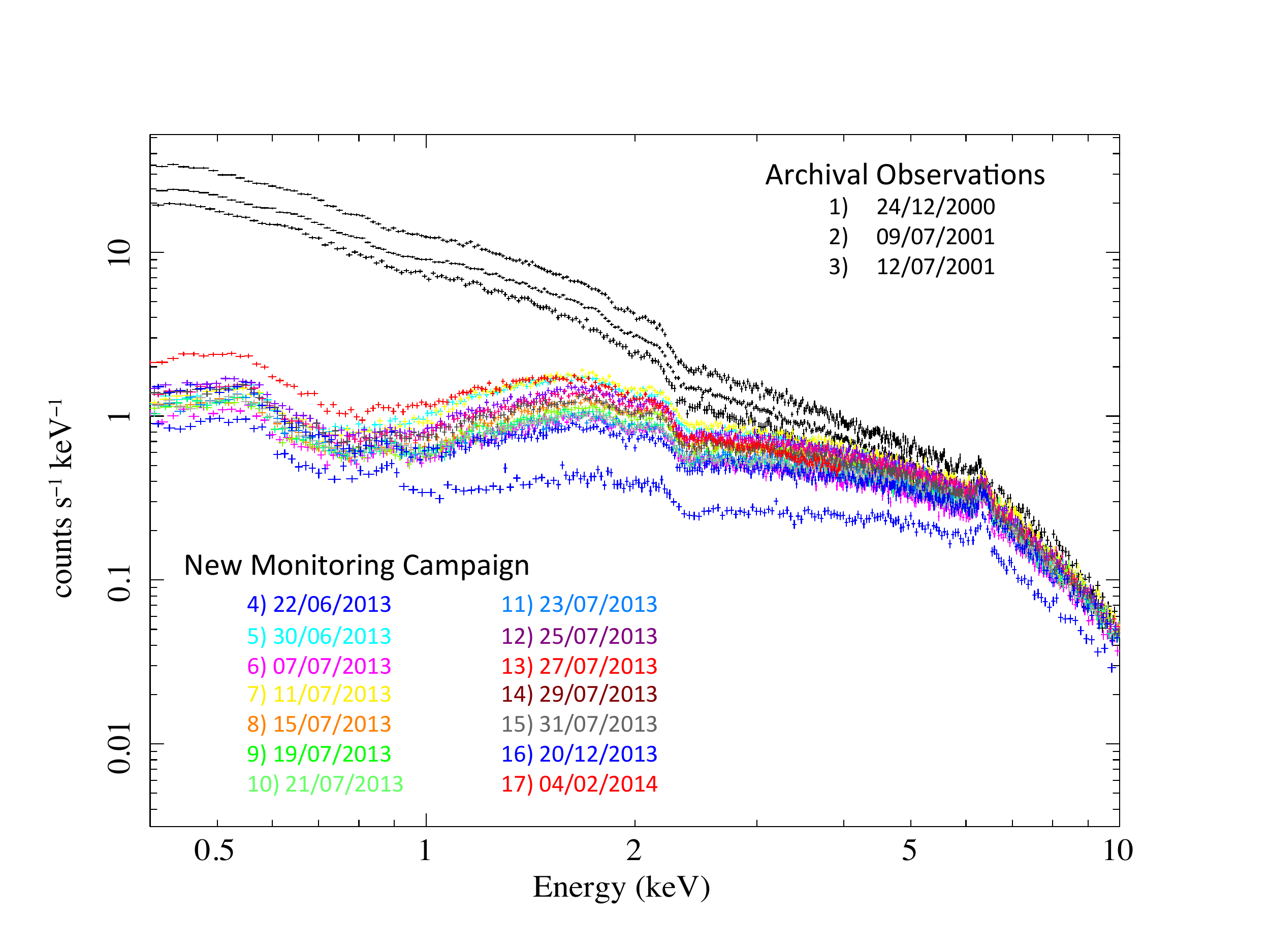}
   \caption{The 0.4-10 keV spectra of NGC5548 obtained with the EPIC pn during the three archival observations (black) and during the 14 observations of the
    2013 summer campaign (color). Observations are summarized in Table 1.}
             \label{f3}
%   \end{figure*}
% 
%         \begin{figure*}[!]
%   \centering
  
  \hspace{-1.5cm}
    \includegraphics[width=9cm,height=7cm,angle=0]{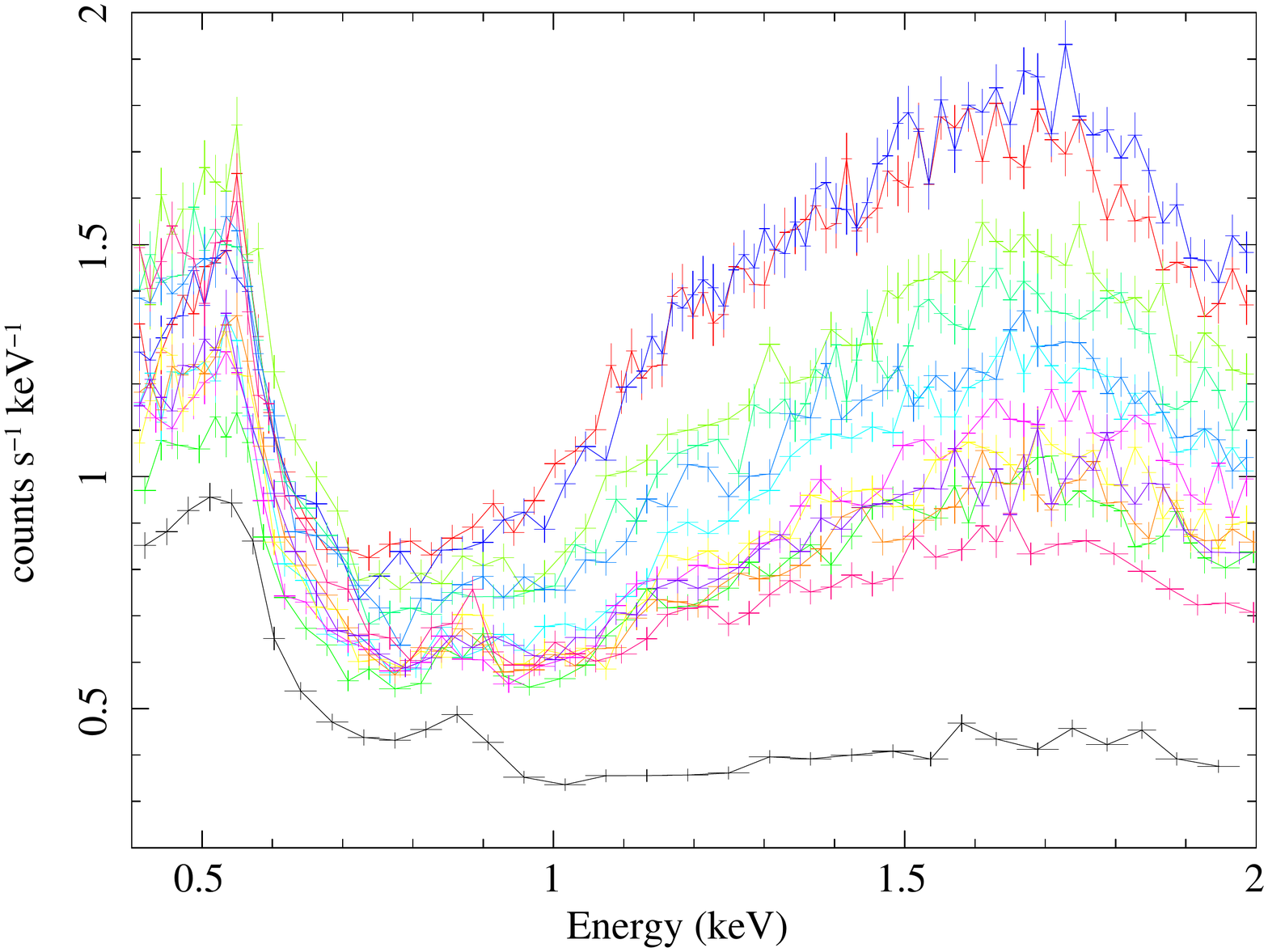}
    \hspace{-0.8cm}
   \includegraphics[width=9cm,height=7cm,angle=0]{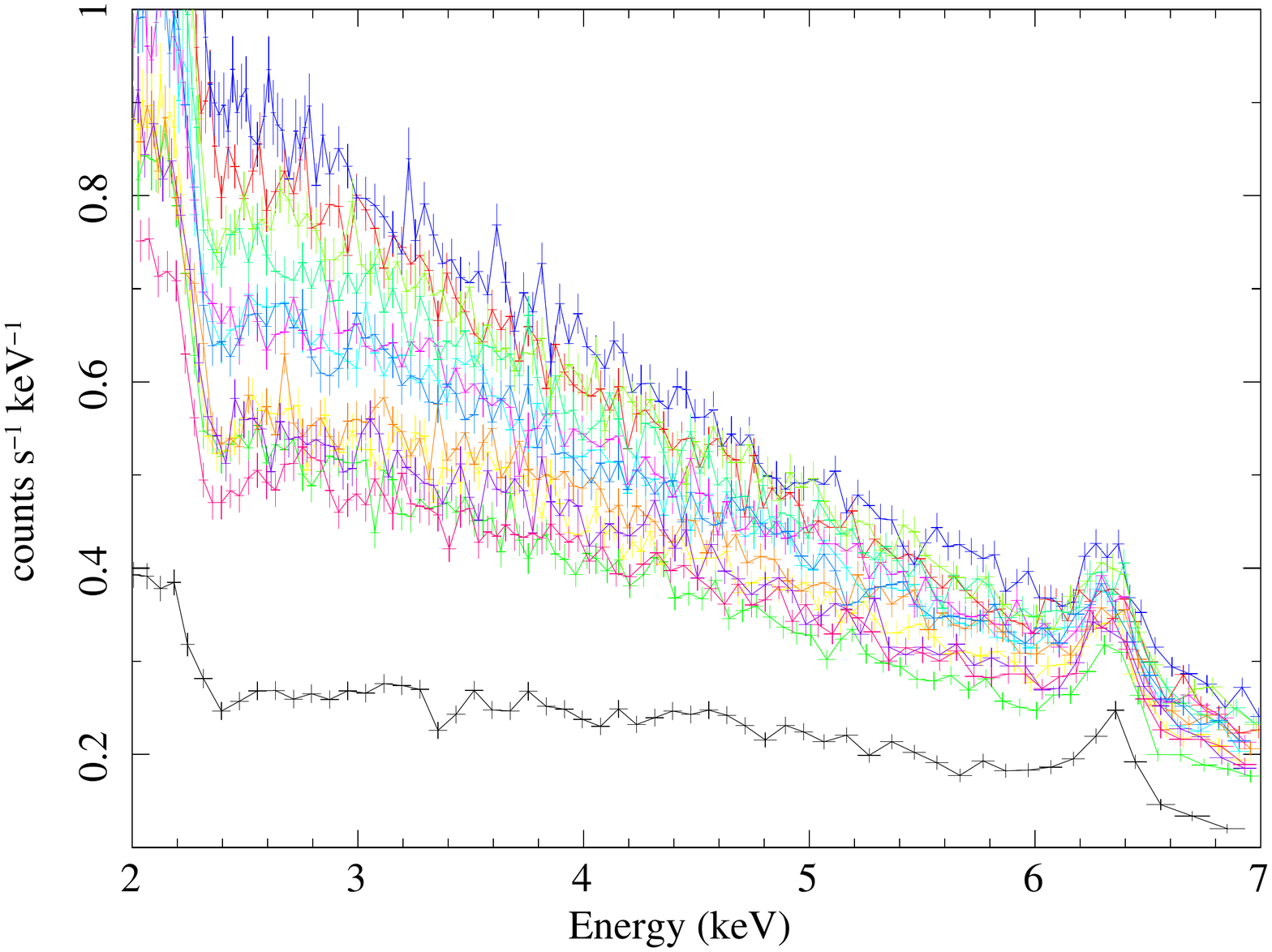}
    \vspace{-1cm}
    \hspace{-1.5cm}{\vspace{0cm}
    \includegraphics[width=9cm,height=7cm,angle=0]{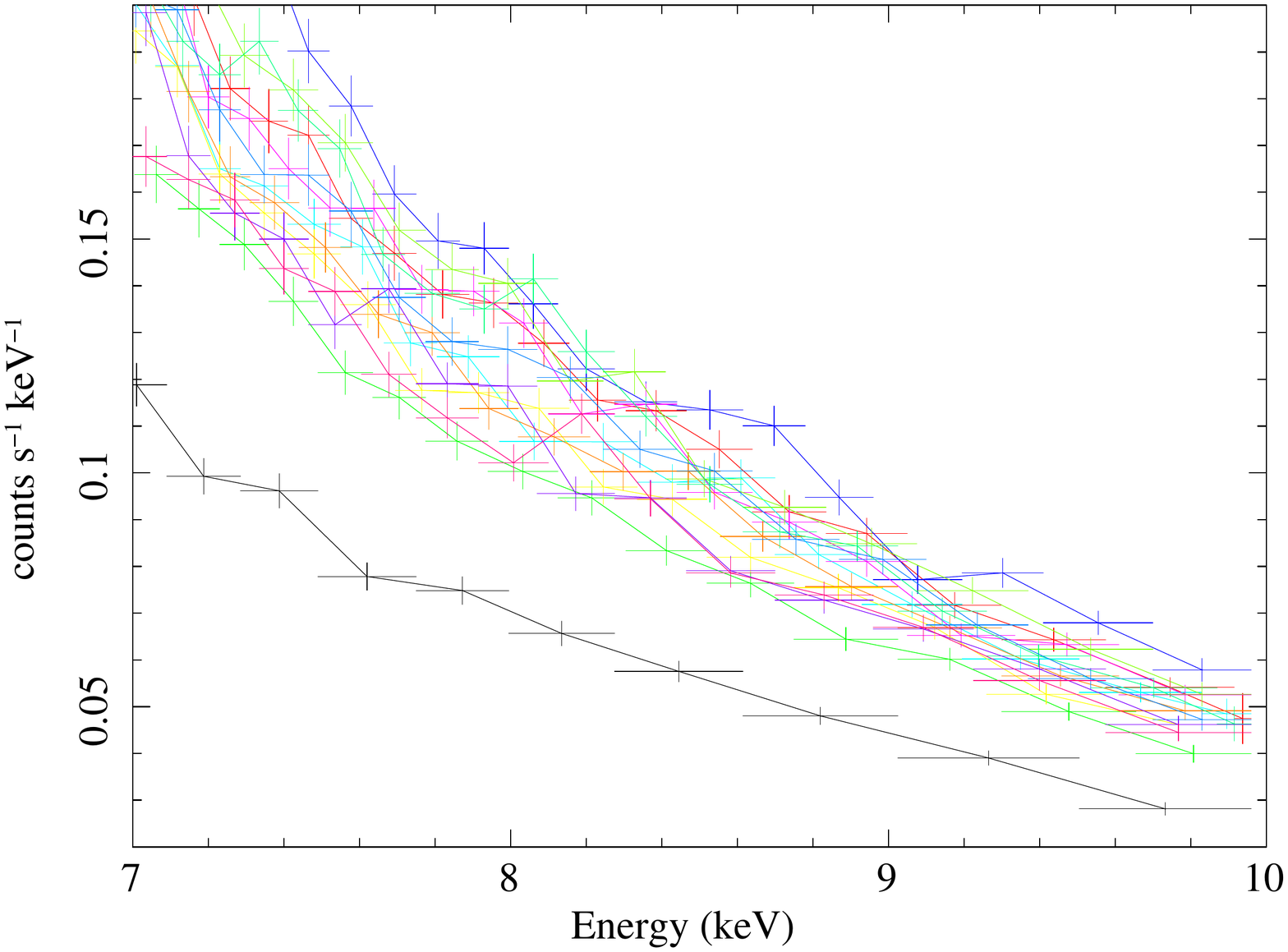}
    \hspace{-0.8cm}
    \includegraphics[width=9cm,height=7cm,angle=0]{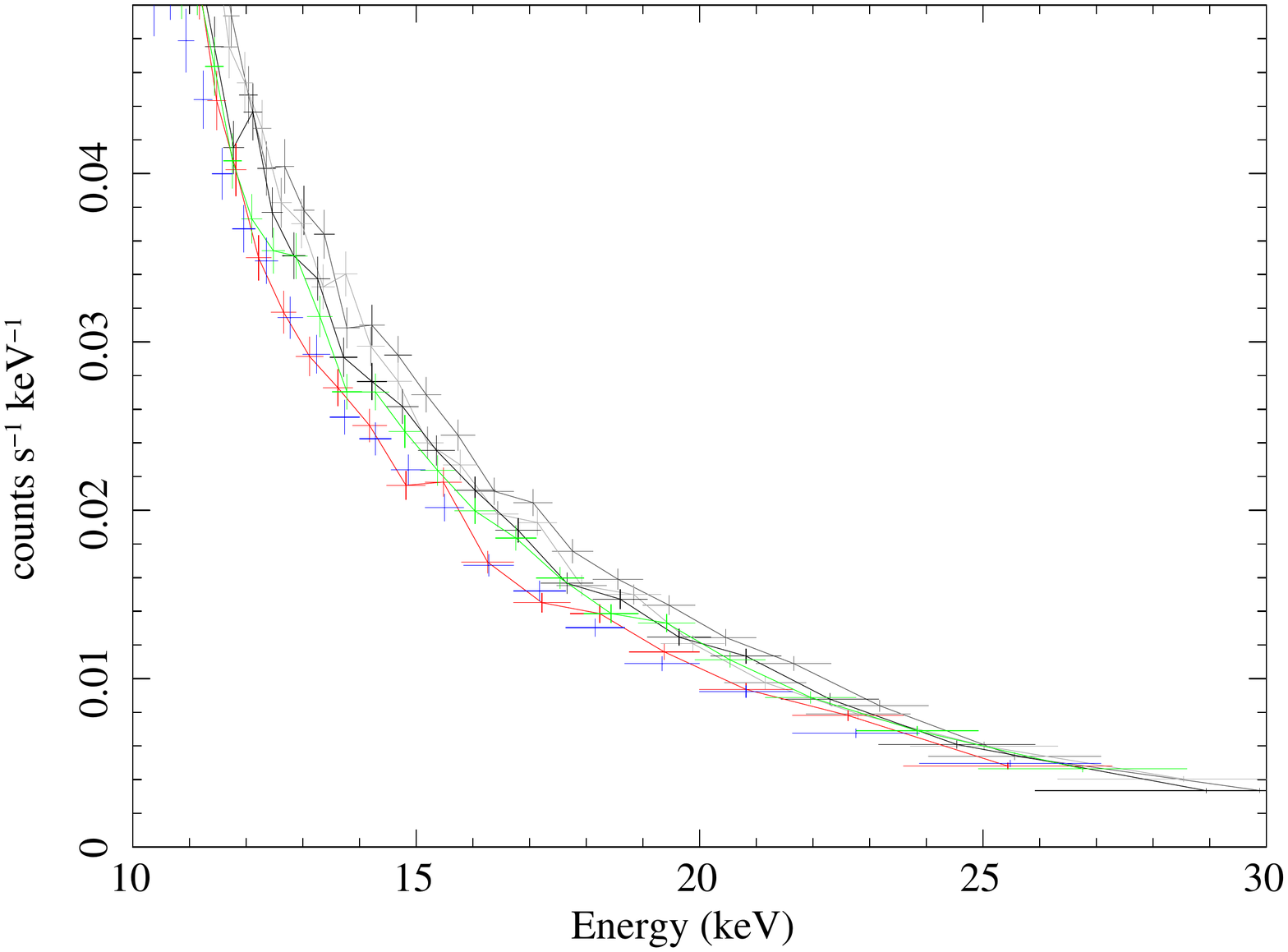}}
    \centering
   \caption{X-ray spectra obtained during the campaign (M1$\rightarrow$M14, including the simultaneous M4N, M8N and M13N \emph{NuSTAR} observations plotted above 10 keV) plotted on linear scales, and with the same factor of $\sim$10 extension range (from bottom to top) in the y-axis scale intensities. These are shown to illustrate the important and complex variability as a function of energy up to 2 keV, and its gradual reduction above 2 keV, and up to 30 keV.}
             \label{f4}
   \end{figure*}

\section{Spectral Analysis}

Within each observation, the source varied only very weakly either in flux (top and medium panels of Fig. \ref{f1}) or in spectral shape (bottom panel of Fig. \ref{f1}), as found 
also from the very low F$_{\rm var}$ value when calculated on short-timescales (Fig. \ref{f2}). We thus extracted the mean pn spectra for each of the 17 observations, grouping the data to a maximum of 10 channels per energy resolution element and 30 counts per channel to apply the $\chi^2$ minimization statistics. In all following fits, the Galactic column density was fixed at the value of $N_{\rm Hgal} = 1.55\times10^{20}$~cm$^{-2}$ (Dickey \& Lockman 1990), and abundances were taken from Lodders (2003).

Fig. \ref{f3} shows the overall  \emph{XMM-Newton}  spectra during all observations (1-17) which clearly show the strong absorption that is affecting all the 0.3-10 keV spectra during the 2013 campaign (obs. 4-17, in color) when compared to the 2000-2001 archival unabsorbed spectra (obs. 1-3, in black). Fig. \ref{f4} further illustrates the complex and strongly energy dependent variability of the source during the campaign: the top two panels show the large and complex variability in the soft 0.4-2 keV (top left panel) and hard 2-7 keV 
(top right panel). The two bottom panels show the reduction in variability going from lower (7-10 keV) to higher (10-30 keV) energies.
Because of the heavy and variable absorption affecting the source, the spectrum is very complex and model parameters (such as the photon index $\Gamma$, the absorption column densities and their covering factors) are often strongly degenerate. Moreover, during the unobscured, archival observations (A1$\rightarrow$3) the source shows a very significant, moderately strong, soft-excess 
(see e.g. Pounds et al. 2003b) which is however not easy to detect during the monitoring campaign (M1$\rightarrow$14) because of the heavy obscuration.

To overcome, or minimize, the above limitations we decided to proceed in the following way:
First, we used the three \emph{XMM-Newton} spectra for which \emph{NuSTAR} was simultaneously available, and used data at first only above 4 keV, in order to obtain 
the best possible constraints on the underlying source continuum components, in particular the reflection component. 
Second, we considered also the data below 4 keV, using all informations from previous papers of this series and from literature, and combine also the information at 
soft X-rays obtained from the archival \emph{XMM-Newton} observations. 
%used the best-fit scattering {\tt Cloudy} model (version 13.03; Ferland et al. 2013) including both lines 
%and Thomson-scattered continuum obtained for fitting the narrow emission components in the time-averaged line dominated RGS spectrum, as in Whewell et al. (2015), and we extended the contribution of this 
%same scattered component up to 80 keV. 
%Third, we used the three \emph{XMM-Newton} archival observations in order to obtain the best possible modeling of the very significant, but moderately strong, soft-excess emission. 
These two different steps and studies (Sect. 4.1 and Sect. 4.2) are preliminary and crucial to then best characterize the obscurer variability using the 
whole dataset of \emph{XMM-Newton} observations.

%%\subsection{The three XMM+NUSTAR simultaneous observations between ~0.4-79 keV:}
%
%\subsection{The three \emph{XMM}+\emph{NuSTAR} simultaneous observations:}
%
%\subsubsection{Underlying continuum, Fe K line and reflection component}

\subsection{Hard ($>$4 keV) X-ray Band: Underlying Continuum and Constant Reflection Component}

\subsubsection{The three \emph{XMM-Newton}+\emph{NuSTAR} simultaneous observations}

As mentioned above, we first considered only the data in the 4--79~keV interval where the data are less sensitive to the precise modeling of the obscurer, in an attempt to minimize degeneracies induced by the absorbers. 
\emph{XMM-Newton} and \emph{NuSTAR} spectra were fitted individually for each of the 3 simultaneous observations, with the same model but letting all parameters free to vary, except for the normalizations
 of the two \emph{NuSTAR} modules (FPMA/B) which were kept tied together. 
For plotting purposes only, and to better identify any systematic deviation from the fits, the spectra were then grouped together (using the \textit{setplot group} command in {\tt XSPEC}) and color-coded with pn spectra in black and \emph{NuSTAR} spectra in red.
Cross-normalization values among different instruments (FPMA vs FPMB vs pn) were always left free to vary but these were never larger than a few percent, typically 2-3 \%, in line with the expectations (Madsen et al. 2015). 
%We also added a WA component in all fits, either as a ``standard" two ionization components\footnote{In order not to over-fit the current low resolution pn data, and consistently with what found previously with {\it XMM-Newton} or \emph{Suzaku} 
%(Pounds et al. 2003, Krongold et al. 2010, Liu et al. 2010, Brenneman et al. 2010), we used a two ionization warm absorber model during the unobscured state which, for the sake of the present analysis, approximates sufficiently well the 5 ionization 
%components of the warm absorber found in higher resolution data (e.g. Papers VI and ref. therein).} during the unobscured archival observations, or as a multi-temperature de-ionized 
%component during the obscured state of the campaign, as obtained in Paper 0. We note that given the low energy resolution of the EPIC pn and 
%\emph{NuSTAR} CZTs, its effects can barely be seen at energies above a few keV because of its low column density of $\log {\rm N}_{\rm H}$=20-22 cm$^{-2}$ (see 
%Table S2 in Paper 0), but the WA is significant at lower energies despite the presence of the heavy obscurer, and measured by the RGS and LETG 
%data (see Fig. S3 in Paper 0). 

\begin{figure}[!]
\centering
\includegraphics[width=10cm,height=6cm,angle=0]{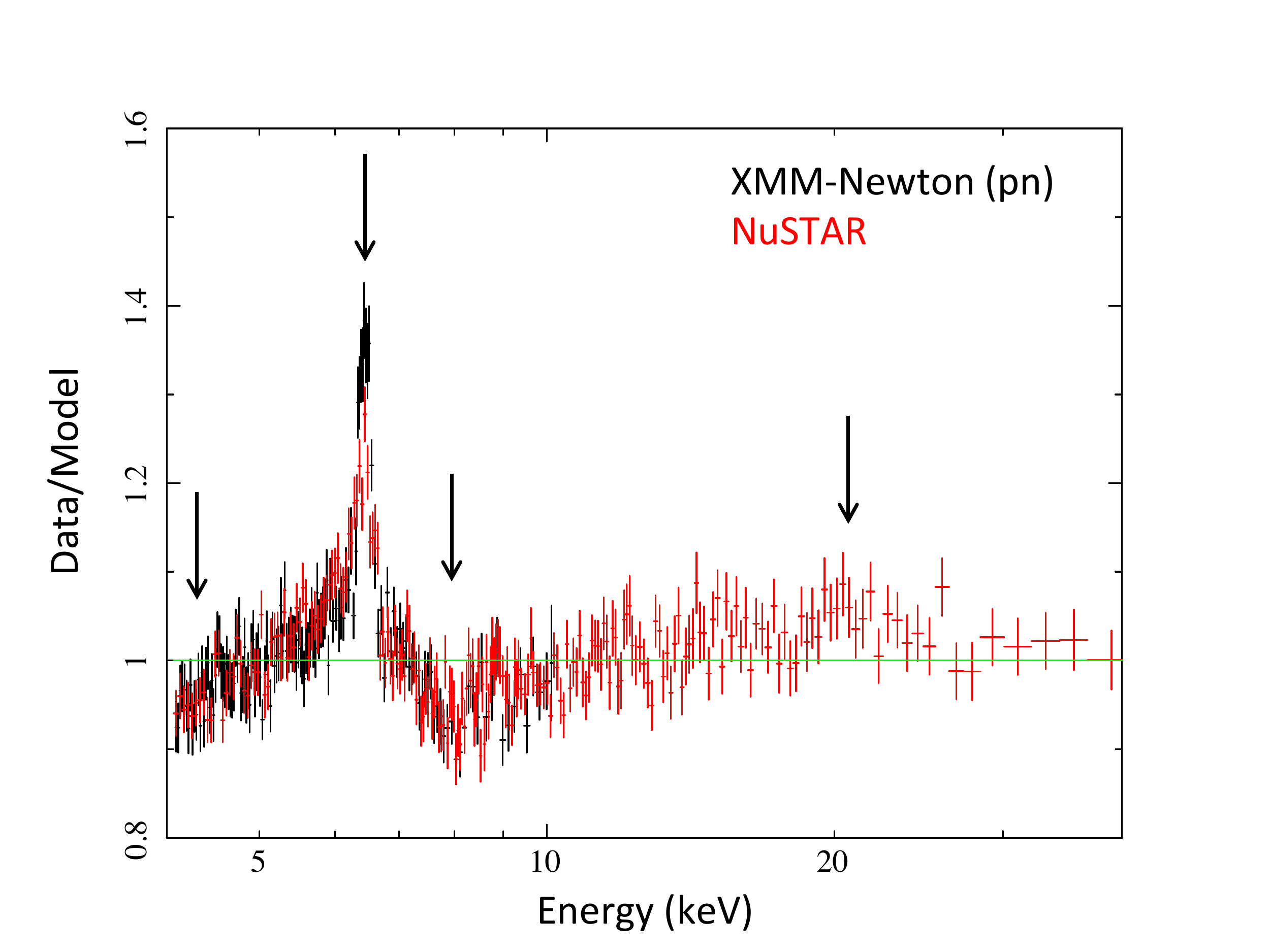}
%\parbox{10cm}{\vspace{-10cm}{\hspace{4cm}}
%\includegraphics[width=5cm,height=2.5cm,angle=0]{pn_nustar_obs1_2_4_4_70_phabs_pexmonfixed_cutoffpl_line.pdf}}
\parbox{10cm}{\vspace{-0.5cm}
\includegraphics[width=10cm,height=6cm,angle=0]{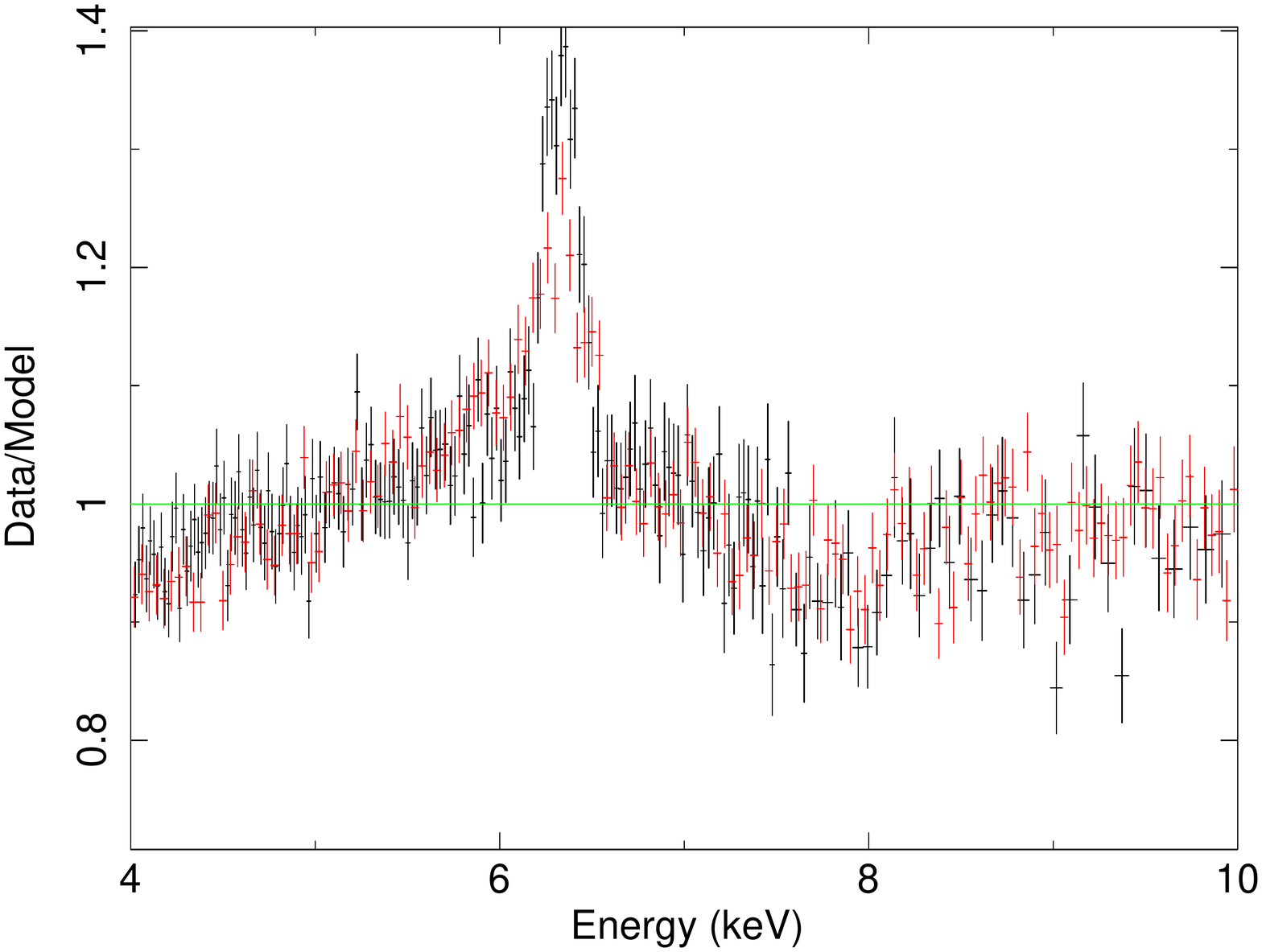}}
\parbox{10cm}{\vspace{-0.5cm}\hspace{-1cm}
%\includegraphics[width=10cm,height=6cm,angle=0]{eeufspec_pn_nustar_pexmon_pcfabscv1_cutoffpl.pdf}}
%\parbox{10cm}{\vspace{-0.5cm}
%\includegraphics[width=10cm,height=6cm,angle=0]{ratio_pn_nustar_pexmon_pcfabscv1_cutoffpl.pdf}}
\includegraphics[width=11cm,height=9cm,angle=0]{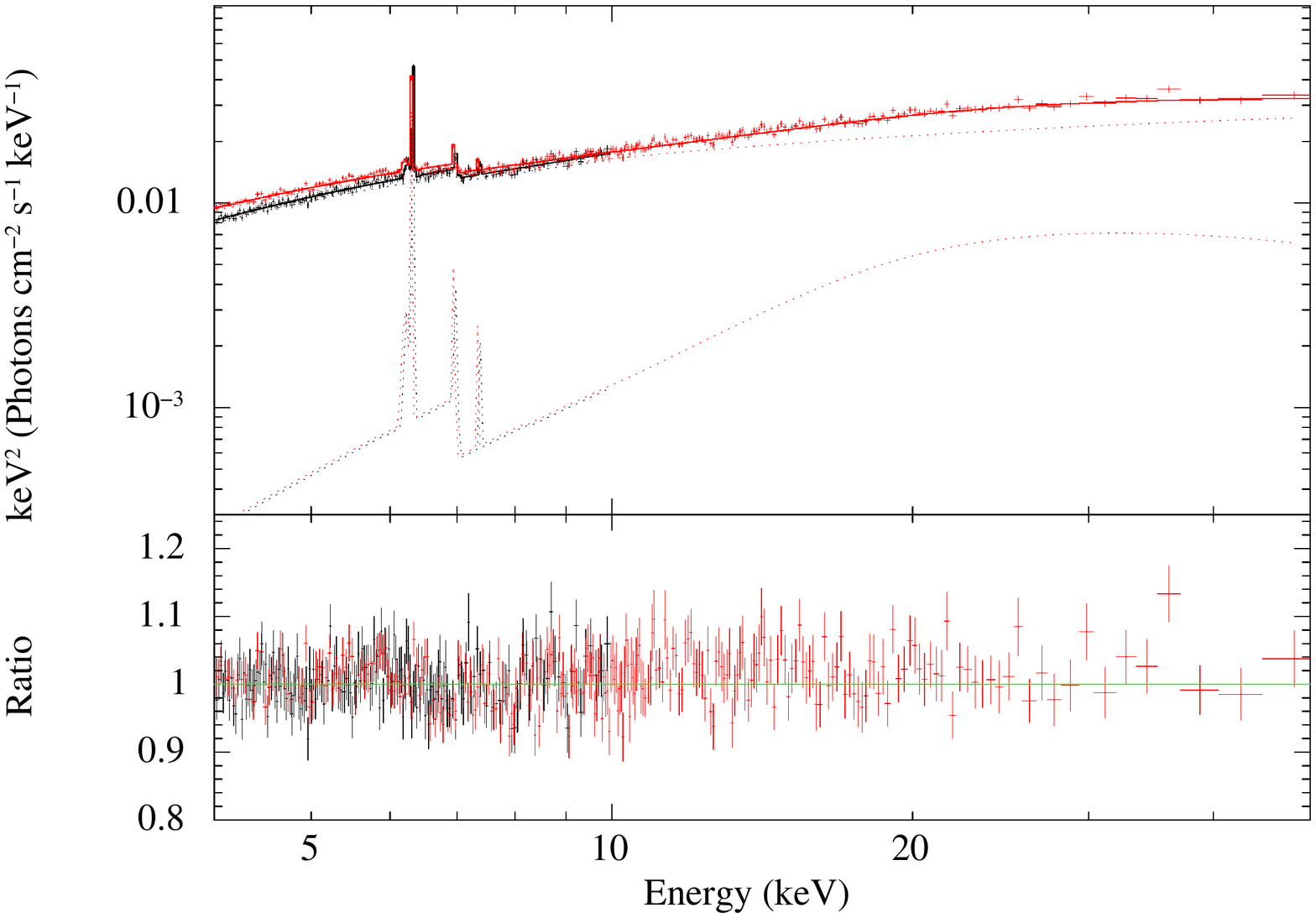}}
\caption{\small (Top): Data are plotted as the ratio to a single power-law continuum model fitted to the grouped observations M4N, M8N, and M13N 
with \emph{XMM-Newton} (black) and \emph{NuSTAR} (red) simultaneous data. As indicated by the arrows, going from low to high energies, 
one can clearly note the presence of: a sharp low-energy
 cut-off, a narrow Fe K line at $\sim$6.4 keV, some absorption feature(s) between 7-8 keV, and a high energy hump between 10-30 keV. (Middle): Same as Top panel, but zoomed between 4 and 10 keV.
 (Bottom): Best-fit spectrum, model and ratios plotted between 4-50 keV (see Sect. 4.1.1 for details).}
\label{f5}
\end{figure}

Fig. \ref{f5} shows the ratios (between $\sim$4-50 keV) obtained from a fit of the 4-79~keV spectra with a single non-absorbed power-law model (with $\Gamma \sim 1.3-1.5$). 
Low energy curvature, a narrow line at $\sim$ 6.4 keV and a high energy hump between 10-30 keV are readily seen in the data (see Fig. \ref{f5}, Top). 
We thus added a cold absorption column density plus a gaussian emission line and a continuum reflection model ({\tt pexrav} in {\tt XSPEC}). 
Despite the possible presence of additional absorption feature(s) between 7-8 keV (see Fig 5, bottom panel, and following analysis in Sect. 4.5), the use of a more 
complex absorber, either partially covering or ionized, was not required here when fitting the continuum above 4 keV, likely because of the strongly 
curved low energy cut-off which requires a substantially cold absorber to be modeled, with N$_{\rm H}$$\simeq$4--7$\times$10$^{22}$ cm$^{-2}$ (Table 2).

Best-fit parameters obtained from \emph{XMM-Newton} only were in very good agreement (typically within a 1-$\sigma$ error) with those obtained with \emph{NuSTAR} only, except for a mildly flatter slope 
(by $\Delta\Gamma$$\sim$0.1) required for \emph{XMM-Newton} wrt \emph{NuSTAR}. We choose to tie, within each observation, all parameters obtained from both instruments, 
except for letting their cross-normalizations and photon index free to vary (to take into account the remaining calibration uncertainties), but report here and below values of photon indices and 
fluxes obtained for the pn only in order to allow better comparison with observations without the {\it NuSTAR} simultaneous data. Best-fit values for the 3 observations are reported 
in Table 2. These values are consistent with those obtained by Papers III and IV.

Given the neutral energy for the Fe K line, its narrow width, and its constant intensity, the line is consistent with being produced by reflection from a cold 
and distant reflector (see Sect. 5.1 for further discussion). This is consistent with our analysis below (Sect. 4.1.2) using the whole set of \emph{XMM-Newton} observations. This agrees also with our previous (model-independent) findings based on the source fractional variability amplitude (Sect. 3). The line equivalent width (EW$\sim$70-110 eV) with respect to the underlying continuum reflection (R$\sim$0.56-0.91, see Table 2) is consistent with the line being produced by a plane parallel 
neutral Compton thick reflector, and solar abundance, thus we decide to choose for simplicity the {\tt XSPEC} model {\tt pexmon} (Nandra et al. 2007) which gives a self-consistent description of both the neutral Fe K$_{\alpha}$ line 
and the Compton reflection continuum. 
This model also self-consistently generates the Fe K$_{\beta}$, Ni K$_{alpha}$ and Fe K$_{\alpha}$ Compton shoulder expected from a Compton-thick 
reflecting medium. Following the results presented in Paper III, we assumed, and fixed, the parameters of the intrinsic continuum illuminating the reflection slab to typical values of $\Gamma$=1.9, $E_c$=300 keV, inclination=30 deg and solar abundances. With this model, we obtained the best-fit parameters listed in Table 2. These values are in agreement with those shown in 
Paper III. 
%More complex analysis of the high energy continuum being beyond the scope of this paper, we refer to Paper III for a thorough analysis of all high-energy observations and for a 
%discussion on thermal Comptonization models applicable in this source.

From now on in this analysis these values will be referred to as our baseline underlying continuum model. Moreover, motivated by the fact that the line intensity
did not vary significantly (neither in intensity nor in energy) during the other 14 observations available (see Sect. 4.1.2), we decided to freeze 
the reflection component to the average value obtained from the 3 \emph{XMM-Newton} + \emph{NuSTAR} simultaneous observations, i.e. a normalization at 1 keV of 5.7$\times$10$^{-3}$ photons keV$^{-1}$ cm$^{-2}$ s$^{-1}$, 
also in agreement with Paper III.

\subsubsection{The whole 17 $\emph{XMM-Newton}$ observations}

Given the above results, we thus proceeded in our analysis by adding to the previous observations (M4N, M8N, and M13N) the other 14 available \emph{XMM-Newton} observations, including the 3 archival observations and the remaining 11 from the 2013 campaign. The spectra were first considered, again, only above 4 keV and focusing on the properties of the Fe K line and reflection component before and during the campaign. 

\begin{figure}[!]
\centering
\parbox{10cm}{\vspace{0cm}\hspace{-0.7cm}
\includegraphics[width=10cm,height=10cm,angle=0]{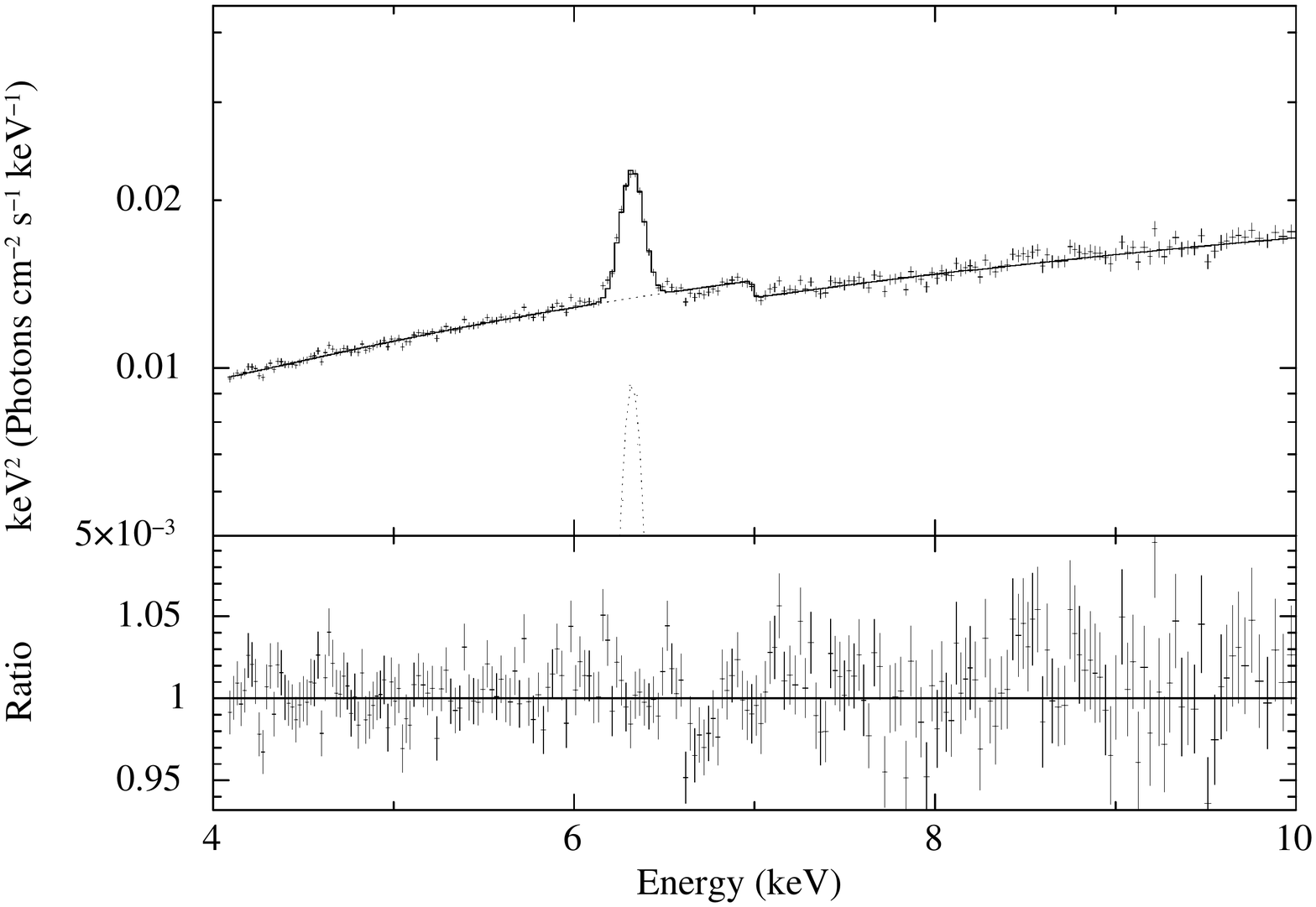}}
\caption{\small The 4-10 keV band spectrum obtained from the individual fitting of all the 17 $\emph{XMM-Newton}$ spectra, 
and the data/model ratios, grouped in a single dataset.}
\label{f6}
\end{figure}

\begin{figure}[!]
\centering
\parbox{10cm}{\vspace{-0cm}\hspace{-0.5cm}
\includegraphics[width=10.5cm,height=12cm,angle=0]{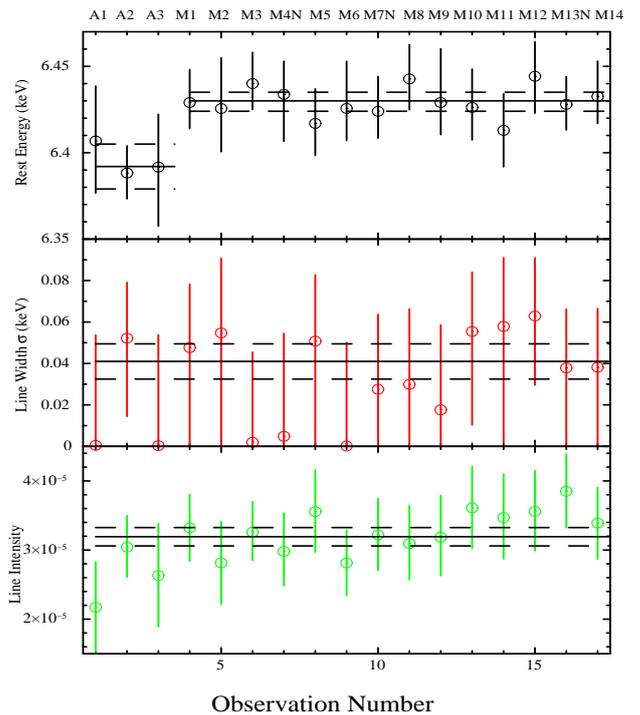}}
\caption{\small Time series of the FeK line best-fit parameters: Rest-frame energy (top, black) in keV, 
line width (middle, red) in keV, and line intensity (bottom, green) in photons cm$^{-2}$ s$^{-1}$ in the line. Best-fit values and 1-sigma (68\%) errors for a fit with a constant value are shown.}
\label{f7}
\end{figure}

%Then spectra were extended down to 0.4 keV in order to perform a full-band analysis and obtain the best possible characterization of the complex absorber.

Following the previous analysis, we first fitted all the 17 \emph{XMM-Newton}  spectra with a single, cold, absorber plus an Fe~K$\alpha$ line, plus a {\tt pexrav} continuum reflection model. This simple model yielded a good characterization of all \emph{XMM-Newton}  spectra, as demonstrated by the grouped spectrum shown in Fig. \ref{f6}.
The time-series for the Fe~K line parameters during all 17 observations are shown in Fig. \ref{f7}. 
We note that the line energy during the campaign was slightly higher ($\simeq$6.43$\pm$0.01 keV) than during the first 3 archival observations ($\simeq$6.39$\pm$0.01 keV), and was also systematically higher than the energy $\simeq$6.40$\pm$0.01 keV
obtained using only the MOS data. As discussed in Sect. 2, we attribute the energy shift, and slight broadening, of the FeK line to remaining CTI response degradation that has not properly been accounted for.

%To investigate this further, as mentioned above, we performed a detailed analysis of the spectral energy scale by comparing our results for the different instruments (MOS1, MOS2 and pn), using different pattern selections (from single-only to single+double events), and for several different CTI response correction files (from CCF v27 to 45) which were released in 2013 and 2014 (CCFv27 until 45). Our results were also compared to the values obtained for the AlK line (at 1486 eV) and the MnK doublet (at 5895 and 6489 eV) obtained for the calibration source during a $\sim$70 ks CALCLOSED observation performed in 2013, August 31st, i.e. not long after the gross part of the campaign and should thus be representative of the absolute energy scale of the instrument. 
%Nevertheless, inspection of the MOS and Nustar spectra, as well as the detailed analysis of the CTI correction to be applied here and discussed in Sect.2 made us conclude 
%that the FeK line slight (20-30 eV) (blue)shift seen during obs. 4 to 17 (Fig. \ref{f7}, top panel)is likely to be attributed to remaining CTI response degradation that was not properly accounted for, despite our use of the most recent calibrations and correction procedures (see Smith et al. 2014).
%We also attribute to the same effect the fact that the line appears to be marginally, but often, broad (with $\sigma \sim$40-50 eV) during the campaign (Fig. \ref{f7}, middle panel). 
%On the other hand, t
The line is consistent with being constant in intensity during all the 17 observations (Fig \ref{f7}, bottom panel), i.e. not only during the campaign, but also after comparison with the ($\sim$13 years) earlier archival observations. Overall, in agreement with our earlier findings based on the fractional variability (Sect. 3), this analysis readily demonstrates that the Fe~K line emission is neutral, narrow, and most importantly constant in time. We thus choose again to model both line and continuum using the self-consistent cold reflection model {\tt pexmon}, and fit all data with the reflection intensity fixed at its average value (5.7$\times$10$^{-3}$ photons keV$^{-1}$ cm$^{-2}$ s$^{-1}$) obtained in Sect. 4.1.1, and freezing then the {\tt zashift} parameter to its best-fit value. We note that this simple analysis readily shows that, during the campaign, the photon index was flatter ($\Gamma$$\sim$1.5-1.7) than typically found during either the 3 archival observations 
or historically in this source ($\Gamma$$\sim$ 1.7-1.9, e.g. Dadina 2007), and despite allowing in the fit for large absorption column densities with values between $\log {\rm N}_{\rm H}$ $\sim$ 22-23 cm$^{-2}$. 

\subsection {Soft ($<$4 keV) X-ray Band: Warm Absorber, Intrinsic Soft-Excess and Scattered Component}

As mentioned above, the soft (E$<$4 keV) X-ray spectrum of this source is known to be rather complex. Historically, it is known to require at least two components to be properly modeled, such as a complex, multi-temperature warm absorber, plus an intrinsic soft X-ray emission component (commonly called ``soft-excess"). In this part of the spectrum, we require at least one additional component, a soft scattered component, in order to be consistent with earlier and present observations. We briefly address below the evidence and need for each of these three components, even before considering the complex (and variable) 
obscurer found during the campaign, and which will be discussed only afterwards. 

{\it Warm Absorber:}
Following the analysis of Paper 0, we have added into our model a constant column density warm absorber model calculated from the same SED as discussed in Papers I and II. 
This component absorbs only the baseline underlying continuum model, and not the soft emission lines introduced below (which are already corrected for absorption from the WA, as discussed below), and it accounts for the historical multi components highly ionized warm absorber that is clearly seen when the source is in its typical unobscured state (Kaastra et al. 2002, 2004; Steenbrugge et al. 2003, 2005; Krongold et al. 2010;
Andrade-Velazquez et al. 2010; Paper VI). Absorption from this warm absorber component was clearly detected during the unobscured archival 
observations and, given the pn low energy resolution, its six WA ionization components could be approximated by only two warm absorber  
ionization components with ionization parameters\footnote{The ionization parameter $\xi$ is defined here as 
$\xi \equiv \frac{L}{{n_{\rm{H}}\,r^2 }}$ where $L$ is the luminosity of the ionising source over the 13.6 eV--infinity band in $\rm{erg}\ \rm{s}^{-1}$, $n_{\rm{H}}$ the hydrogen density in $\rm{cm}^{-3}$ and $r$ the distance between the ionised gas and the ionising source in cm.} $\log \xi$$\simeq$1--2.7 erg cm s$^{-1}$ and low column density ($\log {\rm N}_{\rm H}$$\simeq$21--22 cm$^{-2}$) consistently with previous literature results (Krongold et al. 2010). 
The effect of this multi-component WA is less visible, but still significant, on the spectra during the campaign, when the source is highly obscured
(see the transmission curve of this component in Fig. S3 of Paper 0). 
Following Paper 0 and II, we also know from the UV spectra that the kinematics of the warm absorber has not changed over the last 16 years, and Paper VI shows that all historical data on NGC5548 are consistent with a multi-component WA which is assumed to vary in response to changes in the underlying flux level only. For the obscured states, we thus kept the warm absorber parameters frozen at their average value found by Paper 0, and calculated using an ionization 
balance which assumes illumination from the average obscured SED.

{\it Soft-Excess Intrinsic Emission:}
NGC5548 is known to have a clear and strong soft-excess intrinsic continuum (see e.g. Kaastra \& Barr 1989, Kaastra et al. 2000, 2002, Steenbrugge et al. 2003, 2005). 
Even if there is no direct evidence for the presence of this same soft-excess during the campaign, because of the strong obscuration, it is important to model it to our best to reduce as 
much as possible the parameter degeneracies in our following analysis of the multilayer obscurer variability (see Sect. 4.3.2).
We thus performed a re-analysis of the 3 {\it XMM-Newton} archival observations (A1$\rightarrow$A3), when the source was in its typical unobscured state. 
We first used the same baseline model (Power-law plus reflection component) as for the high energy part of the spectrum (Sect. 4.1.1) and the above warm absorber affecting only the lower energies. 
We confirm the results obtained by Pounds et al. (2003b) whom analyzed the {\it XMM-Newton} 
spectra of A2+A3 to simultaneous data from the MECS+PDS onboard {\it BeppoSAX}: a weak soft excess is seen, after allowing for the over-lying absorption, as a smooth upward curvature in the X-ray continuum below $\sim$2 keV. Unlike Pounds et al. (2003b), we do not attempt here to test different models\footnote{The soft-excess could be modeled by Pounds et al. (2003b) by either two black-body models, a single-temperature Comptonized thermal emission or enhanced highly ionized reflection from an accretion disc.} for the soft-excess component, nor try to constrain in detail its shape nor intensity during A1$\rightarrow$A3.
This would be beyond the scope of this paper. Instead, we choose a Comptonization model, which is able to describe the soft X-ray continuum in a way that is consistent with the source 
UV-to-soft X-ray properties seen before, during, and after the campaign, as shown in Paper VII. In fact, the long-term and broad-band UV-to-soft X-ray analyses presented in 
Papers I and VII using {\it Swift} data indicate a correlation between the far UV and soft-X-ray emission, suggesting the presence of an intrinsic emission component linking the 
UV to the soft-X-rays, similar to the one measured in Mrk509 (Mehdipour et al. 2011, Petrucci et al. 2013), and possibly due to thermal Comptonization. We thus include in our fits the same thermal Comptonization 
model ({\tt Comptt}) as in Paper I, fixing the shape at the best-fit values found in Paper VII, but letting its normalization to be a free parameter in our fits. 
A similar approach was taken in Paper IV, but fixing the normalization to the value expected adopting the correlation measured by Paper I and VII. 

{\it Scattered Component:}
We include in all our models a soft scattered component to account for the narrow emission lines that are clearly detected in the higher resolution data available from 
the RGS instruments between 0.3-2 keV in the obscured states. We make use of the results obtained from the detailed analysis of Paper V. Their analysis shows that the RGS spectrum is clearly dominated by narrow emission lines (see their Fig. 1), and that these are consistent with being constant in flux during the whole campaign. We therefore prefer to use here, and include in our model, the average best-fit model obtained in Paper V using the whole 770 ks RGS stacked spectrum, rather than use the lower-statistics observation by observation, and have to deal with cross-instrument calibration issues (but see Paper IV for addressing some of these issues). This average emission line model was calculated using the spectral synthesis code {\tt Cloudy} (version 13.03; Ferland et al. 2013), the unabsorbed SED calculated in Paper I and was used as a fixed table model in the fit in {\tt XSPEC}. It reproduces well, and self-consistently, all the narrow 
emission lines, including the He-like triplets of Neon, Oxygen and Nitrogen, the radiative recombination continuum (RRC) features, and the (Thomson electron) scattered continuum seen in the RGS spectrum 
(see Fig. 6 in Paper V). 
The best-fit parameters of the emitting gas are $\log \xi$ $=$1.45 $\pm$ 0.05 erg cm s$^{-1}$, $\log {\rm N}_{\rm H}$ $=$ 22.9 $\pm$ 0.4 cm$^{-2}$ and $\log {\rm v}_{turb}$ $=$ 2.25 $\pm$ 0.5 km s$^{-1}$.
The emission model also requires, and includes, absorption from at least one of the six components of the warm absorber found by previous analyses of these and historical data (see Paper V for more details). For the purposes of the present broad-band modeling, this same {\tt Cloudy} model was extended up to E$\sim$80 keV which corresponds to the high energy limit of the 
\emph{NuSTAR}  spectral band, by accounting for Compton and resonant scattering up to these higher energies and including the expected weak FeK emission lines produced by this Compton thin layer of gas.
The contribution from this component to the broad-band model is shown in the unfolded spectrum of Fig. 8 (dashed red line in panel e). Its contribution to the soft (0.5-2 keV) X-ray flux is $\sim$1.8$\times$10$^{-13}$ erg cm$^{-2}$ s$^{-1}$, and corresponds to about 8\% of the total soft X-ray flux, while in the hard (2-10 keV) band it is $\sim$4.8$\times$10$^{-13}$ erg cm$^{-2}$ s$^{-1}$ (i.e. $\sim$2\% of the total flux, and a factor of $\sim$ 3 lower than the reflection component).
As discussed in Paper V, this component is consistent with being produced by (photo-ionized) scattered emission from a distant narrow line region (NLR) at a distance of $\sim$14 pc from the central source.

%All 17 \emph{XMM-Newton}  observations were then fitted independently using the same baseline model.

%\begin{landscape}
\begin{table*}[!htb]
\caption{Hard (4--79 keV) X-ray continuum emission (\emph{XMM-Newton} +\emph{NuSTAR} observations): Power-law plus Reflection Models}
\label{tab: pn_nustar_baseline_par}
\begin{center}
\begin{small}
\begin{tabular}{clccccccccc}
%\hline
\multicolumn{11}{l}{\it ``Phenomenological" reflection model (Cutoff-PL + Fe K$_{\alpha}$ emission line + {\tt pexrav})}\\
\hline
Obs. &  Obs. & N$_{\rm H}$& $\Gamma$    & $E_c$ & Energy$^1$ & EW$^2$     & Int.$^3$   			&  R ($\Omega/2\pi$) & A$_{norm}$	& $\chi^2/\nu$  \\
N.     & Name   &   ($\times$10$^{22}$ cm$^{-2}$)        		& 			& (keV) &(keV)		& (eV)	   &   ($\times$10$^{-5}$)   &        &   ($\times$10$^{-3}$)            &                         \\
\hline
7 & M4N&  4.1$\pm 0.5$   & 1.74$\pm 0.04$ 	& $>$313 & 6.35$\pm 0.02$ & 69$\pm 9$   &  3.0 & 0.56 & 5.9$^{+0.9}_{-1.3}$ & $3535/3471$ \\
11 & M8N &  7.2$^{+0.5}_{-1.0}$  & 1.75$\pm 0.06$  & $>$212 & 6.38$\pm 0.02$ & 83$\pm 10$  & 3.0 & 0.65 & 6.2$^{+1.1}_{-1.4}$  & $3535/3471$ \\
16 & M13N &  5.8$^{+0.6}_{-1.2}$ & 1.60$\pm 0.04$  & 129$^{+48}_{-30}$ & 6.42$\pm 0.02$ & 110$\pm 10$  & 3.7 & 0.91 & 5.9$^{+1.0}_{-0.5}$  & $3535/3471$ \\
\hline
\\
\multicolumn{8}{l}{\it ``Physical and self-consistent" reflection model (Cutoff-PL + {\tt pexmon})}\\
\hline
Obs. &  Obs. & N$_{\rm H}$& $\Gamma$    &    $E_c$ &  R ($\Omega/2\pi$) & A$_{norm}$ 				& $\chi^2/\nu$  \\
 N.     &  Name   	&  ($\times$10$^{22}$ cm$^{-2}$)           		& 			   & (keV)  &                            &   ($\times$10$^{-3}$)            &                         \\
\hline
7 & M4N &  3.8$\pm 0.7$   & 1.73$\pm 0.02$ 	& $>$310 & 0.51 & 5.4$\pm 0.5$ & $1234/1157$ \\
11 & M8N &  5.5$\pm 0.8$  & 1.63$\pm 0.05$  & 157$^{+153}_{-54}$ & 0.79 & 6.0$\pm 0.5$  & $1178/1157$ \\
16 & M13N &  5.7$\pm 1.2$ & 1.60$\pm 0.06$  & 123$^{+109}_{-41}$ & 0.95 & 6.2$\pm 0.5$  & $1153/1157$ \\
\hline
%\hline
\end{tabular}
\end{small}
\end{center}
\tiny{(1) Emission line rest-frame energy centroid, in units of keV; (2) Emission line rest-frame equivalent width, in units of eV. 
The width of the line was fixed to $\sigma$=0.1 keV (see text for details); (3) 2--10 keV flux in units of 
10$^{-11}$ erg~s$^{-1}$~cm$^{-2}$}\\
\label{t2}
\end{table*}

%\end{table*}
%%\end{landscape}
%%\begin{landscape}
%\begin{table*}[!htbp]
%\caption{4--79 keV continuum emission: "Physical and self-consistent" reflection model (Cutoff-PL + {\tt pexmon})}
%\label{tab: pn_nustar_baseline_par}
%\begin{center}
%\begin{scriptsize}
%\begin{tabular}{lccccccccc}
%\hline
%\hline
%Obs. & N$_{\rm H}$& $\Gamma$    &    $E_c$ &  R ($\Omega/2\pi$) & A$_{norm}$ 				& $\chi^2/\nu$  \\
%    	&             		& 			   & (keV)  &                            &   ($\times$10$^{-3}$)            &                         \\
%\hline
%M4N &  3.8$^{+0.7}_{-0.7}$   & 1.73$^{+0.02}_{-0.02}$ 	& $>$310 & 0.51 & 5.4$^{+0.5}_{-0.5}$ & $1234/1157$ \\
%M8N &  5.5$^{+0.8}_{-0.8}$  & 1.63$^{+0.03}_{-0.05}$  & 157$^{+153}_{-54}$ & 0.79 & 6.0$^{+0.5}_{-0.5}$  & $1178/1157$ \\
%M13N &  5.7$^{+1.2}_{-1.2}$ & 1.60$^{+0.06}_{-0.06}$  & 123$^{+109}_{-41}$ & 0.95 & 6.2$^{+0.5}_{-0.5}$  & $1153/1157$ \\
%\hline
%\hline
%\end{tabular}
%\end{scriptsize}
%\end{center}
%\small 
%\end{table*}

\begin{table*}[!htbp]
\caption{\emph{XMM-Newton} +\emph{NuSTAR} observations: 0.4--79 keV best-fit model (Cutoff-PL + Soft-excess {\tt Comptt} emission +  {\tt pexmon} + Scattered component + two partially covering ionized obscurers)}
\label{tab: pn_nustar_baseline_par}
\begin{center}
\begin{small}
\begin{tabular}{clccccccccc}
\hline
\hline
Obs. & Obs. & $\Gamma$  & $\log {\rm N}_{\rm H,1}$&    Cf$_1$ &  $\log \xi_1$ & $\log {\rm N}_{\rm H,2}$ & Cf$_2$ & $\log \xi_2$ &  A$_{comptt}$ & $\chi^2_{red}(\chi^2/\nu)$  \\
N.     &Name &    &  (cm$^{-2}$)     &      &  (erg cm s$^{-1}$)  &  (cm$^{-2}$)    &    & (erg cm s$^{-1}$)  &   &   \\
\hline
7 & M4N& 1.56$\pm 0.01$ &  22.13$\pm 0.01$   & 0.87$\pm 0.01$  & $<$-0.2 & $>$23.1    &  0.21$^{+0.02}_{-0.06}$  & $\equiv$-1& 30.8$^{+21.6}_{-2.9}$  & 1.15(1555/1348) \\
11 & M8N& 1.60$\pm 0.05$ &  22.37$\pm 0.03$   & 0.87$\pm 0.02$  & 0.75$^{+0.08}_{-0.41}$ & 23.2$\pm 0.09$    &  0.44$\pm 0.05$  & $\equiv$-1 & 21.4$^{+14.3}_{-12.3}$ & 1.03(1392/1348) \\
16 & M13N& 1.54$\pm 0.06$ &  22.35$\pm 0.04$   & 0.87$\pm 0.01$  & 0.50$^{+0.18}_{-0.81}$ & 23.2$\pm 0.09$    &  0.46$\pm 0.05$  & $\equiv$-1 & 64.2$^{+26.5}_{-14.3}$ & 1.02(1371/1348) \\
\hline
\hline
\end{tabular}
\end{small}
\end{center}
\small
\label{t3}
\end{table*}
%\end{landscape}

%\begin{landscape}
\begin{table*}[!htbp]
\caption{Whole \emph{XMM-Newton} and \emph{XMM-Newton}+\emph{NuSTAR} observations: 0.4--10 keV best-fit model (Cutoff-PL + Soft-excess {\tt Comptt} emission +  {\tt pexmon} + Scattered component + two partially covering ionized obscurers)}
\label{tab: pn_baseline_par}
\begin{center}
\parbox{19cm}{\hspace{-0.5cm}
\begin{small}
\begin{tabular}{lllllllllll}
%\hline
%\hline
%\multicolumn{11}{l}{\it Baseline model (Scattered component + Soft-excess Comptt emission + Cutoff-PL + Pexmon) + two partially covering ionized obscurers}\\
\hline
\hline
Obs. & Obs. & $\Gamma$  & $\log {\rm N}_{\rm H,1}$&    Cf$_1$ &  $\log \xi_1$ & $\log {\rm N}_{\rm H,2}$ & Cf$_2$ & $\log \xi_2$ &  A$_{comptt}$ & $\chi^2_{red}(\chi^2/\nu)$  \\
N.     &Name &    &  (cm$^{-2}$)     &      &  (erg cm s$^{-1}$)  &  (cm$^{-2}$)    &    & (erg cm s$^{-1}$)  &   &   \\
\hline
1 & A1& 1.76$\pm 0.02$ &  21.38$\pm 0.04$   & $\equiv$1  & 1.07$\pm 0.16$ & 21.82$\pm 0.17$ &  $\equiv$1   & 2.69$^{+0.07}_{-0.15}$  & 72.2$\pm 5.8$  & 1.25($528/423$) \\
2 & A2& 1.79$\pm 0.02$ &  21.32$\pm 0.03$   & $\equiv$1  & 1.05$\pm 0.09$ & 22.02$\pm 0.07$ &  $\equiv$1   & 2.72$\pm 0.04$  & 58.3$\pm 4.0$  & 2.14($904/423$) \\
3 & A3& 1.82$\pm 0.02$ &  21.33$\pm 0.04$   & $\equiv$1  & 1.10$\pm 0.11$ & 21.82$^{+0.13}_{-0.20}$ &  $\equiv$1   & 2.70$^{+0.05}_{-0.19}$  & 92.5$\pm 7.5$  & 1.40($593/423$) \\
\hline
4 & M1& 1.57$\pm 0.07$ &  22.71$\pm 0.06$   & 0.84$\pm 0.02$  & $<$0.66 & $>$23.1  &  0.44$^{+0.08}_{-0.16}$  & $\equiv$-1 & $<$6.6 & 1.23($520/423$) \\
5 & M2& 1.61$\pm 0.07$ &  22.17$\pm 0.03$   & 0.87$\pm 0.03$  & 0.57$^{+0.03}_{-0.16}$ & $>$23.1  &  0.19$\pm 0.02$  & $\equiv$-1 & 7.8$^{+2.3}_{-2.3}$  & 1.41($596/423$) \\
6 & M3& 1.44$\pm 0.08$ &  22.28$\pm 0.07$   & 0.78$\pm 0.04$  & $<$2.5 & $>$23.4  &  $<$0.24  & $\equiv$-1 & $<$11.5  & 1.12($473/423$) \\
7 & M4N& 1.56$\pm 0.07$ &  22.13$\pm 0.01$   & 0.87$\pm 0.01$  & $<$-0.2 & $>$23.1    &  0.21$^{+0.02}_{-0.06}$  & $\equiv$-1 & 30.8$^{+21.6}_{-2.9}$  & 1.15(1555/1348) \\
8 & M5& 1.67$\pm 0.07$ &  22.30$\pm 0.03$   & 0.94$\pm 0.02$  & 0.61$^{+0.14}_{-0.18}$ & 23.33$\pm 0.11$ &  0.31$\pm 0.09$  & $\equiv$-1 & 69.8$^{+25.3}_{-33.8}$  & 1.14($482/423$) \\
9 & M6& 1.55$\pm 0.07$ &  22.35$\pm 0.05$   & 0.88$\pm 0.03$  & 0.02$^{+0.55}_{-0.02}$ & 23.11$^{+0.18}_{-0.42}$ &  0.22$\pm 0.09$  & $\equiv$-1 & 23.3$^{+21.6}_{-12.6}$  & 1.25($530/423$) \\
10 & M7& 1.55$\pm 0.08$ &  22.27$\pm 0.05$   & 0.89$\pm 0.03$  & 0.67$^{+0.10}_{-0.26}$ & 23.30$\pm 0.13$ &  0.37$\pm 0.10$  & $\equiv$-1 & 26.01$^{+19.7}_{-13.5}$  & 1.16($492/423$) \\
11 & M8N& 1.60$\pm 0.05$ &  22.37$\pm 0.03$   & 0.87$\pm 0.02$  & 0.75$^{+0.08}_{-0.41}$ & 23.2$\pm 0.09$    &  0.44$\pm 0.05$  & $\equiv$-1 & 21.4$^{+14.3}_{-12.3}$ & 1.03(1392/1348) \\
12 & M9& 1.70$\pm 0.07$ &  22.29$\pm 0.03$   & 0.93$\pm 0.02$  & 0.77$^{+0.06}_{-0.12}$ & 23.31$^{+0.09}_{-0.12}$ &  0.33$\pm 0.08$  & $\equiv$-1 & 84.1$^{+68.3}_{-42.3}$  & 1.08($458/423$) \\
13 & M10& 1.67$\pm 0.07$ &  22.28$\pm 0.02$   & 0.94$\pm 0.02$  & 0.62$\pm 0.11$ & 23.36$\pm 0.09$ &  0.32$\pm 0.08$  & $\equiv$-1 & 127.0$\pm 48$  & 1.17($503/423$) \\
14 & M11& 1.57$\pm 0.08$ &  22.23$\pm 0.04$   & 0.91$\pm 0.03$  & 0.57$^{+0.15}_{-0.24}$ & 23.26$^{+0.11}_{-0.17}$ &  0.30$\pm 0.09$  & $\equiv$-1 & 74.8$^{+47.2}_{-26.8}$  & 1.15($488/423$) \\
15 & M12& 1.54$\pm 0.08$ &  22.29$\pm 0.05$   & 0.87$\pm 0.04$  & 0.54$^{+0.23}_{-0.54}$ & 23.28$^{+0.12}_{-0.16}$ &  0.35$\pm 0.07$  & $\equiv$-1 & 30.5$^{+19.6}_{-13.4}$  & 1.12($475/423$) \\
16 & M13N& 1.54$\pm 0.07$ &  22.35$\pm 0.04$   & 0.87$\pm 0.01$  & 0.50$^{+0.18}_{-0.81}$ & 23.2$\pm 0.09$    &  0.46$\pm 0.05$  & $\equiv$-1 & 64.2$^{+26.5}_{-14.3}$ & 1.02(1371/1348) \\
17 & M14& 1.61$\pm 0.06$ &  22.01$\pm 0.05$   & 0.89$\pm 0.02$  & 0.85$\pm 0.03$ & $>$23.39 &  0.17$\pm 0.07$  & $\equiv$-1 & 67.7$\pm 13.2$  & 1.43($605/423$) \\
\hline
\hline
\end{tabular}
\end{small}
}
\end{center}
\label{t4}
\end{table*}
%\end{landscape}

\begin{figure}[!]
\centering
%\parbox{10cm}{\vspace{-10cm}{\hspace{4cm}}
%\begin{verbatim}
%    Your text here.
%  \end{verbatim}
\parbox{10cm}{\vspace{-0.7cm}
\includegraphics[width=10cm,height=4.cm,angle=0]{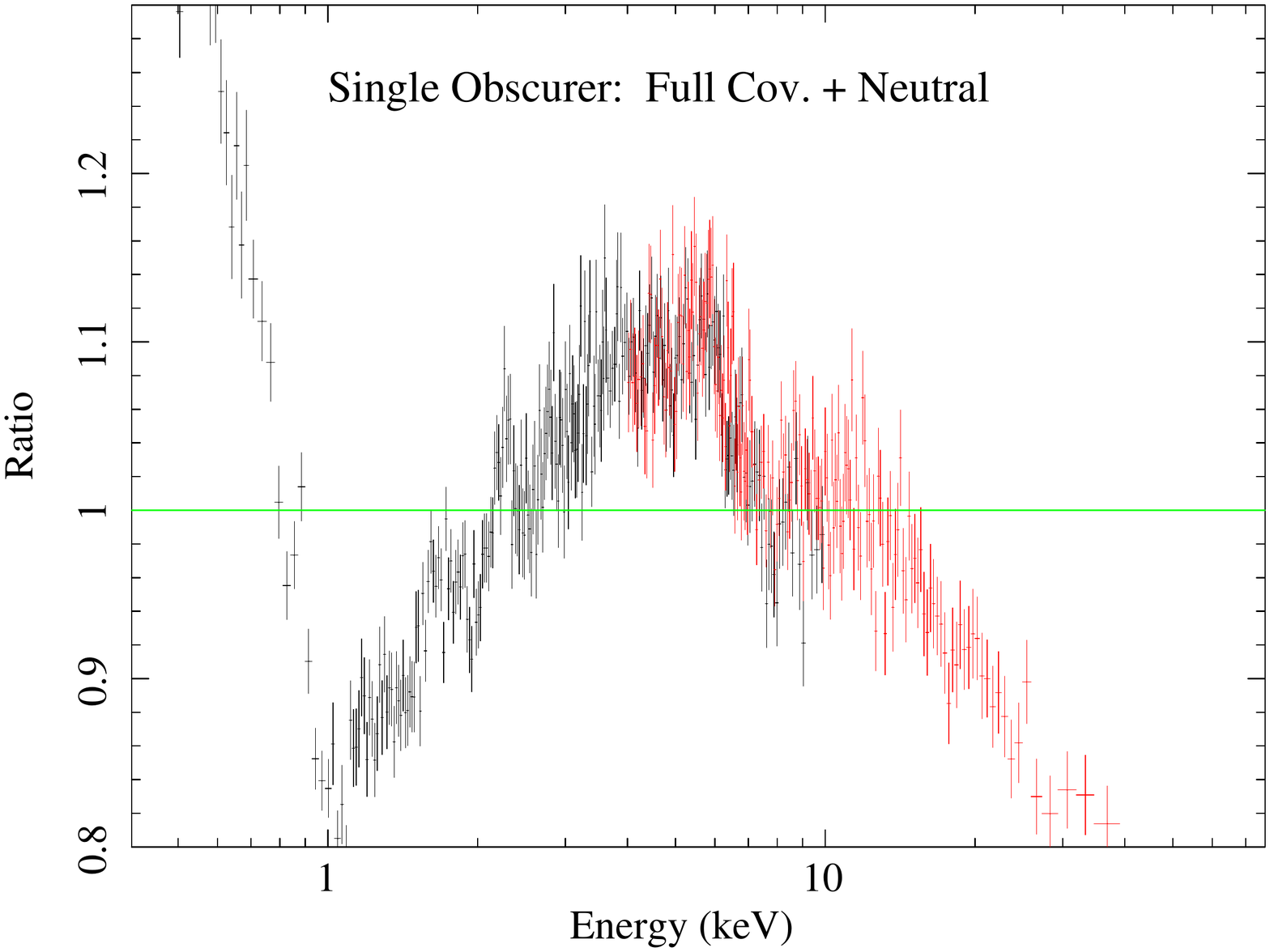}}
\parbox{10cm}{\vspace{-6.5cm} \hspace{8.2cm}a)}
\parbox{10cm}{\vspace{-6.5cm} \hspace{7cm}$\chi^2$ = 13831}
\parbox{10cm}{\vspace{-1.cm}
\includegraphics[width=10cm,height=4.5cm,angle=0]{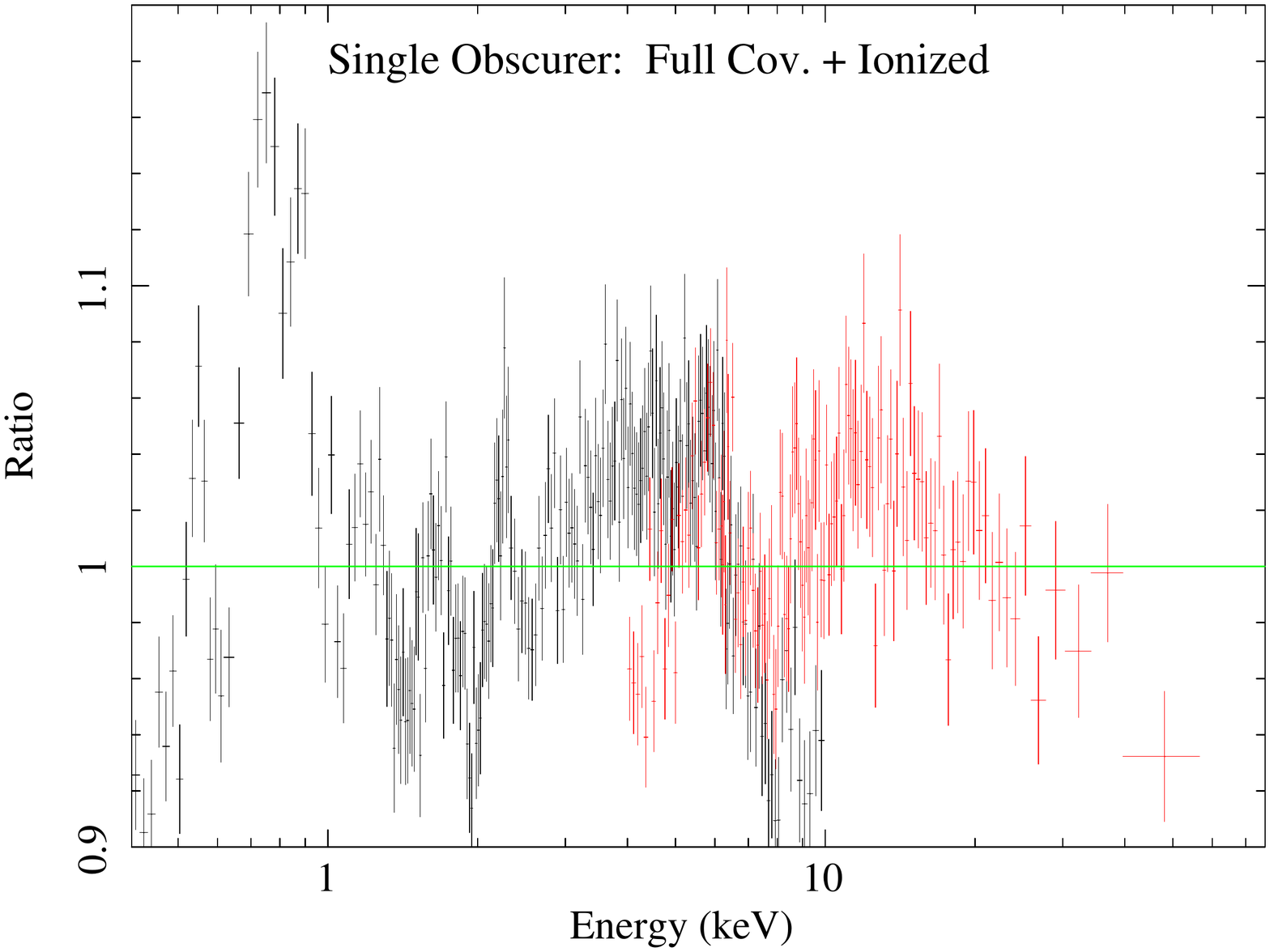}}
\parbox{10cm}{\vspace{-7.2cm} \hspace{8.2cm}b)}
\parbox{10cm}{\vspace{-7.2cm} \hspace{7cm}$\chi^2$ = 5254}
%\parbox{10cm}{\vspace{-1cm}
%\includegraphics[width=10cm,height=3.5cm,angle=0]{ratio_single_partial_neutral_abs_new.pdf}}
%\parbox{10cm}{\vspace{-5.5cm} \hspace{8.2cm}c)}
%\parbox{10cm}{\vspace{-5.5cm} \hspace{7cm}$\chi^2$ = 4514}
\parbox{10cm}{\vspace{-1cm}
\includegraphics[width=10cm,height=4.5cm,angle=0]{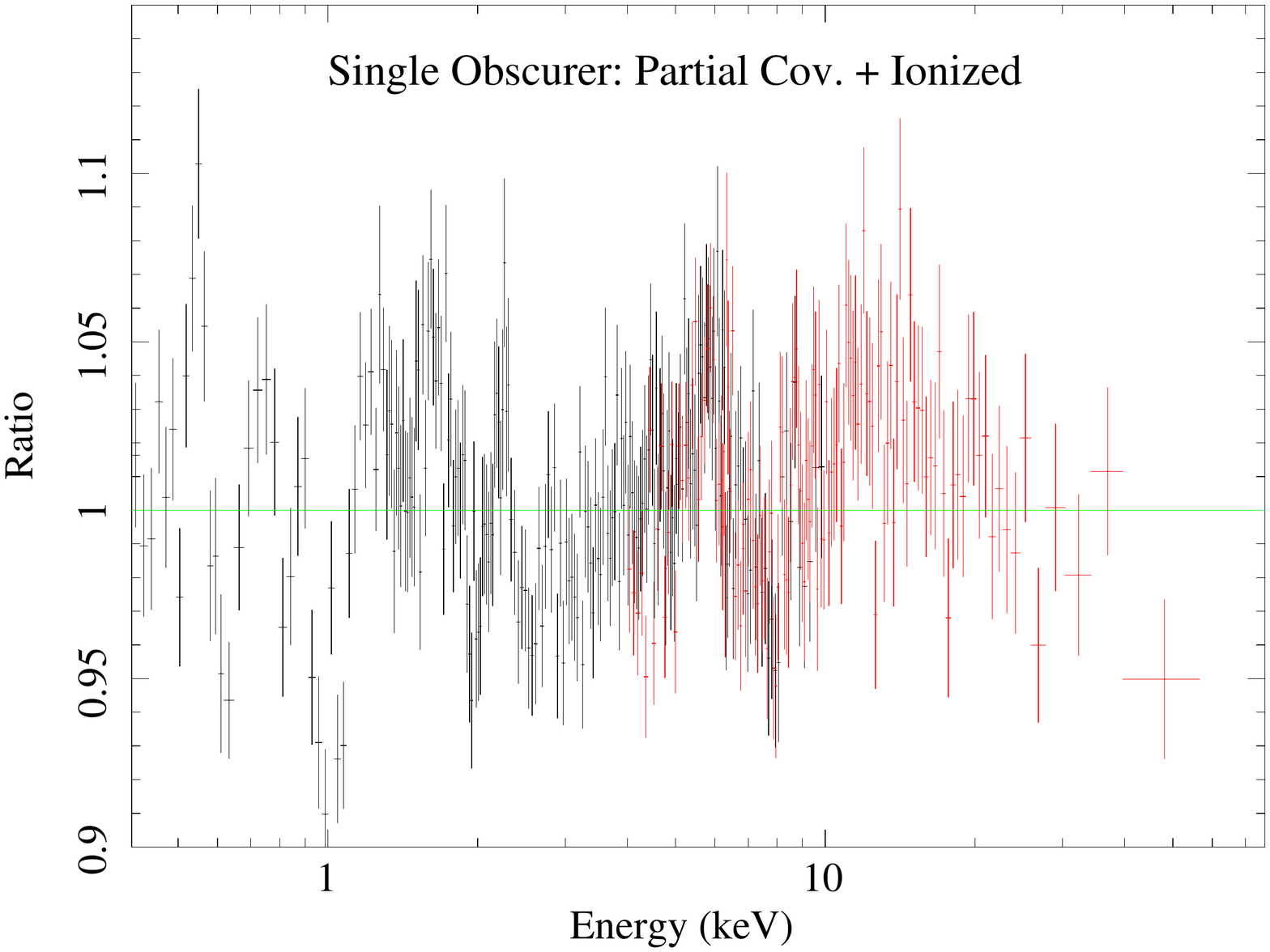}}
%\parbox{10cm}{\vspace{-4.5cm} \hspace{8cm}c)}
%\parbox{10cm}{\vspace{-0.5cm}
%\includegraphics[width=10cm,height=3cm,angle=0]{ratio_double_pcov_full_neutral.pdf}}
\parbox{10cm}{\vspace{-7.2cm} \hspace{8.2cm}c)}
\parbox{10cm}{\vspace{-7.2cm} \hspace{7cm}$\chi^2$ = 4502}
%\parbox{10cm}{\vspace{-1cm}
%\includegraphics[width=10cm,height=3.5cm,angle=0]{ratio_double_pcov_pcov_neutral.pdf}}
%\parbox{10cm}{\vspace{-5.5cm} \hspace{8.2cm}e)}
%\parbox{10cm}{\vspace{-5.5cm} \hspace{7cm}$\chi^2$ = 4318}
\parbox{10cm}{\vspace{-1cm}
\includegraphics[width=10cm,height=4.5cm,angle=0]{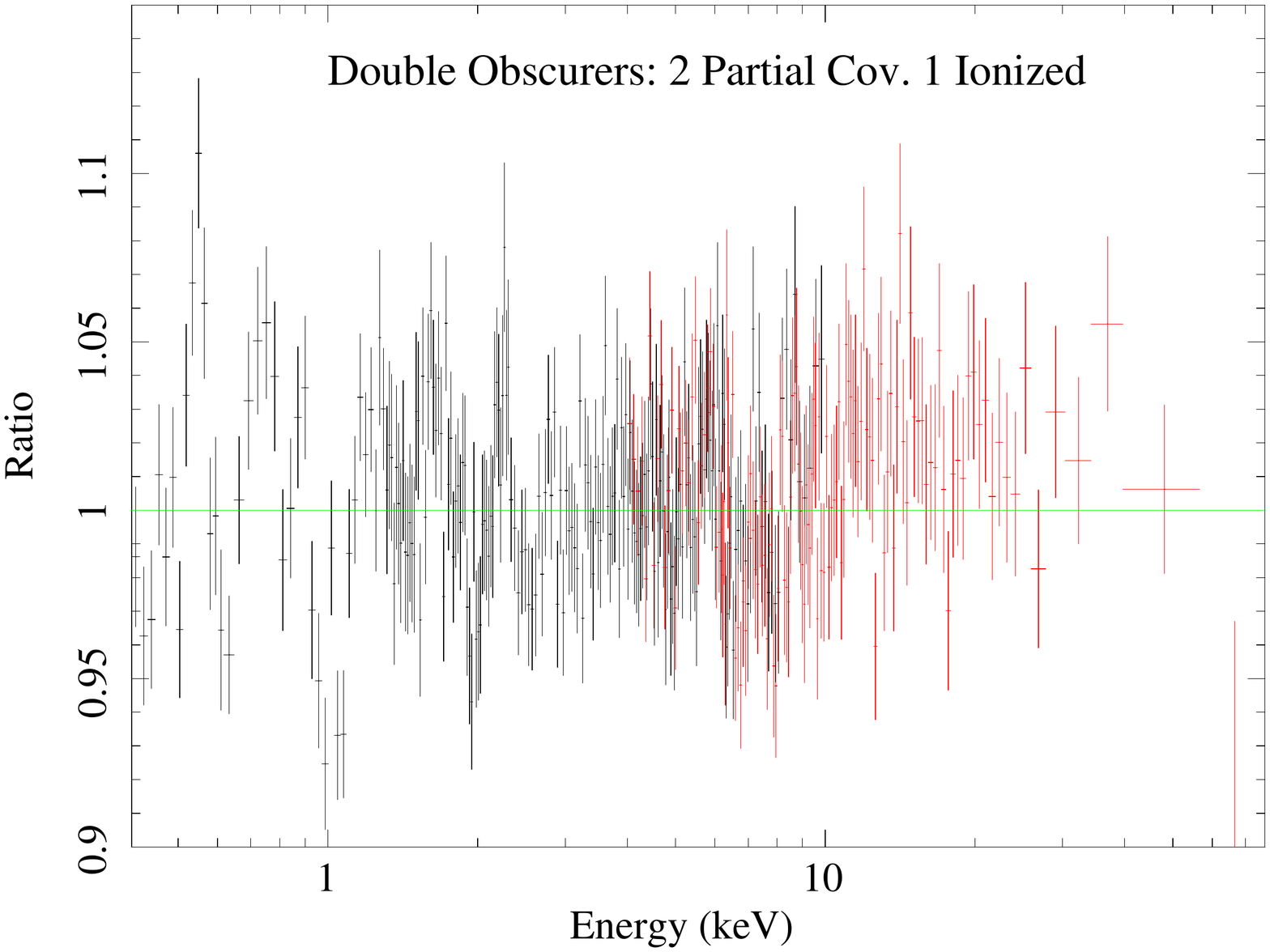}}
\parbox{10cm}{\vspace{-7.2cm} \hspace{8.2cm}d)}
\parbox{10cm}{\vspace{-7.2cm} \hspace{7cm}$\chi^2$ = 4301}
\parbox{10cm}{\vspace{-1cm} \hspace{-1cm}
\includegraphics[width=11cm,height=10cm,angle=0]{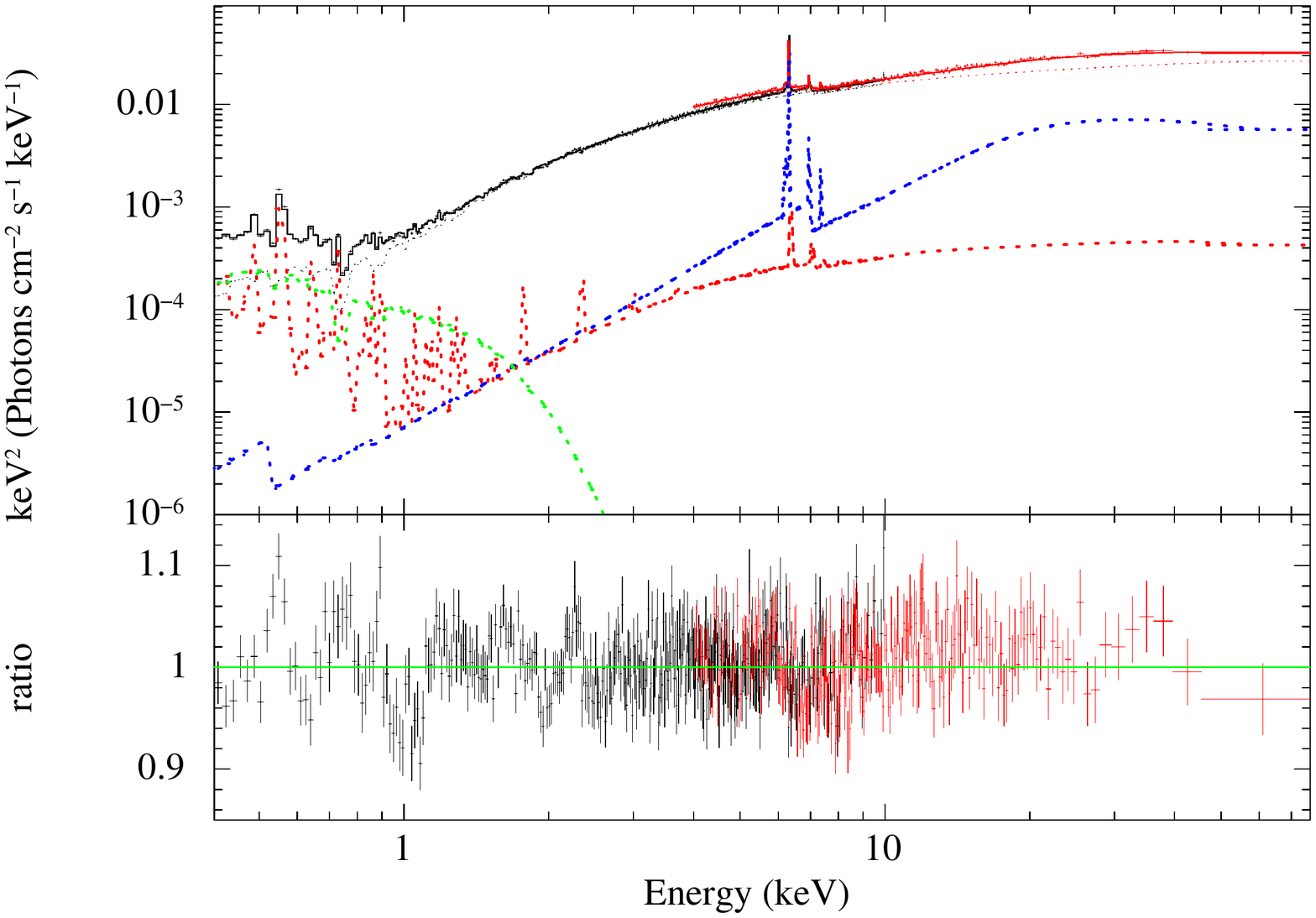}}
\parbox{10cm}{\vspace{-17cm} \hspace{8.2cm}e)}
%\parbox{10cm}{\vspace{-1cm}
%\includegraphics[width=14cm,height=9cm,angle=0]{nustar_setpgroup_eeufspec_rat_nice.pdf}}
%%\parbox{10cm}{\vspace{-8.5cm} \hspace{8cm}g)}
\parbox{9cm}{\vspace{-1cm}
\caption{\small Data-to-model ratios (panels a-d) of the 3 \emph{XMM-Newton}+\emph{NuSTAR} observations for different models (see text for details), and plot of 
the unfolded best-fit spectrum of the grouped spectra (panel e).}}
\label{f8}
\end{figure}

\subsection{Total (0.4-78 keV) X-ray band: The multilayer obscurer(s)}

\subsubsection{The Three \emph{XMM-Newton}+\emph{NuSTAR}  simultaneous observations}

We proceeded by fitting now the whole data down to 0.4 keV starting again with the three \emph{XMM-Newton}+\emph{NuSTAR} simultaneous observations only in order to obtain a best-fit model over the full 
energy band available. As for the previous analysis, we fit the 3 observations independently, but using the same model and grouping the {\it XMM-Newton} spectra together, and then the {\it NuSTAR} spectra together. 
The grouping is intended to maximize the statistical deviations (and residuals) from the adopted model, thereby helping to identify any missing model component/feature that would be 
present in all three spectra, but not included in the model. Following the results discussed above, we incorporate in our broad band model the following emission components: 

\begin{enumerate}[i)]

\item
a power-law continuum;

\item
a cold and constant reflection component (Sect. 4.1.1); 

 \item
a (soft) thermal Comptonization emission model (Sect. 4.2); 

\item
a scattered emission-line dominated component (Sect. 4.2); 

% \end{enumerate}

%This forms our baseline  broad-band emission continuum model which is also absorbed by
% 
% \begin{enumerate}[v)]
% 
\item
the de-ionized WA, as given in Paper 0 (Sect. 4.2), and;
 
% \end{enumerate}
%
% and we then add 
 
% \begin{enumerate}[vi)]
 
 \item
up to two new absorbing column densities (called  ``obscurer" components hereinafter, following the name given by Paper 0 to distinguish them from the WA component) to account for the heavy and complex obscuration 
 seen in the spectra. 
 
 \end{enumerate}

 It should be noted that the obscurer is covering only the power-law continuum plus the thermal Comptonization emission, while it is not covering either the 
 reflection continuum nor the scattered components. 
We started with one single neutral and fully covering obscurer, then left its ionization parameter free to vary, then its covering factor, and then both parameters.
We then added a second obscurer along the LOS, and repeated the procedure until a best-fit was found, checking that every additional model parameter was statistically 
required (using the F-test statistical test). Some of the data-to-model residuals obtained during this procedure are shown in Fig. 8 (panels a-e). 

Large residuals are obtained when fitting the obscurer with a single, fully covering, either neutral absorber (panel a) or an ionized one (panel b).
We made numerous attempts to fit the obscurer with a single, fully covering, ionized absorber using either a ``standard" SED for the source 
(typical of when the source was in its unabsorbed state) or an ``obscured" SED (typical of the source absorbed state during the campaign) as input for our {\tt Cloudy} table model. 
No matter the SED chosen and the wide range of parameters used in these fits ($\log \xi \sim$ 0.1--4 erg cm s$^{-1}$, $\log {\rm N}_{\rm H} \sim$20-24 cm$^{-2}$), a single fully covering ionized obscurer is clearly 
inadequate (panel b of Fig. 8) in producing simultaneously the smooth curvature below 4 keV, followed by the upward emission below 1 keV. Actually, very little difference between the two SEDs 
were recorded in fitting the pn data, in either the best-fit parameters or residuals obtained. Much better fits and residuals were instead obtained when allowing 
for the obscurer to be partially covering the source (panel c), yielding covering factors of ${\rm C}_{f,1}$ $\sim$ 0.84-0.94.

We then proceeded to add a second, independent, absorbing column density covering the same underlying continuum, and again contributing to explain the ``flat" low-energy curvature 
between 1-4 keV. The fit improvement was substantial ($\Delta \chi^{2} \sim$200) 
and we reached satisfactory fits ($\chi^{2}_{\nu} \sim$1-1.2) with a double obscurer partially covering the source, with ${\rm C}_{f,2}$ $\sim$ 0.2-0.4, and for which we let the ionization parameter of the lowest column density 
to be free to vary (Fig. 8, panel d). 
%However we reached an even better fit ($\Delta \chi^{2}$ $\sim$20) if we let the ionization parameter of the lowest column density obscurer to be free to vary. 
Best-fit values are given in Table 3 and indicate that, based on this first analysis of the 3 \emph{XMM-Newton}+\emph{NuSTAR} observations, the obscurer is better described 
by at least two different column densities, one of which mildly ionized and the other one essentially cold, which both partially cover the source (Fig. 8, panels d and e). Using the same best-fit model, we will 
then attempt to fit each of the remaining \emph{XMM-Newton} observations individually in order to understand which of the obscurer parameter(s) is driving the complex variability (Sect. 4.3.2).

We note that despite our efforts to use physically well motivated and sophisticated models (such as {\tt Cloudy}), and apply these to all sets of observational data available during the campaign 
(from the UV to the hard X-rays, i.e. papers I to VII of this series) in a consistent picture, we are left with residuals of emission/absorption line-like features below 1 keV, around 2 keV, and around 6 keV. 
Albeit being rather weak (typically a few eV equivalent width), they are statistically significant, owing to the great statistics ($>$ 1 million counts in total) 
achieved when grouping the 3 \emph{XMM-Newton}+\emph{NuSTAR} spectra. We address these one by one.
 The residuals around 2 keV are probably to be ascribed to remaining systematic calibration uncertainties due to the detector quantum efficiency at the Si K-edge (1.84 keV) and mirror effective area 
at the Au M-edge ($\sim$2.3 keV). This was also found in Papers 0, III and IV, and it was decided there to cut this part of the spectrum out, also 
to avoid inconsistencies with the RGS spectra which suffered less from such calibration effects.
Features at energies lower than $\sim$1.5 keV could be modeled by a combination of a few narrow absorption and/or emission lines at energies around $\sim$0.5-0.6 keV 
and 1-1.1 keV, and EW variable between $\sim$8-15 eV, depending on the line and observation considered. 
We estimate that the origin of these features could be ascribed to either remaining uncertainties in the CTI-energy scale at low energies in the pn data, that we know were important 
during these observations, or to an improper (or approximate) modeling of the emission and absorption lines. In the latter case, uncertainties may be ascribed to the use of the average best-fit models for the 
warm absorber and the scattered component (see Fig. 6 of Paper V), as well as a too approximate calculation for the Fe UTA atomic structures (at $\sim$ 0.7-0.8 keV, Behar et al. 2001) in the {\tt Cloudy} models.
Also a different intrinsic broadening and/or blueshift of either the warm absorber, the scattered component and/or the obscurer itself, which the current pn data does not allow to properly constrain, 
may play a role as well here. 
Lastly, the possible origin of the remaining features at around 6-7 keV will be addressed below (Sect. 4.5) after investigation of also all the remaining {\it XMM-Newton} observations.

%\subsection{Variability of the multi-layer obscurer}
\subsubsection{The Whole 17 \emph{XMM-Newton} observations: Variability of the multi-layer obscurer}

\begin{figure}[!]
\centering
\parbox{10cm}{\vspace{-0.5cm}
\includegraphics[width=10cm,height=3.7cm,angle=0]{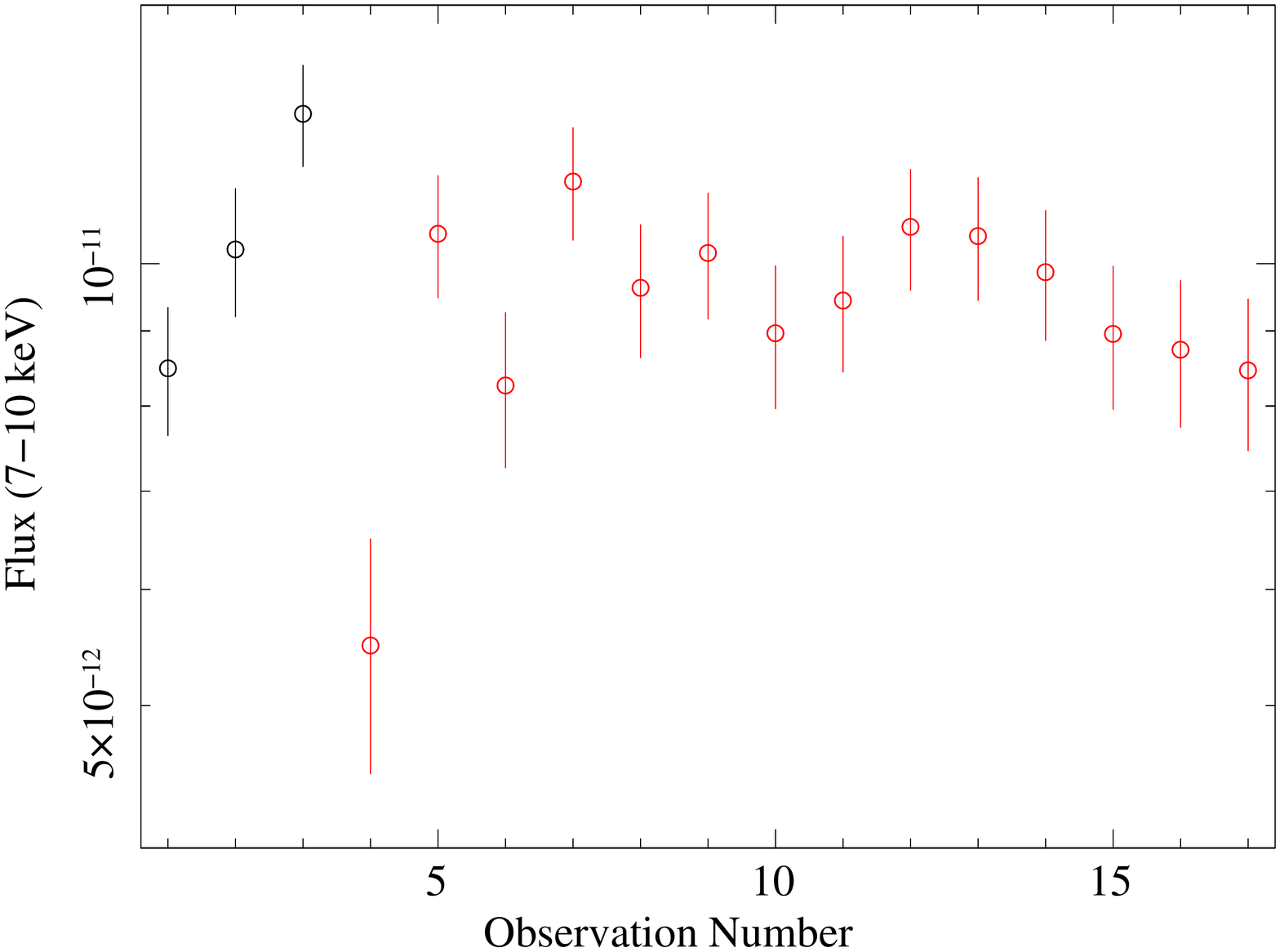}}
\parbox{10cm}{\vspace{-5.5cm} \hspace{8.2cm}a)}
\parbox{10cm}{\vspace{-0.5cm}
\includegraphics[width=10cm,height=3.7cm,angle=0]{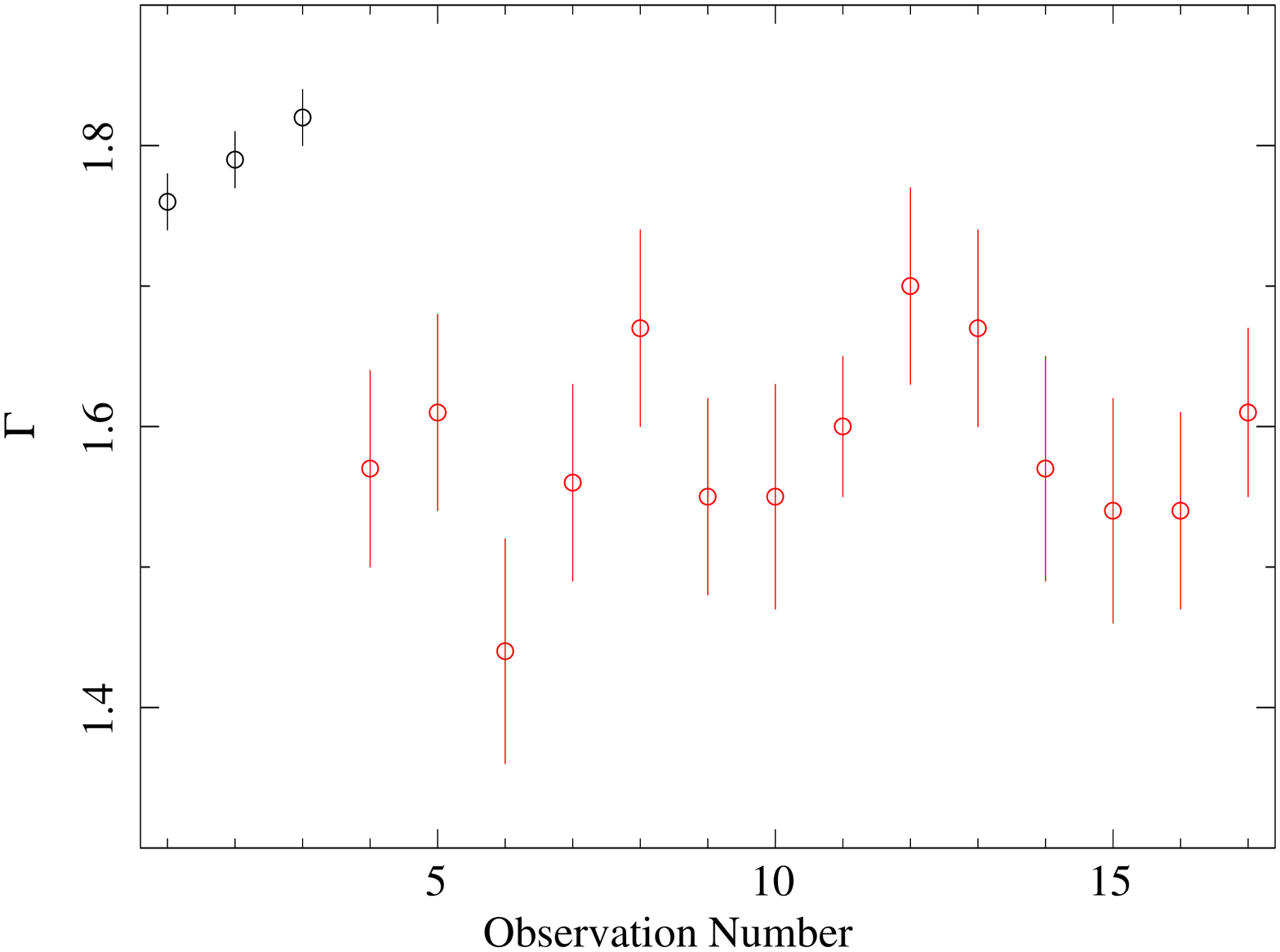}}
\parbox{10cm}{\vspace{-5.5cm} \hspace{8.2cm}b)}
\parbox{10cm}{\vspace{-0.5cm}
\includegraphics[width=10cm,height=3.7cm,angle=0]{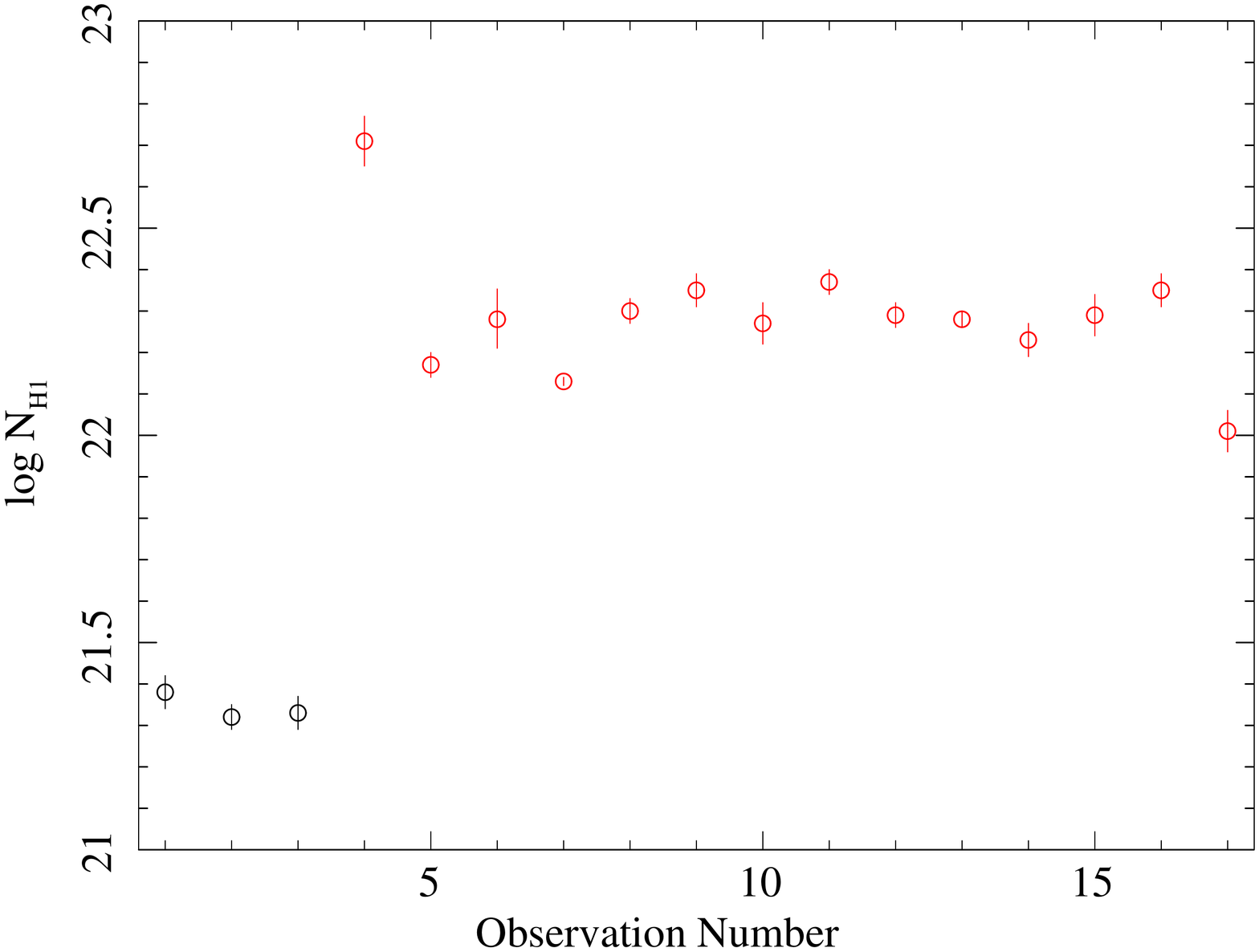}}
\parbox{10cm}{\vspace{-5.5cm} \hspace{8.2cm}c)}
\parbox{10cm}{\vspace{-0.5cm}
\includegraphics[width=10cm,height=3.7cm,angle=0]{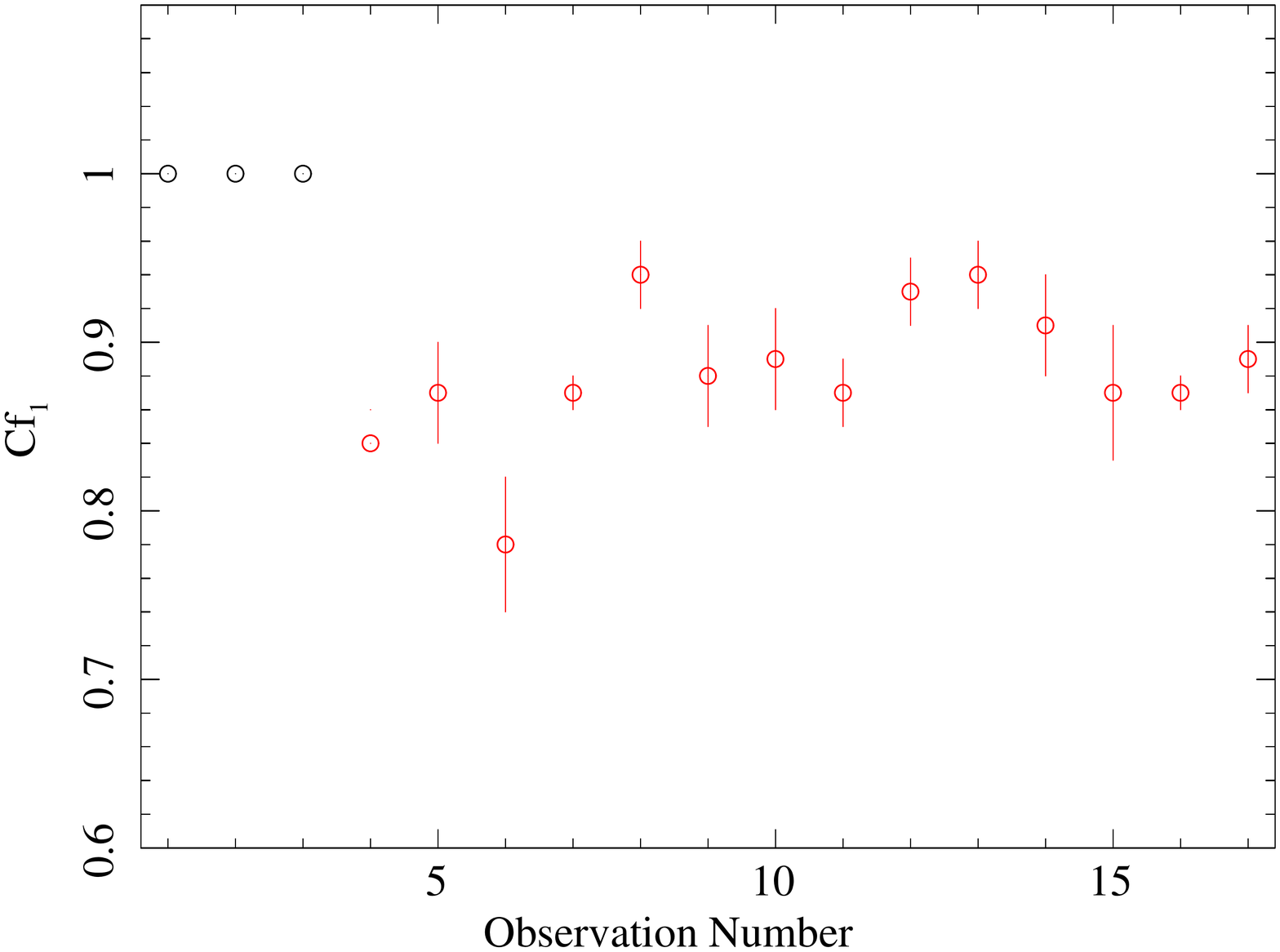}}
\parbox{10cm}{\vspace{-5.5cm} \hspace{8.2cm}d)}
\parbox{10cm}{\vspace{-0.5cm}
\includegraphics[width=10cm,height=3.7cm,angle=0]{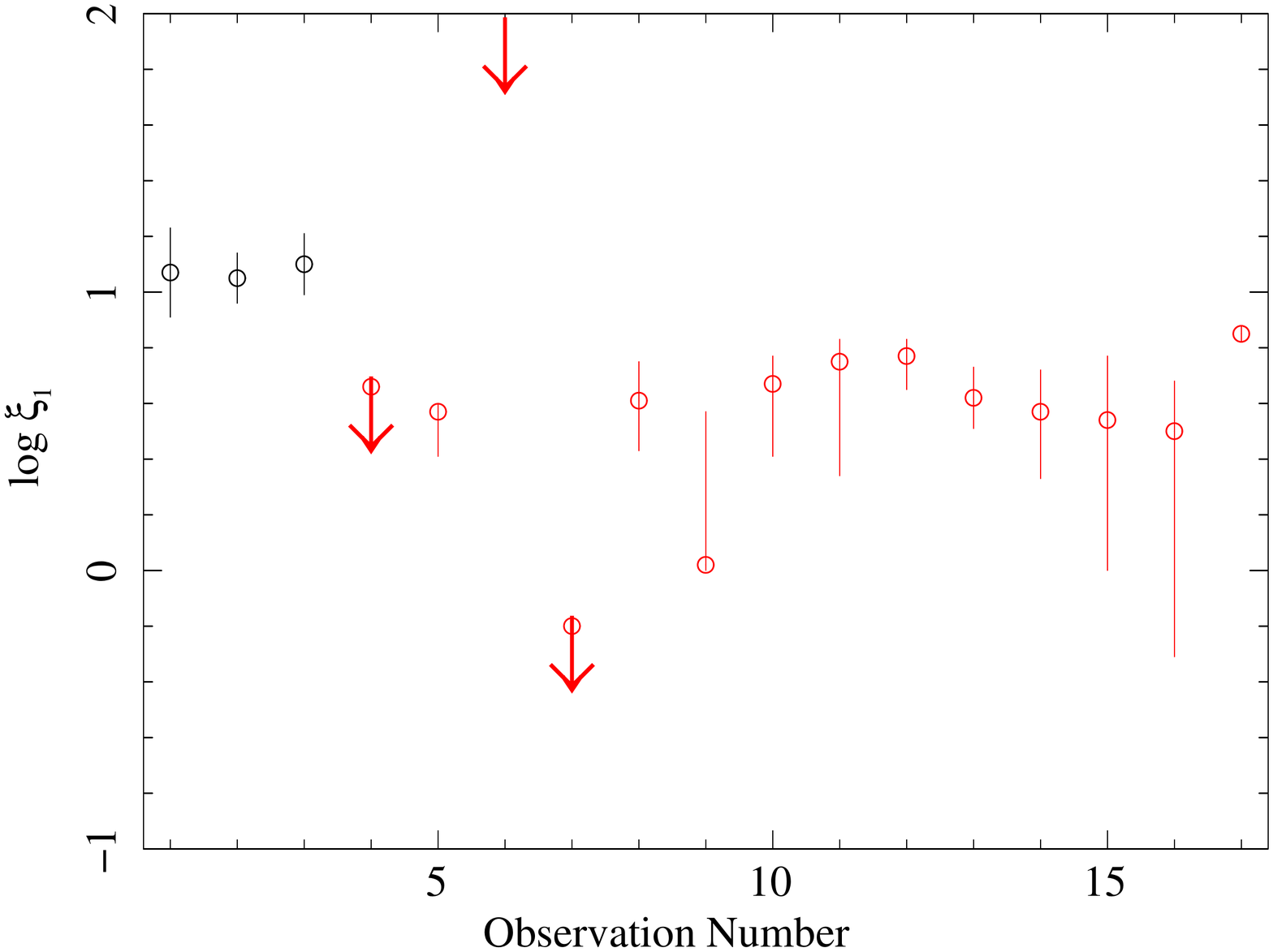}}
\parbox{10cm}{\vspace{-5.5cm} \hspace{8.2cm}e)}
\parbox{10cm}{\vspace{-0.5cm}
\includegraphics[width=10cm,height=3.7cm,angle=0]{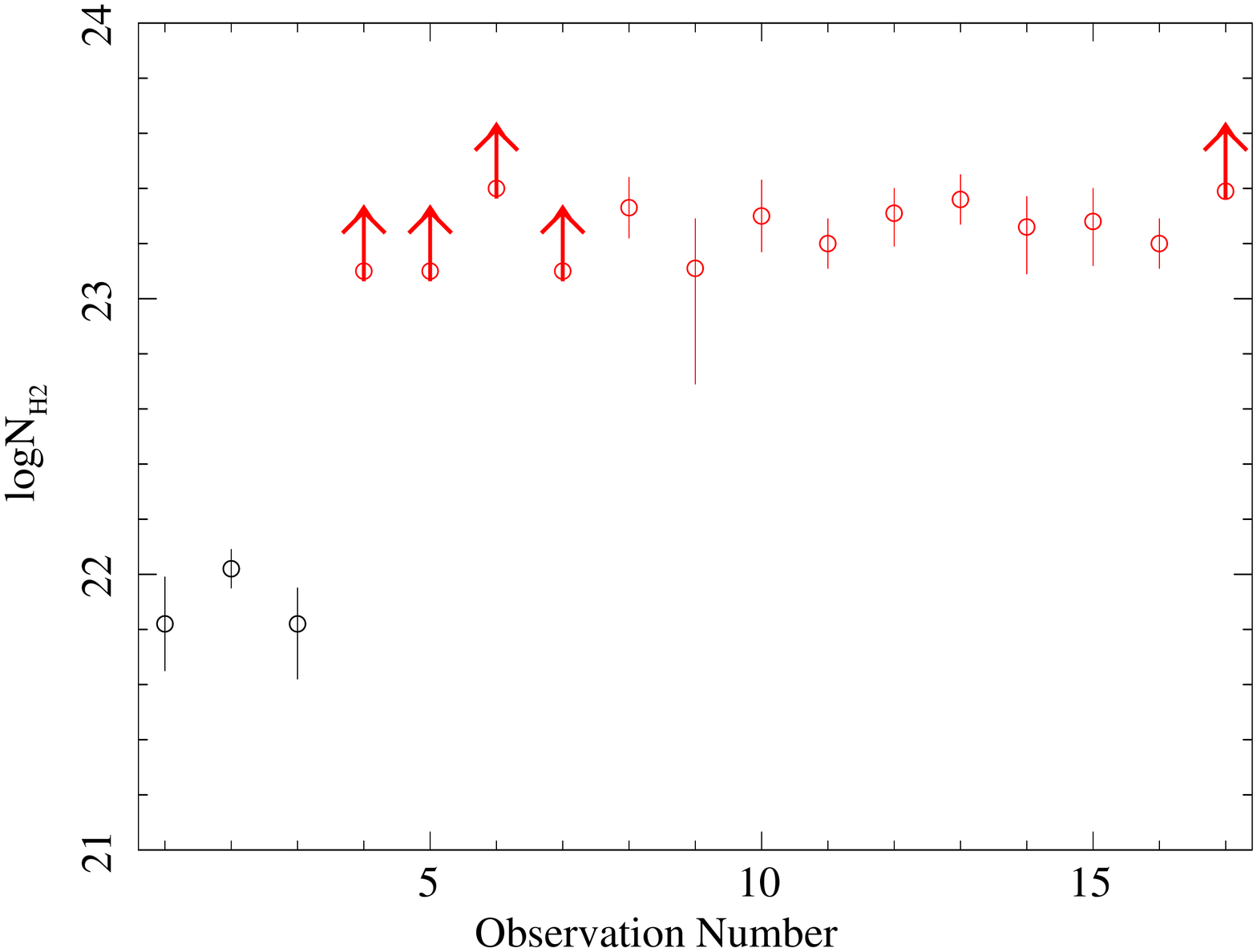}}
\parbox{10cm}{\vspace{-5.5cm} \hspace{8.2cm}f)}
\parbox{10cm}{\vspace{-0.5cm}
\includegraphics[width=10cm,height=3.7cm,angle=0]{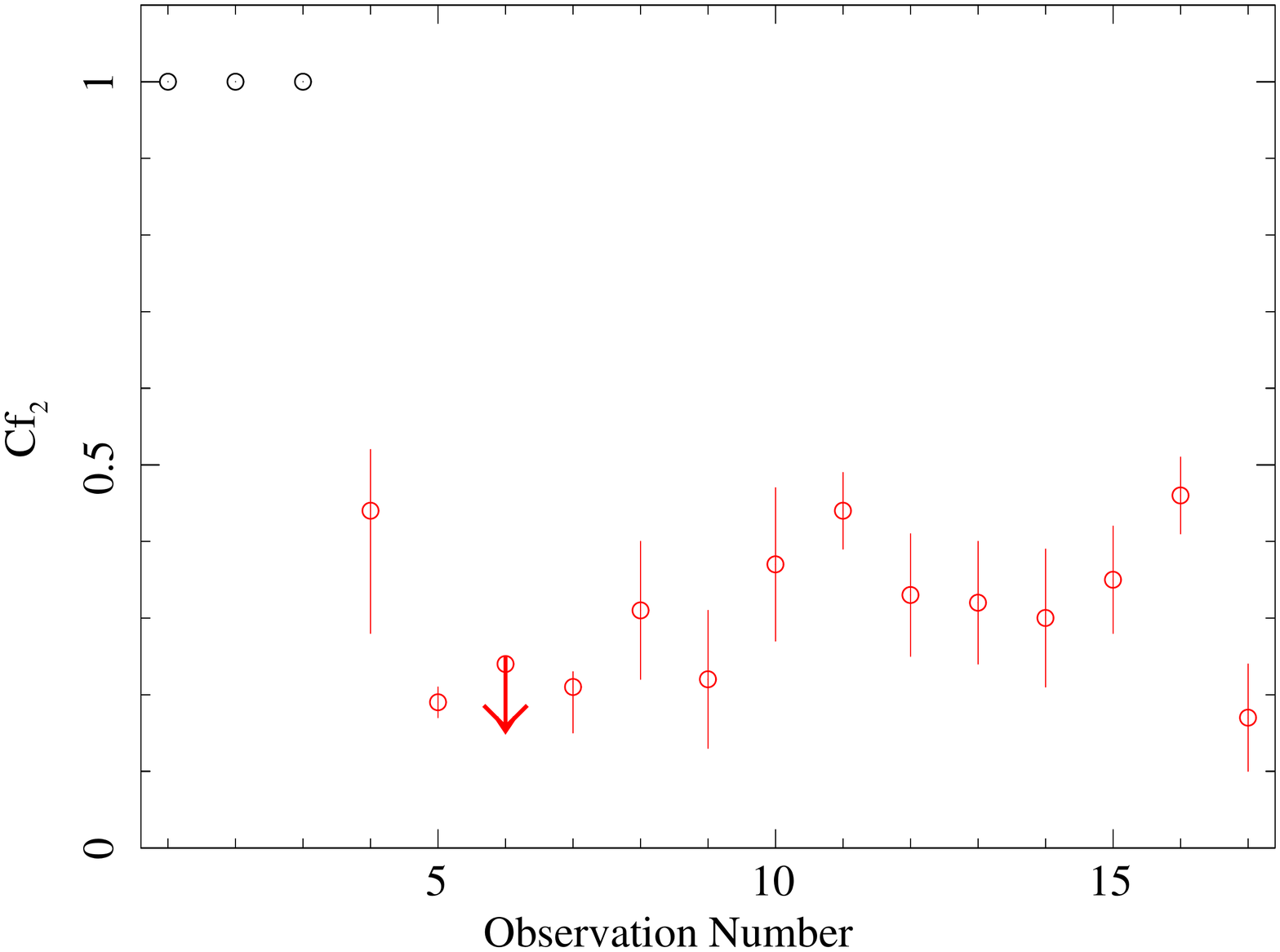}}
%\parbox{10cm}{\vspace{-5.5cm} \hspace{8.2cm}g)}
%\parbox{10cm}{\vspace{-0.5cm}
%%\includegraphics[width=10cm,height=3.7cm,angle=0]{obs_1_17_xi2_ed.pdf}}
\parbox{10cm}{\vspace{-5.5cm} \hspace{8.2cm}h)}
\caption{\small Interesting spectral fit parameters evolution versus the observation number (where, for better clarity, the archival observations are marked in black, while those from the campaign in red). From top to bottom: Power law flux between 7-10 keV, 
$\Gamma$, $\log {\rm N}_{\rm H1}$, Cf$_1$, $\log \xi_{1}$, $\log {\rm N}_{\rm H2}$, Cf$_2$, and $\log \xi_{2}$ versus the observation number.}
\label{f9}
\end{figure}

We then applied to the whole set of 17 observations the best-fit model obtained for the 3 \emph{XMM-Newton} + \emph{NuSTAR} simultaneous observations (Sect. 4.3.1), i.e. including a constant soft scattered 
component, a constant warm absorber, a constant reflection component, plus two absorbing column densities which partially obscure the primary power-law continuum plus the {\tt Comptt} soft-excess component. 
Each observation was fitted independently letting the parameters of the two ionized absorber(s), the power law (its photon index and normalization), and the normalization of {\tt Comptt} free to vary.

%TBD: WANT TO TRY TO FIX THE GAMMA AT 1.8, AND SEE IF CAN STILL FIND A BEST-FIT
%TBD: WANT TO TRY TO FIX THE NH1 AND NH2 AT A FIXED VALUE, EACH, AND LEAVE COVERING FACTOR...SEE IF STILL FIND A BEST-FIT

Best-fit values obtained for all the 17 observations are shown in Table 4, and indicate that each observation was well described by the above best-fit model, where the obscurer is a 
combination of one mildly ionized ($\log \xi_{1}$ $\sim$ 0.5-0.8), almost totally covering (Cf$_1$ $\sim$0.8-0.9) the source with a column density of $\log {\rm N}_{\rm H1}$ $\sim$ 22.2-22.7 cm$^{-2}$, 
plus one cold/neutral ($\log \xi_{2}$ always less than 0.2, thus fixed at -1) absorber with a larger column of $\log {\rm N}_{\rm H1}$ $\sim$ 23.2-23.4 cm$^{-2}$, partially covering (Cf$_2$ $\sim$0.2-0.4) the source. 
We note that during the first 3 archival observations, when the source was unobscured, the best-fit parameters of our {\tt Cloudy} models converged into a two-component 
warm absorber solution that is consistent with the values reported by the \emph{Suzaku} data (Krongold et al. 2010), and which were considered a good approximation, at low energy resolution, of the multi-temperature 
warm absorber detected in grating \emph{Chandra} and \emph{XMM-Newton} spectra (Andrade-Velazquez et al. 2010,  Steenbrugge et al. 2005, Paper VI).
During the campaign, our best-fit values are overall in agreement with the average values found by Paper 0 and the independent measurements from Di Gesu et al. (2015), except for the 
much larger value of $\xi_{1}$ found here with respect to Paper 0 and Paper IV, who found a $\log \xi_{1} =$ -1.2$\pm$0.08. 
There are multiple possible reasons for this apparent discrepancy. 
First, our analysis is performed observation-by-observation and accounts for the strong soft X-ray spectral variability, while our earlier analysis in 
Paper 0 reported the time-average values, and Paper IV fixed their ionization parameters to the average values obtained in Paper 0.
Second, we used {\tt Cloudy} and the latest results from Paper V to model the ionized absorbers/emitters, while previous analysis in Paper 0
 and Paper IV used the {\tt xabs} model in {\tt SPEX}. This may have introduced some systematic differences, 
in particular in modeling the RRC and Fe-UTA (see also Paper V). 
%Third, given our limited pn sensitivity, we did not include any intrinsic
%velocity shifts to the obscurer's absorption models. 
Finally, as mentioned above (Sect. 4.3.1), weak but statistically significant residuals are left below $\sim$ 1 keV, 
and between 1.8-2.5 keV, which may be attributed to remaining calibration uncertainties of the pn spectra. In Papers 0 and III, 
the first were fitted by adding a few emission and absorption lines in the average spectra, while the latter energy band was excluded in their analysis. 
We tested on the three \emph{XMM-Newton}+\emph{NuSTAR} spectra that adding a few ad-hoc emission and/or absorption lines would indeed contribute to decrease the ionization 
parameter $\xi_{1}$ down to values ($\sim$0.3) where it becomes rather unconstrained and degenerate with the other parameters, though 
maintaining an overall best-fit substantially unchanged. 
We thus attribute to at least one of the above reasons the apparent discrepancy in $\xi_{1}$ which should not however have any implication on the
analysis below which focuses on the variability, i.e. the relative intensity, of the most intense features measured observation by observation. 
But we stress that the absolute value of this parameter must be considered model-and calibration-dependent, thus poorly constrained, by the present analysis.

To further compare with Papers I, IV and VII, we tried also to either i) fix the intensity of the Comptonization component to those values predicted 
from the measured UV flux, and following the UV-soft-X correlation found as given in Paper VII or ii) to link any of the free parameters ($\Gamma$, 
N$_{\rm H1}$, Cf$_1$, N$_{\rm H2}$, Cf$_2$, A$_{comptt}$) listed in Table 4 to a same constant value. The fits always returned significantly worse 
statistical values (by at least $\Delta \chi^2 > 10$) in at least a few observations, supporting the need for all those free parameters. The drawback being here that in some 
observations the model is clearly over-fitting the data and yields poorly constrained parameters (e.g. during M3).

%In order to minimize the degeneracies encountered when the power-law wants a flatter index to reproduce a cold 
%and strong absorber that affects the data up to 5-7 keV, as the one found here, we let the photon index vary within its lower and higher limits found from the 
%3 \emph{XMM-Newton} + \emph{NuSTAR} broad-band fits, i.e. 1.54-1.79 (see Table 2). This assumption, less strong than fixing the photon index value at its best-fit value 
%(of $\Gamma \sim$1.7) obtained with the broad-band \emph{XMM-Newton} + \emph{NuSTAR} observations, may nevertheless introduce some systematic error, 
%from observation-to-observation, in the absolute measurements of photon index and absorption parameters. 
%They do not, however, affect strongly any relative measurement and/or variability trends/relations observed 

\begin{figure}[!]
\centering
\parbox{10cm}{\vspace{-0.5cm}
\includegraphics[width=10cm,height=6cm,angle=0]{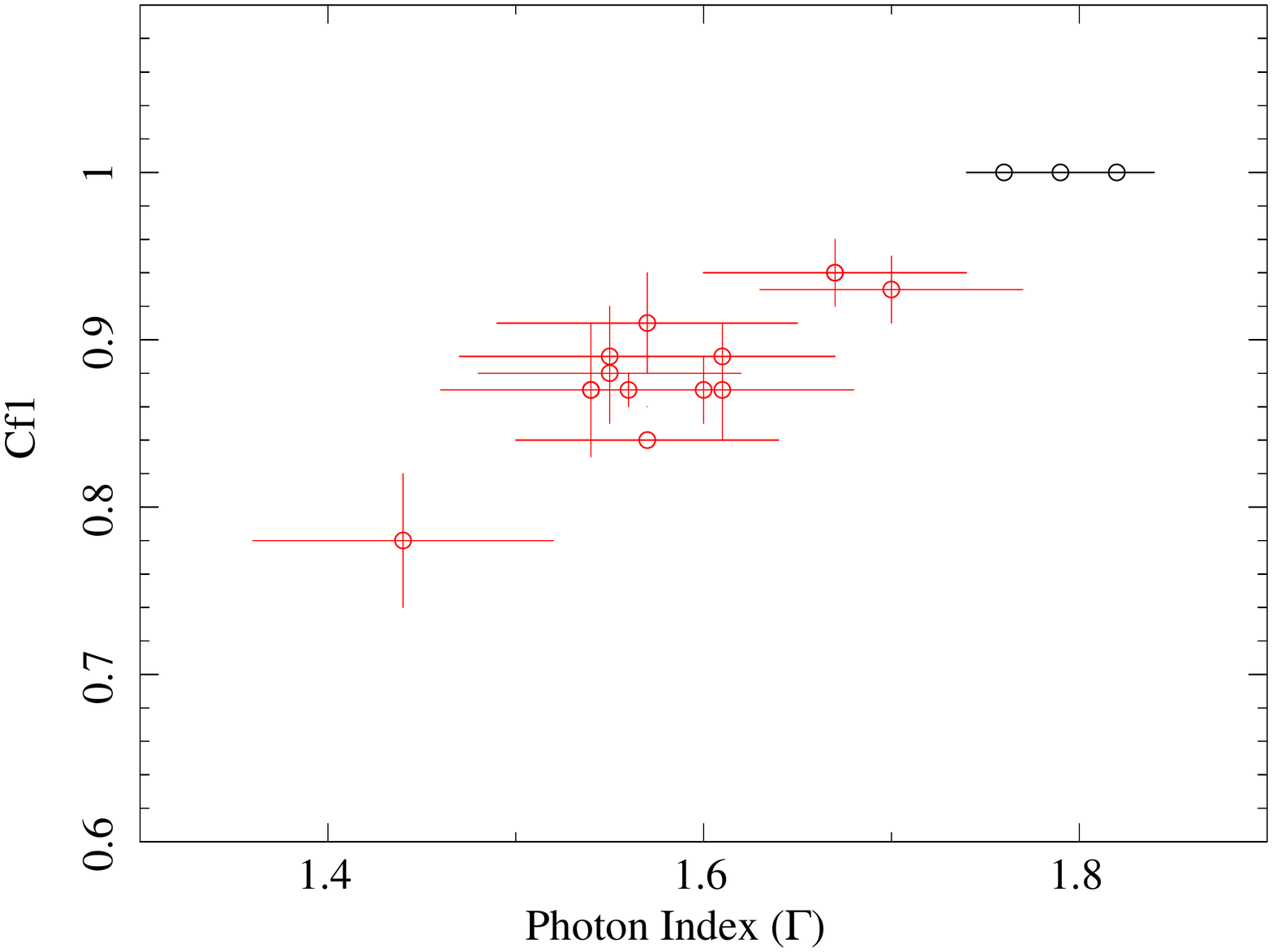}}
\parbox{10cm}{\vspace{-9.5cm} \hspace{8.3cm}a)}
\parbox{10cm}{\vspace{-0.5cm}
%\includegraphics[width=10cm,height=3cm,angle=0]{flux_7_10_nh1.pdf}}
%\parbox{10cm}{\vspace{-4.5cm} \hspace{8.2cm}b)}
%\parbox{10cm}{\vspace{-0.5cm}
%\includegraphics[width=10cm,height=3cm,angle=0]{flux_7_10_cf1.pdf}}
%\parbox{10cm}{\vspace{-4.5cm} \hspace{8.2cm}c)}
%\parbox{10cm}{\vspace{-0.5cm}
%\includegraphics[width=10cm,height=3cm,angle=0]{flux_7_10_xi1.pdf}}
%\parbox{10cm}{\vspace{-4.5cm} \hspace{8.2cm}d)}
%\parbox{10cm}{\vspace{-0.5cm}
%\includegraphics[width=10cm,height=3cm,angle=0]{flux_7_10_nh2.pdf}}
%\parbox{10cm}{\vspace{-4.5cm} \hspace{8.2cm}e)}
%\parbox{10cm}{\vspace{-0.5cm}
%\includegraphics[width=10cm,height=3cm,angle=0]{flux_7_10_cf2.pdf}}
%\parbox{10cm}{\vspace{-4.5cm} \hspace{8.2cm}f)}
%\parbox{10cm}{\vspace{-0.5cm}
%\includegraphics[width=10cm,height=3cm,angle=0]{flux_7_10_xi2.pdf}}
%\parbox{10cm}{\vspace{-4.5cm} \hspace{8.2cm}g)}
%\parbox{10cm}{\vspace{-0.5cm}
\includegraphics[width=10cm,height=6cm,angle=0]{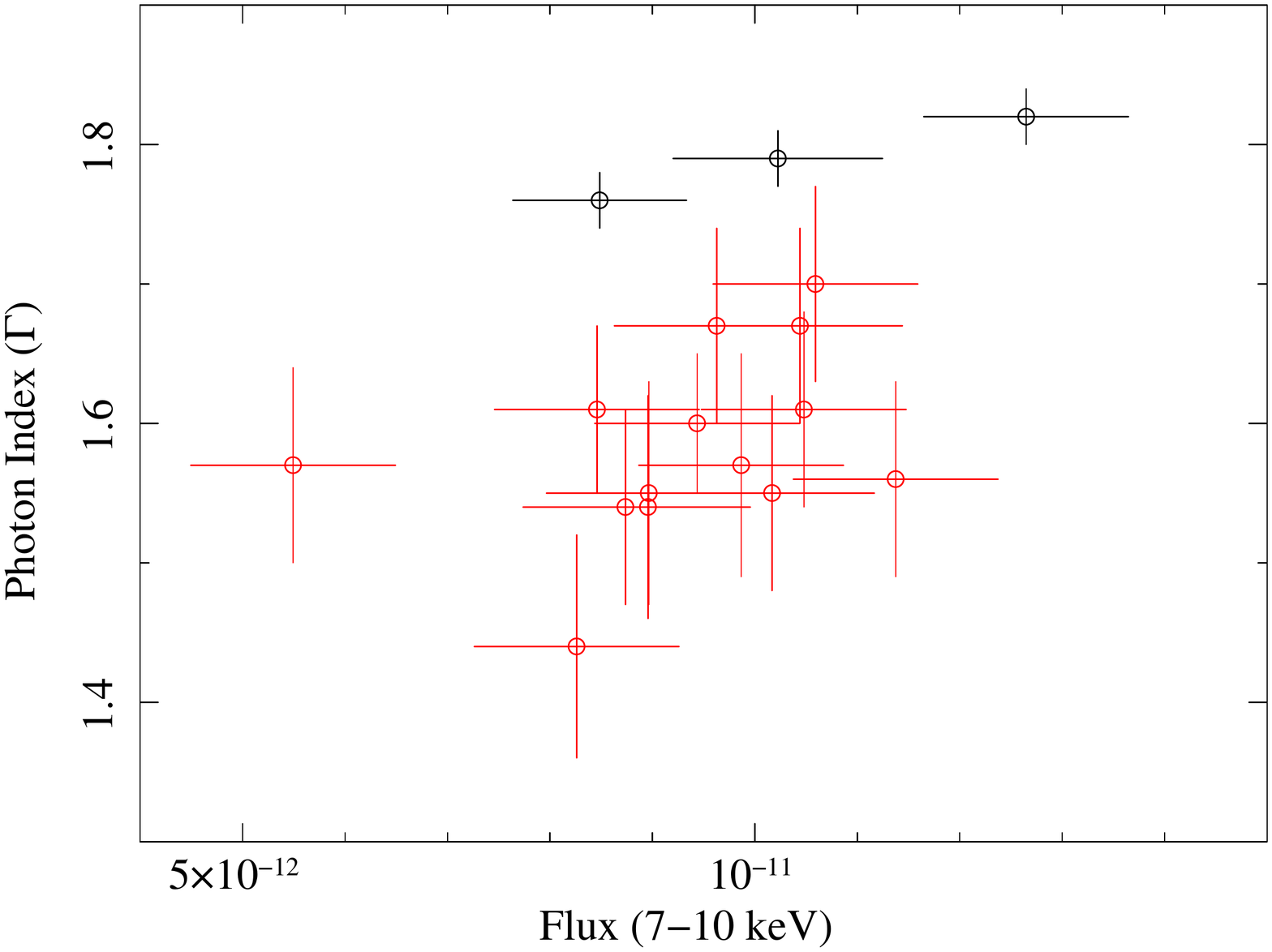}}
%\parbox{10cm}{\vspace{-4.5cm} \hspace{8.2cm}h)}
%\parbox{10cm}{\vspace{-0.5cm}
%\includegraphics[width=10cm,height=3cm,angle=0]{gamma_ed_cf2.pdf}}
\parbox{10cm}{\vspace{-9.5cm} \hspace{8.3cm}b)}
\caption{\small (Panel a) Spectral fit parameters $\Gamma$ versus Cf$_1$ (panel a) and (Panel b) $\Gamma$ versus power law F$_{7-10 keV}$ (panel b). 
Archival observations are marked in black, while those from the campaign in red.}
\label{f10}
\end{figure}

We then investigated the time evolution of all those parameters left free to vary, searching also for trends and correlations among them.
In the top of Fig.\ref{f9}, we plot the 7-10 keV flux light curve, since the flux in this energy band should have very little, if any, sensitivity to the obscuration. Interestingly, the source fluxes between 7-10 keV 
during the archival observations (first 3 observations), the average historical values (Paper VI), and during the campaign (last 14 observations) are all comparable within a factor of $\sim$2.
Moreover, except for observation M3, which shows the flattest of all photon index measurements ($\Gamma$ $\sim$1.44), all other values of the photon index are within the lower and higher limits 
(i.e. 1.54-1.79, see Table 3) found from the 3 \emph{XMM-Newton} + \emph{NuSTAR} broad-band fits, i.e. should really be indicative of the intrinsic power-law continuum underlying shape.
This readily suggests that most of the flux and spectral variability seen at lower energies can be ascribed to the obscurer itself, with only little changes in the intrinsic continuum 
flux and shape. This is confirmed in panels b) to h) of Fig. \ref{f9} which show the time variability of the most interesting best-fit parameters of the two obscurer(s) as a function of the observation number. 
This is also confirmed by our model-independent analysis and modeling of the fractional variability shown below.
Overall, we find only weak (though statistically significant) variations of the ionization parameter $\xi_1$, while $\xi_2$ is consistent with being $\sim$0 (i.e. neutral gas) since the start of the campaign, from 
observation M1. We note that given the complexity of the multi-component spectral model used here, we did not find any single parameter that could be considered to be responsible, alone, for most of the observed spectral 
variability, but several parameters combined to produce the complex spectral variability shown below.

We have then looked for trends and correlations among all the best-fit parameters listed in Fig. \ref{f9}. We find only one weak, but significant, correlation between Cf$_1$  and $\Gamma$, 
with Cf$_1$ $\simeq$ 0.5 $\times$ $\Gamma$ + 0.1 for a Pearson linear correlation coefficient of 0.94, corresponding to a chance probability of $<$10$^{-5}$ (see Fig. \ref{f10}, panel a). 
The correlation also remains significant after deleting the 3 archival observations (correlation coefficient of 0.86, chance probability value of 8$\times$10$^{-5}$).
Why would the covering factor, which is a geometrical factor, depend on the power-law intrinsic shape? This is certainly puzzling, and will be addressed later in the discussion (Sect. 5.2).
Another possibility could be though that the correlation is driven by an intrinsic degeneracy of the model parameters. A quantitative estimate of this effect would require extensive simulations involving complex 
models that are beyond the scope of this paper, but this caveat should be kept in mind. Finally, a general trend (but not a correlation) is also found (see Fig. \ref{f10}, panel b) where the source appears to be systematically intrinsically flatter during the (absorbed) campaign than during the 
archival (unabsorbed) observations. Also this point will be briefly discussed in Sect. 5.2.

\subsection{Modeling of the \emph{XMM-Newton} F$_{\rm var}$ spectrum}

\begin{figure}[!]
\centering
\parbox{10cm}{
\hspace{-0.7cm}
\includegraphics[width=10.5cm,height=7cm,angle=0]{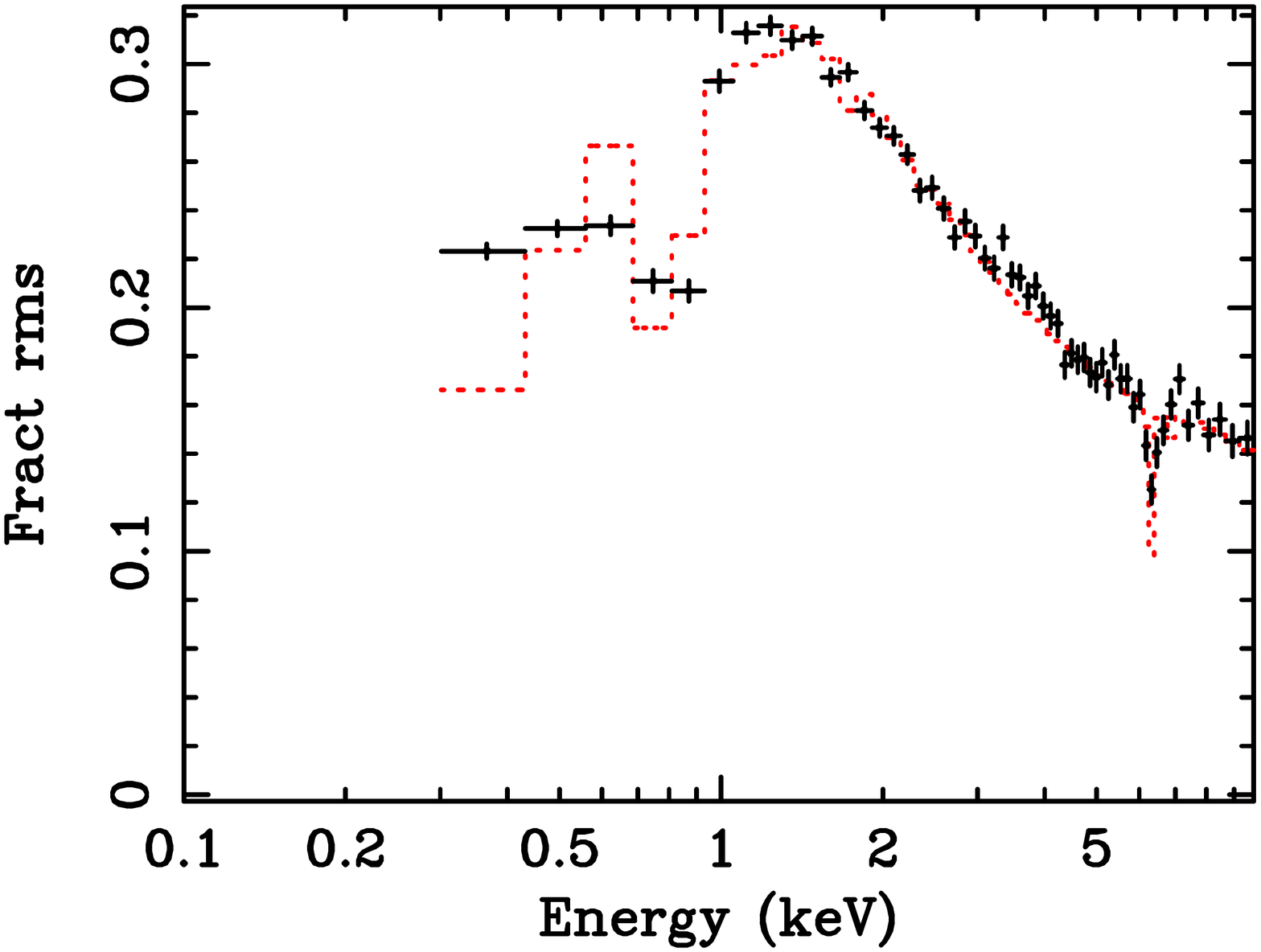}}
\caption{\small Modelling of the F$_{\rm var}$ spectrum using best-fit model and best-fit parameters of Table \ref{t4}.}
\label{f11}
%\end{figure}
%
%\begin{figure}[!htb]
\centering
\parbox{10cm}{
\includegraphics[width=9cm,height=6cm,angle=0]{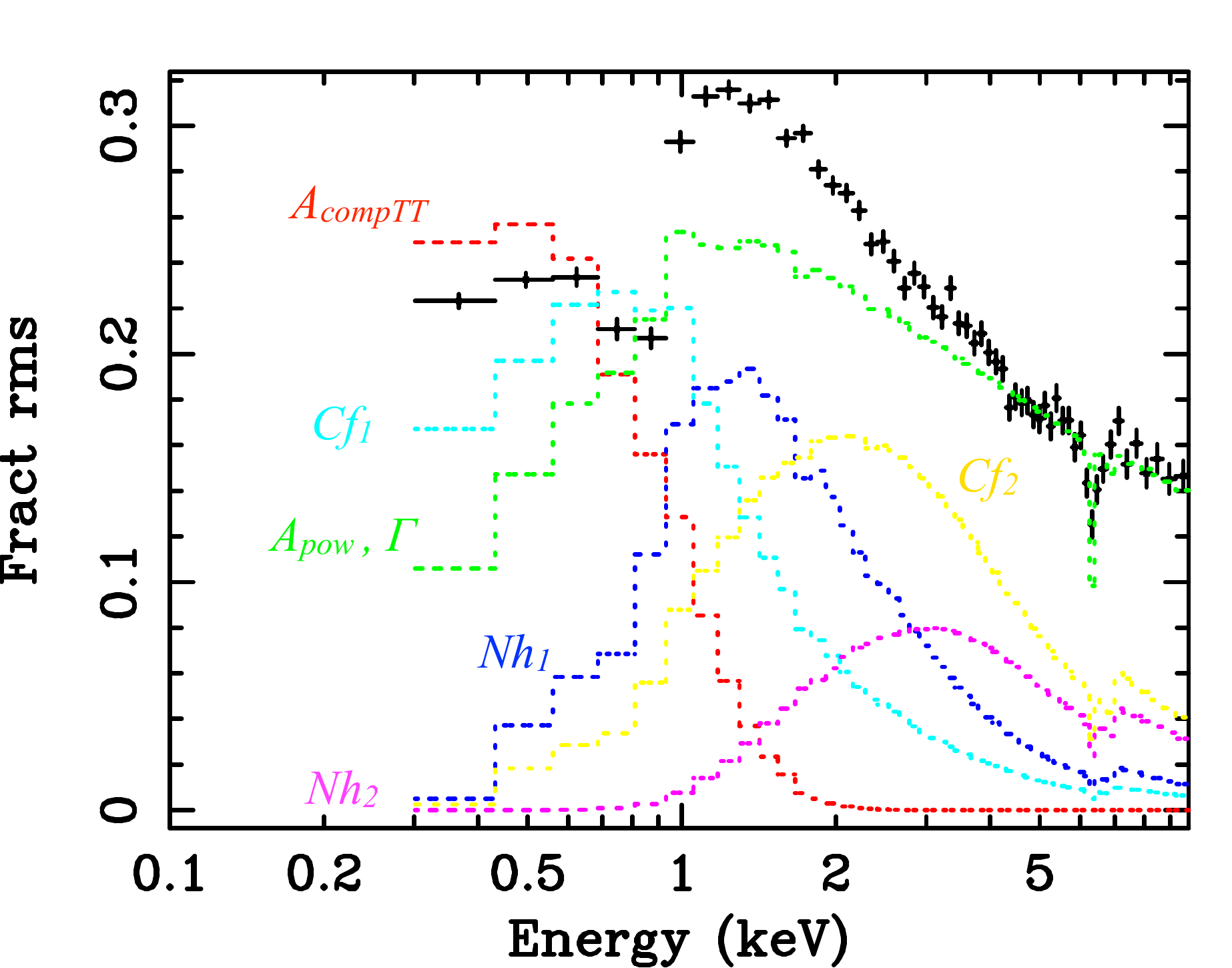}}
\caption{\small Spectral decomposition of the different model components contributing to the total F$_{\rm var}$ spectrum.}
\label{f12}
%\end{figure}
%
%\begin{figure}[!htb]
\centering
\parbox{10cm}{
\includegraphics[width=9cm,height=6cm,angle=0]{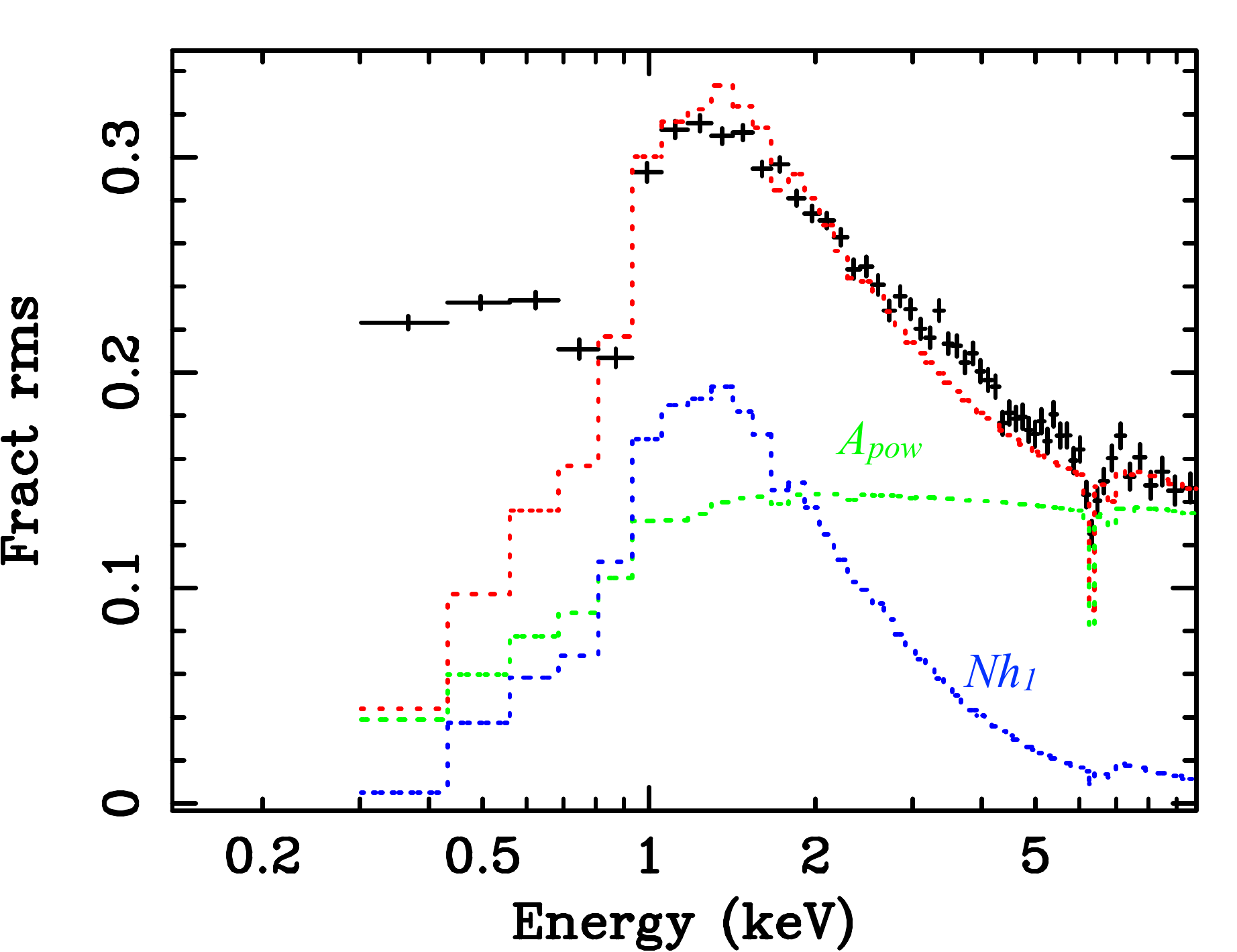}}
\caption{\small Spectral decomposition varying only the column density of the component 1 of the obscurer and the power-law normalization between 7-10 keV (see text for details).}
\label{f13}
\end{figure}

We used the best fit spectral models during the campaign and described in Sect. 4.3.2 to derive the corresponding F$_{\rm var}$ spectra (F$\rm{_{var}^{model}}$ in the following), which we compared with the observed F$_{\rm var}$ spectrum of the campaign shown in Fig. \ref{f2} and Fig. \ref{f11} in an attempt to single out the main parameters responsible for the observed spectral variability.

We first checked whether our best fit model can reproduce the F$_{\rm var}$ spectrum correctly by letting all the parameters of the model vary within the corresponding range of best fit values listed in Table \ref{t4}. The F$\rm{_{var}^{model}}$ curve is shown as a red dotted line in Fig. \ref{f11}. The good agreement between the theoretical and the observed F$_{\rm var}$ demonstrates, as expected given the F$_{\rm var}$ calculation definition, that our best fit spectral model is able to reproduce the correct flux variability at each energy. To obtain the 
F$\rm{_{var}^{model}}$ curve, the excess variance has been normalized by the total flux (which includes also the contribution from constant components) as a function of energy. This explains the net decrease of variability in the soft band and at $\sim$6.4 keV, where the constant soft scattered component (Sect. 4.2) and the narrow FeK emission line (Sect. 4.1.2) give significant contributions.\\

%We note that if all parameters were independent one with each other, and if no degeneracy was present in the fit, the addition of all curves should reproduce the total observed F$_{\rm var}$ spectrum. This is clearly not the case here, indicating some degeneracy between the parameters and/or some intrinsic correlation between them.
%IF ALL THE PARAMETERS WERE INDEPENDENT ONE WITH EACH OTHER AND THAT THERE WAS NO DEGENERACY IN THE FIT THE ADDITION OF ALL THE CURVES SHOULD REPRODUCE THE OBSERVED ONE, CORRECT? SO I THINK THAT YES WE HAVE TO SAY THAT HERE THIS IS NOT THE CASE SUGGESTING SOME DEGENERACY BETWEEN SOME OF THE PARAMETERS OR SOME REAL CORRELATION BETWEEN THEM. KEEPING THIS IN MIND WE CAN ANYWAY REMARK THAT THE NH1 AND POW_NORM FVAR CURVES WELL REPRODUCE THE HIGH ENERGY SHAPE AND THUS STRONGLY SUGGEST THAT THEY ARE THE DOMINANT SOURCE OF VARIABILITY. 

Fig. \ref{f12} shows the theoretical F$\rm{_{var}^{model}}$ spectra obtained by varying just one parameter of the best fit model (or a combination of parameters, normalization and index, in the case of the power law) and leaving all the others fixed to the best fit values of observation M7, chosen as a reference for its similarity to the average values. The variable parameter spans the 14 best fit values listed in Table \ref{t4}. 
These curves show the energy distribution of the variability power of the main parameters of the model. 
%DO WE HAVE TO SAY THAT IT IS NOT POSSIBLE TO SIMPLY ADD THESE CURVES TO OBTAIN THE OBSERVED FVAR? I AM WORRIED THAT THIS MIGHT CONFUSE THE READER, SINCE LATER WE SHOW THAT THE SUM OF THE NH1 AND POW_NORM FVAR CURVES CAN EXPLAIN THE ENTIRE FVAR..
We note that the only components contributing to the soft band variability are the normalization of the intrinsic soft excess (A$_{comptt}$), and the covering fraction of the mildly ionized obscurer(Cf$_1$). On the other hand, the main parameters contributing to the variability above $\sim$1 keV are the column density of the mildly ionized obscurer, N$_{\rm H1}$, and the power law normalization and spectral index.
%1. DO I HAVE TO LEAVE THIS PLOT AS IT IS, OR DO I HAVE TO CHANGE THE GREEN CURVE IN ORDER TO SHOW THE FVAR CORRESPONDING TO VARYING THE 7-10 KEV NORM OF THE POWER LAW ONLY, RATHER THAN THE NORM(AT 1KEV)+GAMMA?
%2. WHAT OTHER INFOS DO WE WANT TO ADD ABOUT THIS PLOT?
%Most of the variability could be attributed to the N$_{\rm H1}$ parameter, 
%which produces most of the "peak" variability between 1-4 keV and the sharp drop at $\sim$ 1 keV, but also all other parameters contributed significantly in some energy bands (see Fig.\ref{12})
%
Interestingly, the F$_{\rm var}$ obtained by varying only the N$_{\rm H1}$ parameter is characterized by a shape very similar to the observed F$_{\rm var}$, with a peak at ~1-1.5 keV and a sharp drop in the soft band. We note also that the colder 
component of the obscurer (N$_{\rm H2}$ and Cf$_2$) brings only weak variability power, and mostly concentrated between 2-4 keV.\\

We verified whether variations of the N$_{\rm H1}$ parameter, plus variations of the normalization of the power law alone (i.e. without variation of $\Gamma$, the corresponding F$_{\rm var}$ being then constant over the entire energy range), can account for most of the observed variability above $\sim$1 keV. To this aim we combined the theoretical F$_{\rm var}$ curves obtained by varying the N$_{\rm H1}$  parameter only, and the power law flux in the energy range 7-10 keV (see Fig. \ref{f13}). 
%SOME PEOPLE MIGHT OBJECT THAT IT IS NOT POSSIBLE TO COMBINE THESE CURVES BY SIMPLY SUMMING THEM, BECAUSE THE FUNCTIONAL FORM OF THIS FUNCTION IS SOMETHING LIKE 
%exp{-Nh1} A_pow E^{-Gamma}
%SO, IF I LET VARY BOTH NH1 AND A_POW, THE FVAR THAT I OBTAIN IS NOT EQUAL TO THE SUM OF THE FVAR OBTAINED BY VARYING NH1 AND A_POW SEPARATELY. 
The 7-10 keV energy range was chosen (rather than the normalization at 1 keV, as given by {\tt XSPEC} fits) so as to better constrain the intrinsic variations of the power law normalization and avoid spurious contribution from other parameters (see Fig. \ref{f12}). Fig. \ref{f13} shows that most of the observed variability at E$>$1 keV can be explained by variations of N$_{\rm H1}$  and power law normalization/flux. The residual variability in the soft band might be attributed either to the {\tt Comptt} component and/or the covering fraction of the mildly ionized obscurer. 
%GABRIELE SUGGESTED TO RIGOROUSLY QUANTIFY HOW MUCH OF THIS SOFT VARIABILITY IS DUE TO THE RESPONSE FUNTION TRANSFERRING VARIABILITY FROM E>1 KEV TO LOWER ENERGY (SEE FILE ÒRESPONSE1.PDFÓ AND ÒRESPONSE2.PDFÓ, WHICH SHOW NARROW GAUSSIANS WITH WIDTH=0, APPEARING LARGER AND WITH A RED TAIL ONCE  CONVOLVED WITH THE PN RESPONSE FUNCTION). I CAN DO IT, BUT I WAS WONDERING THAT IN THE FVAR WE SHOULD NOT HAVE THIS EFFECT SINCE WE DIVIDE BY THE TOTAL FLUX AND SO WE REMOVE THE CONTRIBUTION FROM THE RESPONSE FUNCTION..AM I RIGHT?
%
%According to these results, and in line with results from NuSTAR spectral fits, strong variations of the power law spectral index are not required...
%WHAT ELSE DO WE NEED TO SAY AT THIS POINT?

A similar model-independent analysis is presented in Paper VII for the whole {\it Swift} long-term monitoring, which do probe typically longer timescales of variability than here, i.e. 10 days up to $\sim$ 5 months.
For those periods when the {\it Swift} monitoring included also the {\it XMM-Newton} campaign, the F$_{\rm var}$ recorded by {\it Swift} was consistent in shape with the {\it XMM-Newton} one, although with significantly lower statistical quality.
{\it Swift} was not sensitive enough to constrain variations in column density, which was thus fixed at a constant value of 1.2$\times$10$^{22}$ cm$^{-2}$ as obtained from the time-averaged data presented in Paper 0, 
and most of the spectral variability was attributed to variations of the covering fraction of the obscurer only. Moreover, the signature of the N$_{\rm H1}$ variability, the very sharp ``drop" below $\sim$ 1 keV in the pn F$_{\rm var}$ spectrum, was 
not apparent in the {\it Swift} data either. Given the lower S/N data and the longer timescales probed by {\it Swift}, we consider their results in agreement with the more detailed ones presented here.

%As you mention, in our paper we also require (clear) Nh variability, while you don't need it, but as you say it is maybe a combination of different timescales (and either different physical reasons, or different contributions from the continuum variability for example), and different S/N.
%So overall I think we'll can manage to present our results as consistent, or at least understand if/where there are differences.
%I am happy if you submit the paper at your earliest convenience.

\subsection{Additional emission and absorption complexities in the Fe~K energy band}

\begin{figure}[!]
\centering
\parbox{10cm}{\hspace{-1.5cm}\vspace{-0.5cm}
\includegraphics[width=11.5cm,height=13cm,angle=0]{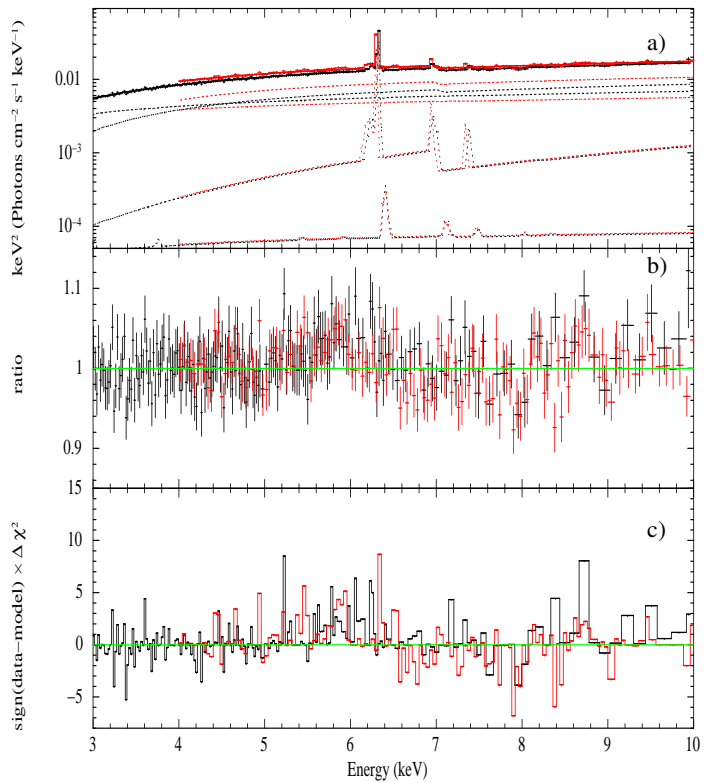}}
\parbox{10cm}{\vspace{-20.5cm} \hspace{8.2cm}a)}
\parbox{10cm}{\vspace{-15.3cm} \hspace{8.2cm}b)}
\parbox{10cm}{\vspace{-9cm} \hspace{8.2cm}c)}
\caption{\small The 3-10 keV band unfolded spectrum obtained from the spectra of the 3 $\emph{XMM-Newton}$ (black) + \emph{NuSTAR} (in red) simultaneous observations.
spectra (panel a), the data/model ratio (panel b) and the residuals in units of $\Delta \chi^2$ (panel c).}
\label{f14}
\end{figure}

\begin{figure}[!]
\centering
\parbox{10cm}{\hspace{-0.5cm}\vspace{-0.5cm}
\includegraphics[width=10cm,height=5cm,angle=0]{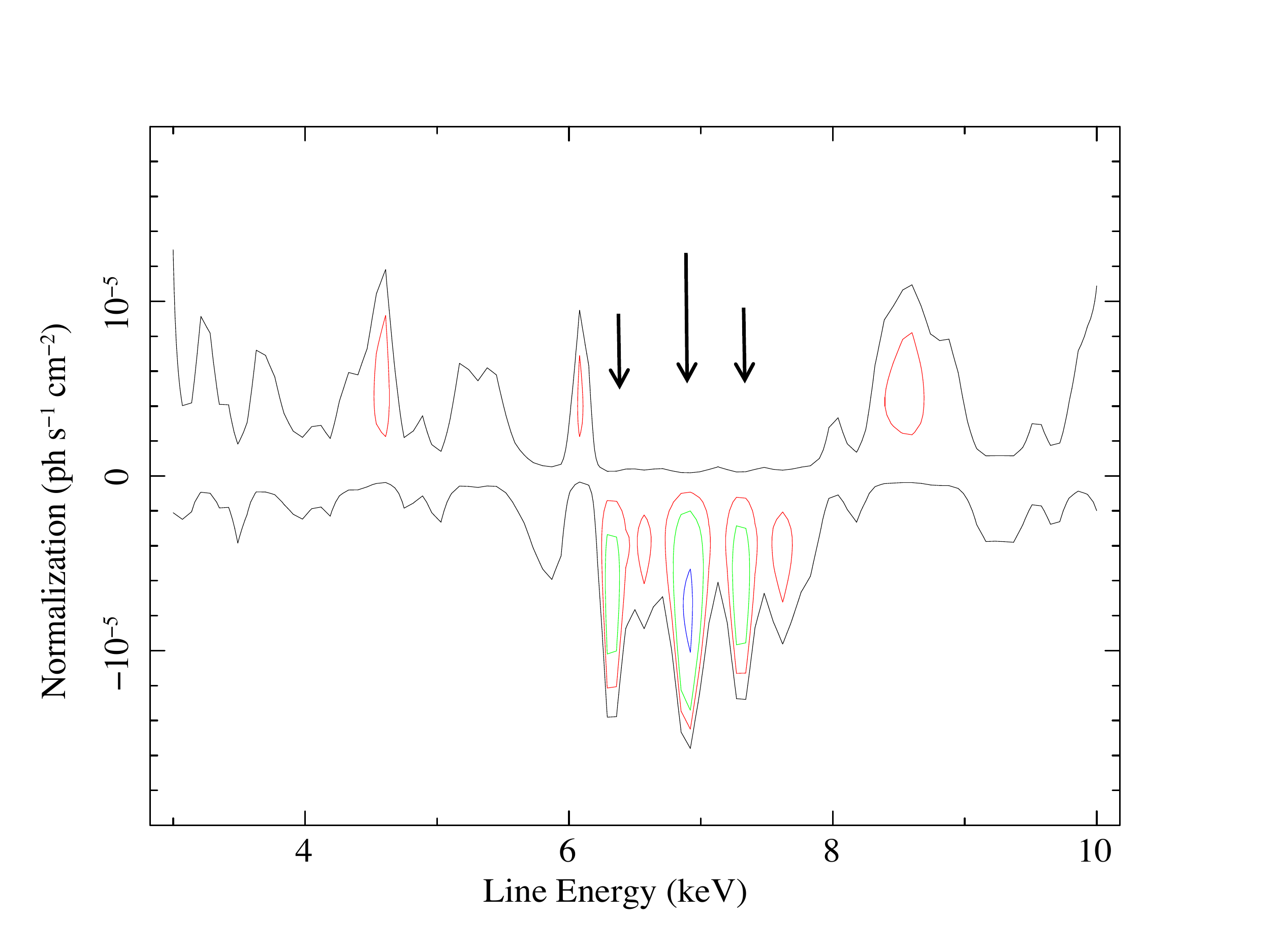}}
\parbox{10cm}{\hspace{-0.5cm}\vspace{-0.5cm}
\includegraphics[width=10cm,height=5cm,angle=0]{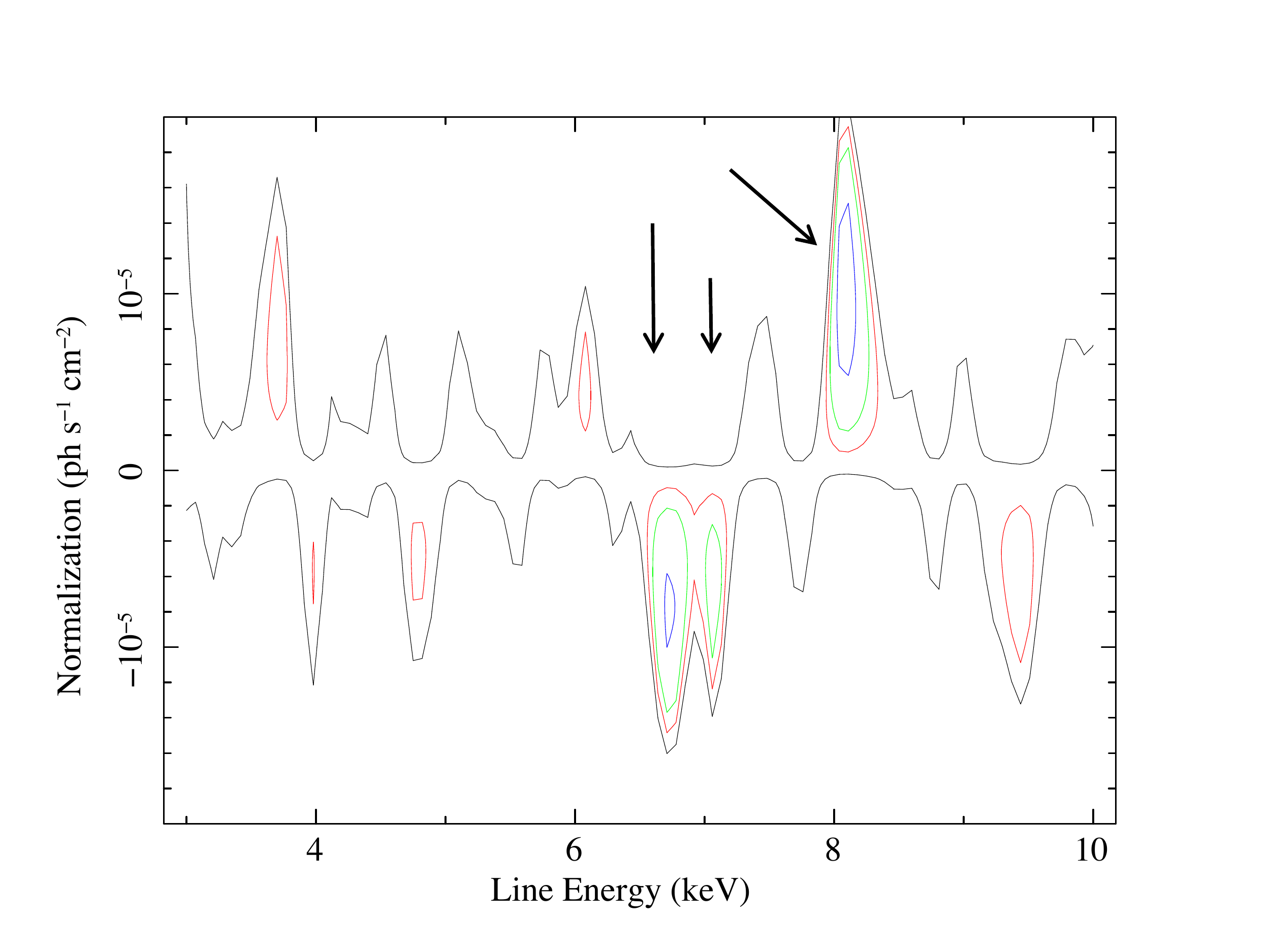}}
\parbox{10cm}{\hspace{-0.5cm}\vspace{-0.5cm}
\includegraphics[width=10cm,height=5cm,angle=0]{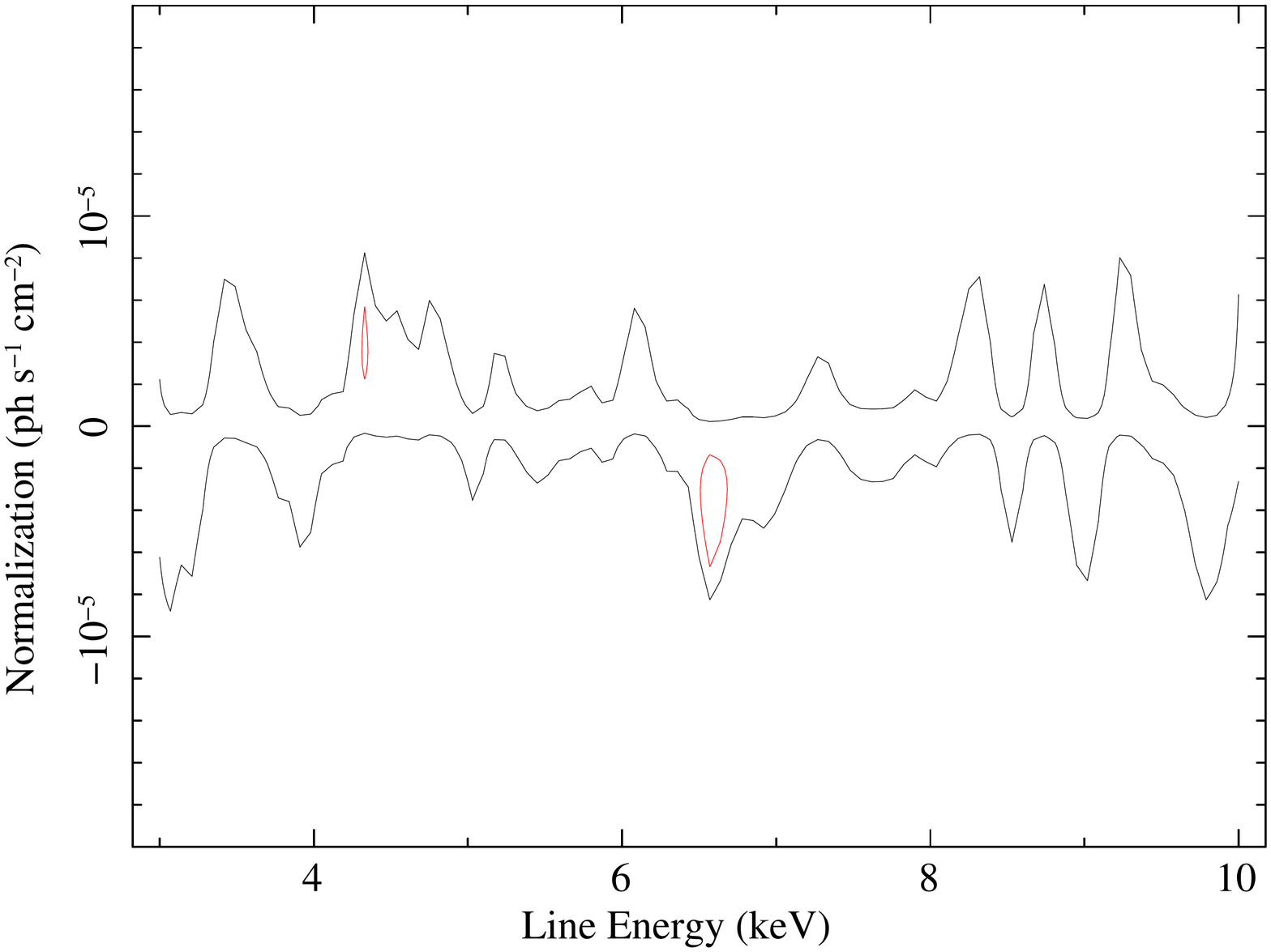}}
\caption{\small Three examples of confidence contours ($\Delta \chi^2$=+0.5, -2.3, -4.61 and -9.21 for the black, red, green and blue contours, respectively) of energy vs. intensity of a narrow emission/absorption line between 3-10 keV 
during observations M2 (Top panel), M10 (Middle panel) and M3 (Bottom panel). These illustrate the typical intensity of residuals left in the observation-by-observation 
best-fit spectra around 6-8 keV with (Top and Middle panels) and without residuals left (Bottom panel).}
\label{f15}
\end{figure}

As mentioned above (Sect. 4.3), after reaching a best-fit broad-band model of the 3 {\it XMM-Newton}  + \emph{NuSTAR} observations, we are still left with additional, albeit weak, features in the Fe~K energy band, namely one moderately broad emission line feature below 6 keV, and a set of at least two absorption features, around 6.7-6.9 keV and $\sim$ 8 keV (see Fig. \ref{f14}). 
As shown in Fig. \ref{f14}, these features are seen in both pn (black) and \emph{NuSTAR} (red), the latter having lower energy resolution but greater effective area than the pn at energies above 7-8 keV.
The 6.7-6.9 keV feature was detected also in the MOS, while at higher energies the MOS statistics are not sufficient to either confirm or disprove the 8 keV line as well. 
The features are also seen, and at the same significance, if the background is not subtracted from the source+background spectra.
F-tests indicate that both absorption lines are significant (at $>$99\%), with $\Delta \chi^{2}$ $\sim$ 25-30 and EW $\sim$15-20 eV, each.
A careful investigation of these features in the other 14 single \emph{XMM-Newton} observations indicates, however, that none of them remains always significant during the campaign. 
Absorption features between 6.4-7.1 keV are seen in $\sim$10 out of the 17 observations, while the emission below 6 keV and the feature around 8 keV are seen only in a handful of observations. 
In all cases, their EWs and statistical significance are low, typically $\lsimeq$ 20 eV and $\Delta \chi^{2}$ between 3 and 10, each. 
To illustrate further these remaining features, three examples of residuals obtained during observations M2, M10 and M3 are shown in Fig. \ref{f15}, where confidence contours 
($\Delta \chi^2$=+0.5, -2.3, -4.61 and -9.21 for the black, red, green and blue contours, respectively) are shown for a narrow emission/absorption feature ``scanned" through the best-fit spectra.

There could be several different explanations for these remaining weak features, including either a wrong, or incomplete, modeling of the complex underlying continuum which includes several emission lines in the FeK band from both the reflection and scattered components, or some weak contribution from the pn background which has some strong emission lines at high energies.
Another obvious explanation could be the presence of an additional outflowing absorption component, at such a high ionization state to contribute only/mostly with FeXXV (He-$\alpha$ and He-$\beta$) and FeXXVI (Ly-$\alpha$ and Ly-$\beta$) absorption lines. Alternatively, both these remaining blueshifted absorption features and red-shifted emission features could be signatures of a same Fe P-Cygni type emission and absorption profile originating in an outflowing, highly ionized wind
(Dorodnitsyn 2010; Hagino et al. 2015; Gardner \& Done, 2015; Nardini et al. 2015). Finally, another possibility for the 6 keV emission line could be a (yet unmodelled) contribution from a weak, and significantly redshifted and broadened reflection component. 
Overall, given the weakness and marginal statistical significance of any of these additional emission and absorption complexities, we refrain here attempting to model them with 
either a proper physical wind model (which would include the expected P-Cygni FeK line profile) or a relativistically blurred reflection model since these would go beyond the scope of this paper, 
and the quality of these datasets.

\section{Discussion}

\subsection{On the origin of the reflector}

A first estimate on the location of the Fe K emitting region can be obtained directly from the measured line width of the Fe K$\alpha$ emission line.
After proper correction for the pn CTI (which produced both energy shift and broadening of the line, see Sect. 2), we obtain an upper-limit 
of 2340 km/s on the line width, namely v$_{FWHM}$ $\lsimeq$ 5500 km s$^{-1}$. Thus, assuming Keplerian motion, R $\sim$ GM$_{BH}$/${\rm v}^2$, where we define 
the velocity width as v=$\sqrt 3\over{2}$ ${\rm v}_{\rm FWHM}$. For an estimated black hole mass of NGC5548 of M$_{BH}$ $\sim$ 3.24 $\times$10$^{7}$ M$_{\odot}$ (Pancoast et al. 2015), 
this corresponds to $R$ $>$ 1.89 $\times$ 10$^{16}$ cm (0.006 pc, or light days). We note that this limit is consistent with the upper values in the velocity 
found for the UV broad absorption line components (Paper 0), consistent with a possible common origin.

The limits obtained from the lack of variability of the line intensity, despite some weak, but significant variability in the 7-10 keV illuminating component, place even stronger constraints 
on the minimum distance of the Fe K reflector from the source. We note that the model-independent F$_{\rm var}$ spectrum clearly shows weaker line variability compared to the underlying 
continuum variability, indicating that the FeK line line does not follow the underlying continuum variations on timescales as long as $\sim$ months. This would thus imply a production site greater than a $\sim$ months light-scale
(i.e. $\gsimeq$0.026 pc), though a detailed estimate of its distance would also depend on the exact reflector geometry and on the source light-curve history which are both unknown.
This is nevertheless also consistent with the only other finding of variability of the FeK line in NGC5548 which indicated in long RXTE monitoring observations a weak but significant correlation on long- (months to years) 
timescales between the FeK line flux and the continuum intensity (Markowitz, Edelson, and Vaughan, 2003). This is also consistent with the limits placed by Liu et al. (2010) based on the lack of variability 
of the FeK line during a $\sim$ 50 days monitoring with the \emph{Suzaku} satellite performed in 2007.

\subsection{On the origin of the obscurer}

We attempt to put constraints on the location and the physical origin of the obscurer, based on its observed ionization state and variability properties.
In agreement with, and extending previous results (Papers III and VII), the present analysis (Sect. 4.3 and Sect. 4.4) provides evidence that most of the spectral variability 
can be explained by changes in column density (and to a lesser extent covering factor) occurring along the LOS, and due to transverse motions of gas across the emission region, on top of a power-law that is 
weakly variable in intensity and shape.

Assuming that the obscuring cloud(s) follow(s) a Keplerian orbit, given the measured ionization state and the variability timescale and amplitude of the absorption variations, we can estimate a rough
 location of the obscurer by using Eq. (8) in Svoboda et al. (2015) (analogous to Eq. (3) in Lamer et al. 2003). Given the BH mass in NGC5548 of 
 3.24$\times$10$^{7}$ M$_\odot$, the average ionizing luminosity between 13.6 eV and 13.6 keV of $\simeq$8.6$\times$10$^{43}$ erg/s during the campaign, the 
 sampled minimum timescale of $\sim$2 days over which we find a significant change of column density ($\Delta$$\log {\rm N}_{\rm H1}$$\simeq$21.7 cm$^{-2}$), and the ionization parameter $\log \xi \sim$ 0.6 erg cm$^{-2}$s, 
 we obtain a rough estimate of the distance of the obscuring cloud(s) of R$\approx$ 3$\times$10$^{17}$ cm $\approx$ 0.1 pc $\approx$116 light days.
Despite being recorded here for the first time in this source, this kind of ``obscuration event" may not be unique to NGC5548. Albeit with less quality data, other fast and variable obscuration events have been 
recorded in the past in a number of intermediate-type or type 1 sources, such as NGC3227 (Lamer et al. 2003), NGC1365 (Risaliti et al. 2005), NGC4388 (Elvis et al. 2004), NGC4151 (Puccetti et al. 2007), 
PG1535+547 (Ballo et al. 2008), NGC7582 (Bianchi et al. 2009), H0557-385 (Longinotti et al. 2009; Coffey et al. 2014), Mrk766 (Risaliti et al. 2011), SwiftJ2127.4+5654 (Sanfrutos et al. 2013), Mrk335 (Longinotti et al., 2013), 
NGC5506 (Markowitz et al. 2014), NGC985 (Ebrero et al. 2016), Fairall 51 (Svoboda et al. 2015) and ESO 323-G77 (Sanfrutos et al. 2016), not to mention the systematic surveys by Malizia et al. (1997), 
Markowitz et al. (2014) and Torricelli-Ciamponi et al. (2015). Based on various arguments, such as the ones above, and including the spectral properties and variability timescales of the absorbers, most authors have associated the 
origin of the absorbing clouds as either broad line region (BLR) clouds, a clumpy torus, or the inner boundary of a dusty torus. 
However there is no proof yet that the clouds probed in X-rays are actually in an orbital Keplerian equilibrium, as usually assumed, and there is no clear explanation yet on what is the physical origin 
for the structure and dynamics of the BLR clouds either. In fact, the clouds detected in X-rays, and possibly also the BLR clouds, could both be facets of 
a same phenomenon, namely a large-scale accretion disc outflow (see e.g. Emmering, Blandford \& Shlosman 1992; Elitzur \& Ho 2009; Takeuchi, Ohsuga \& Mineshige 2013; Waters et al. 2016).

The uniqueness of our findings for NGC5548 is that they support an origin of the obscurer as being part of (or within) such an outflowing accretion disc wind, and where the BLR could possibly 
be identified with the virialized and terminal part of it. The evidences in support of this hypothesis are that i) simultaneously to the above X-ray obscurer measurement, clear and strong UV broad (up to $\sim$5000 km/s) 
absorption lines were detected for the first time in this source (Papers 0 and I); ii) the obscuration event, despite being variable in time, down to a few days timescale, 
has remained systematically present since $\sim$ year 2012, i.e. for the last 4 years (as shown by the analysis of the {\it Swift} long-term light curves, Paper VII), 
and it is still present at the end of year 2015 (Kriss et al., in prep.); iii) detailed modeling of the BLR in NGC5548 (Pancoast et al. 2014) 
constrains the inclination of our viewing angle to NGC5548 to a very low value of 30$\pm$1 deg, i.e. almost face on, that makes any equatorial-only geometrical configuration 
(such as with the BLR or torus structures) difficult to reconcile with this data. This would instead favor more polarly directed components such as those found in MHD-driven outflows (Blandford \& Payne 1982; Fukumura et al. 2014), or 
vertically expanded super-critical outflows (Takeuchi, Ohsuga, \& Mineshige, 2013), although this source definitely is not a super-critical system.

Intriguingly, we find no particular property of the source during our multi-frequency campaign that could help us understand the origin, and reason, for the onset of
this long-lasting obscuration event. For example the measured luminosities at UV, soft and hard X-rays are absolutely in line with historical values for this source.
The only unusual property we found is that the photon index of the underlying power law appears to be systematically and significantly flatter compared to historical 
values ($\Gamma \sim 1.7-1.9$) for this source, even considering only the data above 7 keV where the effect of any residual absorption left should be negligible. 
This may be indicative of a particular ionization state, or physical phenomenon, that favors the formation of an outflow from the accretion disc. 
The apparent state-dependence of winds in Galactic binaries where they seem to be preferentially detected when in their (steeper) high-soft state (e.g. Ponti et al. 2012b) would suggest that this is not the case though.

We also find a significant correlation between Cf$_1$  and the photon index $\Gamma$ (ect. 4.3.2 and Fig. \ref{f10}), that would be puzzling if not related to remaining parameter degeneracies in the fit procedure.
Still, to explain it, one possibility could be that we are witnessing changes in the corona geometry which are linked to its cooling rate, and/or vice-versa. 
For example, assuming a disk-corona geometry similar to the one assumed for Mrk 509 (Petrucci et al. 2013), with an outer accretion disk and an inner corona (within r$_{cor}$). An increase of $\dot{\rm M}_{acc}$ could plausibly 
produce a decrease of r$_{cor}$ (due to e.g. disk recondensation, e.g. Meyer-Hofmeister \& Meyer (2006)), thereby reduce the size of the corona, and thus increase the effective Cf of the obscurer. At the same time, the higher $\dot{\rm M}_{acc}$ would
also increase the source UV flux,  producing an increase of cooling in the X-ray corona, and then a softening of the X-ray spectrum. This would also be consistent with the ``softer when brighter" 
behavior of this source, as previously found in Papers III and VII. In turn, the higher $\dot{\rm M}_{acc}$ would also allow to explain the higher outflow rate, as expected for most disc-wind models 
(Elitzur \& Ho 2009, Nicastro 2000), and predicts that the width of the BLR becomes narrower with respect to historical measurements (Nicastro 2000).

\section{Conclusions}

The \emph{XMM-Newton} monitoring campaign performed in the summer of 2013 revealed NGC5548 in a heavily obscured state (Paper 0). 
This obscuration was unexpected because this source, often taken as an example of a prototypical Seyfert 1 galaxy, 
had never been found strongly absorbed within the last 40-50 years of observations (Paper VI).

Here we have focused on the analysis of the EPIC pn data, the one which clearly shows the obscuration of the X-ray spectra and the three archival \emph{XMM-Newton} observations, and also combined 
the simultaneous \emph{NuSTAR} data which were available during 3 of the 17 \emph{XMM-Newton} observations. Very importantly for our analysis, we took advantage of our previous papers on this campaign, i.e. Papers 0 to VII,
to obtain a best-fit model that is consistent, and physically well motivated, taking into account all available informations from the many datasets of the campaign (including RGS data, {\it Swift}, and HST/COS).

The main results of our spectral analysis have been the following:

\begin{itemize}

\item A best-fit is found, first based on the 3 \emph{XMM-Newton}  + \emph{NuSTAR} observations, and then applied to all 17 \emph{XMM-Newton}  observations which includes: 
i) a power law continuum with $\Gamma$ $\sim$ 1.5-1.7; ii) a cold and constant reflection component that produces a narrow Fe K emission line plus Compton hump; iii)  a soft-excess modeled by thermal Comptonization and contributing only below $\sim$ 1 keV; iv) a scattered emission line component dominating the RGS energy band; v) a constant, multi temperature warm absorber, consistent with its historical values, after accounting for de-ionization due to the obscurer itself; and vi) a multilayer obscurer at mild-to-neutral ionization state, and with at least two components which partially cover the source. This obscurer is in addition 
to the multi component warm absorber that is always present in this source (Paper VI).

\item The Fe~K line properties (energy, width, equivalent width and intensity) are consistent with being part of a reflection component that is produced by a cold reflecting medium
located at distances greater than $\sim$ 0.006 pc ($\sim$ 7 light days) from the source.

\item The intrinsic high energy continuum did not vary strongly (neither in flux nor in shape) during the observations, implying that most of the soft X-ray 
variations can be ascribed to changes in the properties of the new obscurer(s) found along the LOS to NGC5548. 

\item The curvature of the continuum at lower energies and its variability clearly require a complex and intrinsically variable multi-layer absorber. 
From the analysis of the spectral variability, we find a satisfactory explanation in terms of variability of mostly its column density, and to a lesser extent covering factor.

\item Consistent with previous results, the picture that has emerged suggests the presence of a large-scale/elongated (persistent over years) but inhomogeneous absorbing structure appearing along the LOS of NGC5548.
We argue that the full (multi-frequency and multi-epoch) data favor an origin in an accretion disc wind.

\item Unfortunately, we could not constrain the velocity of the X-ray obscurer directly from the X-ray data, but its appearance simultaneous to the UV broad absorption lines 
(with v up to $\sim$5000 km/s) makes a strong case in favor of a direct physical association between the two, coherent with Paper 0.

\end{itemize}

Finally, we have also shown (Sect. 4.5) marginal evidence for additional emission and absorption features around the FeK energy band, indicating further complexities in either the reflecting or absorbing media in this source.
These could be related to an additional outflowing component, potentially at an even higher ionization state and velocity than measured here.
Given the little statistical significance of these features, summed to the complexity of the multiple component model and to the calibration uncertainties of the present dataset, 
we have not discussed these to a great extent. Overall, these observations and theoretical models have reached a level at which we are really pushing the limits to the calibration uncertainties, 
which are hard to resolve. Certainly, a better understanding of the obscurer origin and properties would benefit from more sensitive and 
higher energy resolution as foreseen with the future Athena Space Observatory. 

\begin{acknowledgements}

This paper is based on observations obtained with the \emph{XMM-Newton} satellite, an ESA funded mission with contributions by ESA Member States and USA. 
MC and SB acknowledges financial support from the Italian Space Agency under grant ASI-INAF I/037/12/P1. 
GP acknowledges support via an EU Marie Curie Intra- European fellowship under contract no. FP-PEOPLE-2012-IEF- 331095.

This work is based on observations obtained with \emph{XMM- Newton}, an ESA science mission with instruments and contributions directly funded by ESA Member States and the USA (NASA). This research has made use of data obtained with the \emph{NuSTAR}  mission, a project led by the California Institute of Technology (Caltech), managed by the Jet Propulsion Laboratory (JPL) and funded by NASA.  We thank the International Space Science Institute (ISSI) in Bern for their support and hospitality. SRON is supported financially by NWO, the Netherlands Organization for Scientific Research. M.M. acknowledges support from NWO and the UK STFC. This work was supported by NASA through grants for HST program number 13184 from the Space Telescope Science Institute, which is operated by the Association of Universities for Research in Astronomy, Incorporated, under NASA contract NAS5-26555. M.C. acknowledges financial support from contracts ASI/INAF n.I/037/12/0 and PRIN INAF 2011 and 2012. P.-O.P. acknowledges financial support from the CNES and from the CNRS/PICS. K.C.S. acknowledges financial support from the Fondo Fortalecimiento de la Productividad Científica VRIDT 2013. 
E.B. received funding from the EU Horizon 2020 research and innovation programma under the Marie Sklodowska-Curie grant agreement n. 655324, and from the iCORE program of the Planning and Budgeting Committee (grant number 1937/12).
S.B., G.M. and A.D.R. acknowledge INAF/PICS financial support. G.M. and F.U. acknowledge financial support from the Italian Space Agency under grant ASI/INAF I/037/12/0-011/13. B.M.P. acknowledges support from the US NSF through grant AST-1008882. G.P. acknowledges support via an EU Marie Curie Intra-European fellowship under contract no. FP-PEOPLE-2012- IEF-331095 and Bundesministerium f\"ur Wirtschaft und Technologie/Deutsches Zentrum f\"ur Luft- und Raumfahrt (BMWI/DLR, FKZ 50 OR 1408). F.U. acknowledges PhD funding from the VINCI program of the French-Italian University. M.W. acknowledges the support of a PhD studentship awarded by the UK STFC.

\end{acknowledgements}

{}

\end{document}